\documentclass[english,12pt]{article}
\usepackage{authblk}

\title{Large $N$ Renormalization Group Flows in 3d $\mc{N}=1$ Chern-Simons-Matter Theories}
\author{Ofer Aharony and Adar Sharon}
\affil{Department of Particle Physics and Astrophysics, Weizmann Institute of Science, Rehovot 7610001, Israel}
\usepackage[titletoc,title]{appendix}
\usepackage{enumitem}	
\usepackage{mathtools}
\usepackage[width=1.05\textwidth]{caption}
\usepackage{subcaption}
\usepackage{mathrsfs}
\usepackage{amsmath}

\numberwithin{equation}{section}
\makeatletter
\let\runtitle\@title
\let\runauthor\@author
\makeatother

\usepackage{hyperref}
\hypersetup{
	colorlinks,
	citecolor=blue,
	filecolor=blue,
	linkcolor=blue,
	urlcolor=blue,
	linktocpage=true
}

\usepackage{amsmath}
\usepackage{mathptmx}
\usepackage{amsthm}
\usepackage{amssymb}
\usepackage{amsfonts}
\usepackage{cancel}
\usepackage{verbatim}
\DeclareMathAlphabet{\mathcal}{OMS}{cmsy}{m}{n}

\usepackage[bottom=1in,top=1in,left=0.7in,right=0.7in]{geometry}

\usepackage{graphicx}

\usepackage[rgb,dvipsnames]{xcolor}

\newcommand\mb[1]{\mathbb{#1}}
\newcommand\mc[1]{\mathcal{#1}}
\newcommand{\tr}{\text{Tr}}

\newcommand{\gra}{\alpha}

\newcommand{\grb}{\beta}

\newcommand{\grd}{\delta}
\newcommand{\grD}{\Delta}
\newcommand{\gre}{\epsilon}

\newcommand{\grF}{\Phi}

\newcommand{\grl}{\lambda}
\newcommand{\grL}{\Lambda}

\newcommand{\grS}{\Sigma}

\newcommand{\grk}{\kappa}
\newcommand{\grw}{\omega}

\newcommand{\beq}{\begin{equation}}
\newcommand{\eeq}[1]{\label{#1}\end{equation}}
\newcommand{\ber}{\begin{eqnarray}}
\newcommand{\eer}[1]{\label{#1}\end{eqnarray}}

\newcommand{\al}{\alpha}
\newcommand{\be}{\beta}
\newcommand{\ga}{\gamma}
\newcommand{\G}{\Gamma}
\newcommand{\de}{\delta}
\newcommand{\ta}{\theta}

\newcommand{\pa}{\partial}
\newcommand{\na}{\nabla}

\newcommand{\ab}{{\alpha\beta}}

\newcommand{\then}{~~~\Rightarrow~~~}

\newcommand\dpi[1]{\frac{d^3 {#1}}{(2\pi)^3}}

\usepackage{bbm}

\newcommand\wt{\tilde{\omega}}

\usepackage[backend=bibtex,sorting=none,style=numeric-comp,url=false]{biblatex}
\renewbibmacro{in:}{}
\usepackage{babel}
\bibliography{refs}{}

\begin{document}

\maketitle

\abstract{
We discuss $3d$ $\mc{N}=1$ supersymmetric $SU(N)$ and $U(N)$ Chern-Simons-matter theories, with $N_f$ matter superfields in the fundamental representation of $SU(N)$ or $U(N)$. In the large $N$ \rq t Hooft limit with fixed \rq t Hooft coupling $\lambda$ these theories have one (for $N_f=1$) or two (for $N_f > 1$) exactly marginal deformations in the superpotential. At finite $N$ these couplings acquire a beta function. We compute the beta function exactly for $\lambda=0$, at leading order in $1/N$. For $N_f=1$ we find four fixed points, one of which is triply-degenerate. We show that at large $N$ there are at most six fixed points for any $\lambda$, and conjecture that there are exactly six, with three of them stable (including a point with enhanced $\mc{N}=2$ supersymmetry). The strong-weak coupling dualities of $\mc{N}=1$ Chern-Simons-matter theories map each of these fixed points to a dual one. We show that at large $N$ the phase structure near each of the three stable fixed points is different. For $N_f>1$ we analyze the fixed points at weak coupling, and we work out the action of the strong-weak coupling duality on the marginal and relevant superpotential couplings at large $N$ (which was previously known only for $N_f=1$). In addition, we compute in these theories the 2-point and 3-point functions of the lowest gauge-invariant singlet superfield at large $N$, for all values of $\lambda$ and of the superpotential couplings, and use them to test the large $N$ dualities. This computation is one of the ingredients needed for a computation of the beta function at order $1/N$ for all $\lambda$, which we leave for future work. We also discuss Chern-Simons-matter theories with extra Hubbard-Stratonovich type singlet fields, and suggest dualities between them.
}

\newpage

\setcounter{tocdepth}{2}
\tableofcontents

\section{Introduction and Summary}

Gauge theories in $2+1$ dimensions exhibit rich dynamics, and in many cases flow to interesting fixed points, some of which appear in condensed matter applications. When the theories are not parity-invariant, generically the low-energy physics may be described as a Chern-Simons-matter theory. In recent years it was found that in many cases different gauge theories flow to the same conformal Chern-Simons-matter (CS-matter) theory at low energies, so that they are IR-dual. For theories with $\mc{N}=2$ supersymmetry (SUSY) this duality \cite{Giveon:2008zn,Benini:2011mf,Aharony:2013dha,Aharony:2014uya} 
is similar to Seiberg dualities that were found in other dimensions, and indeed the duality may be related to the $4d$ duality by compactification on a circle \cite{Aharony:2013dha}. However, in $3d$ dualities of this type appear also with no supersymmetry, or with $\mc{N}=1$ supersymmetry. 

In this paper we study the dynamics of $3d$ $\mc{N}=1$ supersymmetric CS-matter theories, with $U(N)$ and $SU(N)$ gauge groups and with $N_f$ matter superfields $\Phi$  in the fundamental representation. The dynamics of these theories is more subtle than that of $\mc{N}=0$ CS-fermion theories, or of $\mc{N}=2$ CS-matter theories, because they have classically marginal deformations corresponding to $W = \omega |\Phi|^4$ terms in their superpotential (there is one such deformation for $N_f=1$, and two for $N_f > 1$). They are similar in this respect to $\mc{N}=0$ CS-scalar theories recently studied in \cite{Aharony:2018pjn}, except that in the CS-scalar case, where there is a classically marginal $\phi^6$ interaction, there is also a $\phi^4$ interaction which generically dominates the renormalization group (RG) flow, so that two fine-tunings are required for the $\phi^6$ interaction to be important in the IR; in our $\mc{N}=1$ theories only the mass needs to be fine-tuned to zero.

When the Chern-Simons level is non-zero, the coupling $\omega$ is generated even if it is not present at high energies, and its flow must be analyzed to understand the IR dynamics. For weak gauge coupling (large CS level) there are non-trivial weakly coupled fixed points for $\omega$ near the origin \cite{Avdeev:1991za,Avdeev:1992jt}, but nothing is known about the behavior when the gauge coupling or the superpotential couplings are strong. In this paper we analyze the dynamics in the \rq t Hooft large $N$ limit, of large $N$ and large Chern-Simons level with a fixed \rq t Hooft coupling $\lambda$, where many computations can be explicitly performed (using methods developed in \cite{Giombi:2011kc,Aharony:2011jz,Aharony:2012nh,Jain:2012qi,Aharony:2012ns,GurAri:2012is,Jain:2013gza,Inbasekar:2015tsa,Choudhury:2018iwf,Dey:2018ykx}). We will see that there are several different fixed points for the marginal couplings at finite large values of $N$.

In general, unlike higher supersymmetries, $3d$ $\mc{N}=1$ supersymmetry does not enable any exact computations to be performed, and does not provide many constraints on the dynamics. In particular the superpotential is not protected from quantum corrections. The main constraint is that supersymmetric vacua always have zero energy density, so that phase transitions between them are always of second order rather than first order. In some special cases, the combination of $\mc{N}=1$ supersymmetry with extra symmetries (in particular time-reversal invariance) may be used to show the existence of exact moduli spaces of vacua \cite{Gaiotto:2018yjh,Gremm:1999su}, and these will arise also in some of the theories that we will discuss. See \cite{Inbasekar:2015tsa,Gomis:2017ixy,Bashmakov:2018wts,Benini:2018umh,Eckhard:2018raj,Benini:2018bhk,Choi:2018ohn,Dey:2019ihe} for some recent investigations of 3d $\mc{N}=1$ CS-matter theories.

Another motivation for studying $3d$ $\mc{N}=1$ CS-matter theories is that these theories can appear at low energies on BPS domain walls of $4d$ $\mc{N}=1$ theories, such as supersymmetric QCD \cite{Acharya:2001dz,Bashmakov:2018ghn}. A priori it is not clear which value of the marginal superpotential couplings arises in this context, but in some cases the phase structure near the IR fixed points of the domain wall theory may be understood from $4d$. As we will discuss, this may be enough to determine which IR fixed point for the marginal couplings arises.

Various dualities have been suggested in the literature for nonabelian $\mc{N}=1$ CS-matter theories with matter in the fundamental representation (see, for instance, \cite{Jain:2013gza,Bashmakov:2018wts,Benini:2018umh,Choi:2018ohn}). At large $N$ their form is precisely known and they are supported by many computations, at least for $N_f=1$ 
\cite{Jain:2013gza,Inbasekar:2015tsa,Dey:2019ihe}. However, at finite $N$ the evidence for them is mostly circumstantial (identifying symmetries, anomalies and phase structures), and their formulation is not precise since the flow of the marginal couplings was not discussed. For infinite $N$, these marginal couplings are actually exactly marginal, so there is a large family of CFTs, and a non-trivial action of duality on this family. For $N_f=1$ this was analyzed in \cite{Jain:2013gza}, and we generalize this analysis to $N_f>1$ here. For finite $N$, in order to specify a duality, one has to specify which fixed point of the marginal couplings it applies to. For large values of $N$ we will use our large $N$ analysis of the beta function to classify these fixed points, and to make precise conjectures about their duality relations.

In addition to the simplest CS-matter theories, it is natural to consider also theories with additional singlet fields $H$ coupled to $|\Phi|^2$ through $W = H |\Phi|^2$-type superpotentials; such descriptions are useful in particular for analyzing the limits where the $|\Phi|^4$ coupling becomes large. We use the formalisms with extra singlet fields to argue that infinite values for the coupling $\omega$ are actually at finite distance away, so that the space of couplings of these theories is naturally compact (for $N_f=1$ it is a circle, and for $N_f>1$ a torus). For infinite $N$ the theories with additional singlets are manifestly equivalent to the original CS-matter theories, but for finite $N$ they are not, and we conjecture that, at least for large $N$, they provide IR-dual descriptions of these theories.

In the first part of this paper we consider theories with a single matter superfield ($N_f=1$). These theories have a single classically marginal operator, for which we attempt to calculate the beta function in the \rq t Hooft large $N$ limit. We begin in section \ref{background} by formulating our theories, and describing the general form of correlation functions in $\mc{N}=1$ superspace. In section \ref{sec:Beta_function_lambda0} we compute this beta function at order $1/N$ for $\lambda=0$, namely in a large $N$ $W=\omega |\Phi|^4$ theory, and we show that it has four fixed points.
For $\lambda\neq 0$ we could not compute the beta function exactly, but we discuss some of the steps towards computing it in section \ref{first_steps}, and we show in section \ref{sec:qualitative_lambda_positive} that there are (for large $N$) six fixed points for small $\lambda$, and at most six fixed points for all values of $\lambda$. Using the duality and some additional symmetry arguments, we then provide a conjecture for the qualitative form of the fixed points of these theories for all $\lambda$ (see Figure \ref{fig:RootPlot}). Our results are analogous to the results recently found for the $\mc{N}=0$ CS-scalar theories (`quasi-bosonic theories') in \cite{Aharony:2018pjn}.

The calculation of the beta function at order $1/N$ requires the two, three and four-point functions of the operator $J=\bar{\Phi}\Phi$ at leading order in $1/N$, as well as its anomalous dimension (at order $1/N$). We calculate the two and three-point functions explicitly at large $N$, as a function of $\lambda$ and of the marginal couplings (some of these results were found independently by the authors of \cite{Karthikunpublished}). These are duality-invariant, providing further evidence for the dualities. In addition, the three-point function allows us to conjecture (in Appendix \ref{app:MZ}) a possible generalization of the results of Maldacena and Zhiboedov \cite{Maldacena:2012sf} to $3d$ $\mc{N}=1$ theories with almost-conserved high-spin symmetries; it would be interesting to confirm this conjecture by directly generalizing their analysis to the $\mc{N}=1$ case. 

We can relate our results to many ideas appearing in the literature; in particular, we discuss their relation to previous statements about dualities in $\mc{N}=1$ CS-matter theories. Our results provide further evidence for some conjectures of theories with emergent $\mc{N}=2$ SUSY (in particular, in the context of domain walls of $4d$ $\mc{N}=1$ $N_f=1$ SQCD \cite{Bashmakov:2018ghn}). In addition, we find a series of theories with a fixed point with emergent time-reversal symmetry, where the fixed point theory has an exact moduli space even at finite $N$.

In section \ref{sec:Nf_many} we discuss theories with $N_f>1$. We start by finding the duality transformation at infinite $N$ for these theories (which is a generalization of the duality transformation found in  \cite{Jain:2013gza} for the case $N_f=1$). We then find the beta functions for the two classically marginal operators for $\lambda=0$ at order $1/N$. We can generalize some of the fixed points to finite but small $\lambda$ as well, and the duality then gives some of the fixed points for $\lambda$ close to 1. Again we find examples of theories with emergent time-reversal symmetry, some of which have an exact moduli space even at finite $N$.

There are various open questions left by our analysis. It would be interesting to complete the computation of the beta function for $\lambda \neq 0$ at leading order in $1/N$. As we discuss in the paper, one thing this requires is control of the 4-point function $\langle J J J J \rangle$ away from co-linear momenta, along the lines of recent investigations for non-supersymmetric theories in \cite{Turiaci:2018nua}. It also requires computing some additional correlation function (beyond the $2$-point function), including at least one $J$, at subleading order in $1/N$, which should be possible (though technically complicated) by generalizing our analysis. 

We discuss the phase structure of the fixed points that we find for $N_f=1$ (using known results from \cite{Dey:2019ihe}), and it should be possible to understand it by similar methods also for $N_f>1$. It should be possible to study other large $N$ theories, such as ones with orthogonal, symplectic or product gauge groups, by similar methods, and to understand their dynamics and fixed points. 

Last but not least, it would be interesting to find methods to analyze the fixed points away from the large $N$ limit, but it is not clear at the moment how to do this. For $N_f>1$, some of the fixed points we discuss can be obtained by an RG flow starting from the $\mc{N}=2$ theories with the same matter content, so the known $\mc{N}=2$ dualities (which are well-established also for finite $N$) imply their validity\footnote{The corresponding flow is much simpler than the ones connecting $\mc{N}=2$ dualities to $\mc{N}=0$ dualities \cite{Jain:2013gza,Aharony:2018pjn}, since no fields need to acquire a mass. For $N_f=1$ the deformations away from the $\mc{N}=2$ fixed point are irrelevant, at least for large $N$, so there is no such flow.}. This is known to be true for large enough $N$, and it would be nice to understand if it is true also for small $N$, or if the fixed points and the flows between them are modified. For other fixed points, and for the $N_f=1$ case, it is not clear how to flow to the $\mc{N}=1$ fixed points from theories with higher supersymmetry, and it would be interesting to study this.

\subsection{Main Results of this Paper}

This paper presents results which have implications to different areas of research, and 
since it is quite long, we now summarize our main results.

\begin{itemize}
\item
The main result of this paper is the analysis of the RG flows of $3d$ $\mc{N}=1$ CS-matter theories in the \rq t Hooft limit at leading order in $1/N$. 
We compute the beta functions of the marginal couplings exactly for $\lambda=0$, in section \ref{sec:Beta_function_lambda0} for $N_f=1$, and in section  \ref{sec:Nf_many} for $N_f>1$, and use this to obtain the fixed points at small $\lambda$. For $N_f=1$ we find a bound on the number of fixed points for all $\lambda$, and present a conjecture for the qualitative form of the RG flows at general $\lambda$ in Figure \ref{fig:RootPlot}.

\item
For infinite $N$ and $N_f>1$ there is for each $\lambda$ a family of CFTs labeled by two exactly marginal superpotential terms. In section \ref{sec:Nf_many_duality} we find how duality acts on these families of CFTs, generalizing the $N_f=1$ duality transformation 
found in \cite{Jain:2013gza}.

\item
We compute the two and three-point correlation functions of the operator $J=\bar\Phi\Phi$ at leading order in $1/N$ in the \rq t Hooft limit, for all values of the marginal couplings. The results are presented in Section \ref{sec:corr_funcs_of_J}. The result for the three-point function leads us to conjecture an extension of the results of Maldacena and Zhiboedov \cite{Maldacena:2012sf} to supersymmetric three-point functions of approximately-conserved higher-spin currents. The extension (and a discussion of the evidence for this conjecture) appears in Appendix \ref{app:MZ}.

\item
Combining calculations in the large-$N$ limit with finite-$N$ results, we conjecture that the CS-matter theories appearing in equation \eqref{eq:emergent_T_dualities} all have exact moduli spaces at a specific fixed point.
\end{itemize}

\section{Background}\label{background}

\subsection{Lagrangian and Duality}\label{sec:lagrangian}

In Euclidean space, the action for 2+1d $\mc{N}=1$ supersymmetric $U(N)$ Chern-Simons theory\footnote{We work at large $N$, keeping only the leading order and sometimes the first subleading order in the $1/N$ expansion. Thus, all our results are also applicable to $SU(N)$ gauge theories, and are independent of the value of the level of the $U(1)$ factor in $U(N)$.} coupled to $N_f$ fundamental matter fields is \cite{Avdeev:1992jt,Ivanov:1991fn}
\begin{equation}\label{eq:action}
S=\int d^3xd^2\theta\left(\mc{L}_{\text{CS}}+\mc{L}_{\text{matter}}\right). \end{equation}
Our superspace conventions are summarized in Appendix \ref{app:conventions}. In terms of $\mc{N}=1$ superfields, the Chern-Simons term for the gauge field is given by
\begin{equation}
\mc{L}_{\text{CS}}=-\frac{\kappa}{2\pi}\text{Tr}\left(-
\frac14 D_\alpha\Gamma^\beta D_\beta \Gamma^\alpha-\frac16 D^\alpha \Gamma^\beta\{\Gamma_\alpha,\Gamma_\beta\}-
\frac{1}{24}\{\Gamma^\alpha,\Gamma^\beta\}\{\Gamma_\alpha,\Gamma_\beta\} \right). \label{eq:L_gauge}
\end{equation}
In components, the action of the gauge field becomes
\begin{equation}
\mathscr{L}_{\text{CS}}=-\frac{\kappa}{2\pi}\gre^{\mu\nu\rho}\text{Tr}\left(A_\mu\partial_\nu A_\rho-\frac{2i}{3}A_\mu A_\nu A_\rho\right),
\end{equation}
while the gaugino is an auxiliary field with no kinetic term, which we can integrate out. Our conventions for the Chern-Simons level are summarized in Appendix \ref{app:conventions}.

We now discuss the additional matter term $\mc{L}_{\text{matter}}$. We separate the discussion into the cases $N_f=1$ and $N_f>1$.

\subsubsection{One matter field ($N_f=1$)}
For $N_f=1$, a single matter superfield $\Phi^a$ ($a=1,\cdots,N$) in the fundamental representation of $U(N)$, the most general renormalizable action takes the form (in superfield notation, suppressing gauge indices)
\begin{equation}\label{eq:L_matter}
\mc{L}_{\text{matter}}=-\frac{1}{2}
\left(D^\alpha\bar{\Phi}+i\bar{\Phi}\Gamma^\alpha \right)
\left(D_\alpha\Phi-i\Gamma_\alpha\Phi \right)+m_0\bar{\Phi}\Phi +\frac{\pi\omega}{\kappa}\left(\bar{\Phi}\Phi\right)^2.
\end{equation}
Note that $\left(\bar{\Phi}\Phi\right)^2$ is classically marginal, and we chose a convenient normalization for its dimensionless coefficient $\omega$. In components we find that $\mc{L}_{\text{matter}}$ splits into three contributions (after integrating out the auxiliary fields):
\begin{align}
\mathscr{L}_{\text{boson}}&=\mathcal{D}^\mu \bar\phi\mathcal{D}_\mu \phi+m_0^2 \bar\phi\phi+
\frac{4\pi\omega m_0}{\kappa}(\bar\phi\phi)^2+
\frac{4\pi^2\omega^2}{\kappa^2}(\bar\phi\phi)^3,\\
\mathscr{L}_{\text{fermion}}&=-\bar \psi \left(i\cancel{\mathcal{D}}+m_0\right)\psi,\\
\mathscr{L}_{\text{int}}&=-
\frac{2\pi(1+\omega)}{\kappa}(\bar\phi\phi)(\bar\psi\psi)-
\frac{2\pi\omega}{\kappa}(\bar\psi\phi)(\bar\psi\phi)+
\frac{\pi(1-\omega)}{\kappa}\left((\bar\phi\psi)(\bar\phi\psi)+h.c.\right).\label{eq:L_int}
\end{align}
We suppress gauge indices, and use brackets to denote gauge index contractions. For $\omega=1$ the Lagrangian has enhanced $\mc{N}=2$ SUSY.

In the \rq t Hooft large $N$ limit, of large $N$ and $\kappa$ with fixed \rq t Hooft coupling $\lambda=\frac{N}{\kappa}$, $\omega$ is exactly marginal so there are consistent QFTs \eqref{eq:action} for every value of $-1 \leq \lambda \leq 1$ and $\omega$. These theories were conjectured
to obey a duality under the following transformation \cite{Jain:2013gza}:
\begin{equation}\label{eq:duality}
	\lambda'=\lambda-\text{sign}(\lambda),\quad \omega'=\frac{3-\omega}{1+\omega},\quad
	m_0'=-\frac{2m_0}{1+\omega}.
\end{equation}
This transformation implies that the gauge-invariant singlet operator $J=\bar\Phi^a\Phi_a$ transforms under the duality as $J'=-\frac{1+\omega}{2}J$. Note that the value $\omega=1$ (where the theory has enhanced $\mc{N}=2$ SUSY) is fixed under the duality. Note also that $\lambda$ is parity-odd, while $\omega$ is parity-even.

We will mainly be interested in theories with $m_0=0$, and will be working in light-cone gauge $\Gamma_-=0$. The $\mc{N}=1$ Lagrangian then becomes
\begin{align}
\mc{L}_{\text{CS}}&=-\frac{\kappa}{8\pi}\tr \left(\Gamma^- i\partial_{--}\Gamma^-\right), 
\label{eq:gauge_Lagrangian}\\
\mc{L}_{\text{matter}}&=-\frac12 D^\gra \bar\grF D_\gra \grF -\frac{i}{2}\Gamma^-\left(\bar\Phi D_- \Phi - D_- \bar{\Phi}\Phi\right)
+\frac{\pi\grw\lambda}{N}(\bar{\Phi}\Phi)^2.
\label{eq:matter_Lagrangian}
\end{align}

For $\lambda\to 0$ the gauge fields decouple, but we can still keep the superpotential interacting by taking $\lambda \to 0$ and $\omega\to \infty$ while keeping $\tilde{\omega}\equiv \pi\omega \lambda$ fixed. This gives the usual $\Phi^4$ theory (in $\mc{N}=1$ superfield notation):
\begin{equation}
\mc{L}=\bar\grF^a  D^2 \grF_a + \frac{\tilde\omega}{N}(\bar\grF^a\grF_a)^2.
\end{equation}

\subsubsection{Many matter fields ($N_f>1$)}\label{sec:many_matter_fields}

We now have $N_f$ superfields in the fundamental representation of $U(N)$, $\Phi_{ia}$ ($i=1,\cdots,N_f$,\; $a=1,\cdots,N$).
The action is still of the form \eqref{eq:action}, with the same $\mc{L}_{\text{CS}}$ as in \eqref{eq:L_gauge}. The general matter Lagrangian $\mc{L}_{\text{matter}}$, assuming an $SU(N_f)$ global symmetry rotating the matter superfields, is now:
\begin{equation} 
\mc{L}_{\text{matter}}=-\frac{1}{2}
\left(D^\alpha\bar{\Phi}^i+i\bar{\Phi}^i\Gamma^\alpha \right)
\left(D_\alpha\Phi_i-i\Gamma^\alpha\Phi_i \right)+m_0(\bar{\Phi}^i\Phi_i) +\frac{\pi\omega_0}{\kappa}\left(\bar{\Phi}^i\Phi_i\right)^2+\frac{\pi\omega_1}{\kappa}\left(\bar{\Phi}^i\Phi_j\right) \left(\bar{\Phi}^j\Phi_i\right),
\end{equation}
where we have again denoted gauge index contractions using brackets. 
Note the existence of two classically marginal superpotential couplings (which coincide in the case $N_f=1$ or when the gauge group is $U(1)$). The theory has $\mc{N}=2$ SUSY for  $\grw_0=0,\;\grw_1=1$. 

The generalization of the duality transformation \eqref{eq:duality} to $N_f>1$ has not yet appeared in the literature. We propose a generalization in Section \ref{sec:Nf_many_duality}, and give some evidence for our proposal.

Again, we will be mostly interested in light-cone gauge with $m_0=0$, so that the matter Lagrangian becomes
\begin{equation}\label{eq:matter_lagrangian_Nf}
\mc{L}_{\text{matter}}=-\frac12 D^\gra \bar\grF^i D_\gra \grF_i -\frac{i}{2}\Gamma^-\left(\bar\Phi^i D_- \Phi_i - D_- \bar{\Phi}^i\Phi_i\right)
+\frac{\pi\grw_0\lambda}{N}(\bar{\Phi}^i\Phi_i)^2+\frac{\pi\grw_1\lambda}{N}\left(\bar{\Phi}^i\Phi_j\right) \left(\bar{\Phi}^j\Phi_i\right).
\end{equation}
We can again take the limit $\lambda\to 0$, where the Lagrangian becomes a $\Phi^4$ theory:
\begin{equation}\label{eq:matter_lagrangian_Nf_lambda0}
\mc{L}=\bar\grF^{ai}  D^2 \grF_{ai}  + \frac{\tilde\omega_0}{N}(\bar\grF^{ai} \grF_{ai} )^2+ \frac{\tilde\omega_1}{N}(\bar\grF^{ai} \grF_{aj} )(\bar\grF^{bj} \grF_{bi} ),
\end{equation}
where $\tilde\omega_n\equiv \pi\lambda\omega_n$ for  $n=0,1$ are kept fixed. 

\subsection{General Form of $\mc{N}=1$ Supersymmetric Correlation Functions}\label{sec:general_3pt_func}

We review some results from \cite{Inbasekar:2015tsa}, where the general form of an $\mc{N}=1$ supersymmetric $n$-point correlation function was studied. Consider an $n$-point function of $n$ scalar superfields $Z_i$:
\begin{equation}
\Gamma_n(p_i,\theta_i) = \langle Z_1(p_1,\theta_1) Z_2(p_2,\theta_2) \cdots Z_n(p_n,\theta_n) \rangle,
\end{equation}
with $p_i$ the momenta and $\theta_i$ the anticommuting superspace coordinates (note that momentum conservation requires $\sum_i p_i^\mu=0$). SUSY imposes the following constraints:
\begin{equation}\label{eq:SUSY_ward}
\sum_{i=1}^n Q_{p_i,\theta_i}\Gamma_n(p_i,\theta_i)=0,
\end{equation}
where $Q$ are the supercharges
\begin{equation}
Q_\gra^{k,\theta}=i\left(\frac{\partial}{\partial\theta^\gra}-\theta^\grb k_{\gra\grb}\right).
\end{equation}
We now review the results for two and three-point functions.

For the 2-point function $\Gamma_2(p,\theta_1,\theta_2)$, the general form of the solution to \eqref{eq:SUSY_ward} is
\begin{equation}
\Gamma_2=e^{-\theta_1^\alpha p_{\alpha\beta }\theta_2^\beta}F_2(X_{12},p),
\end{equation}
where we have defined $X_{ij}\equiv \theta_i-\theta_j$. $F_2$ can be expanded in superspace as
\begin{equation} 
F_2=C_1(p^\mu)-C_2(p^\mu)(\theta_1-\theta_2)^2 .
\end{equation}
Equivalently, we can write
\begin{equation}\label{eq:general_2pt}
\Gamma_2(p,\theta_1,\theta_2)=(C_1(p^\mu)D^2+C_2(p^\mu))\delta^2(\theta_1-\theta_2),
\end{equation}
which is the general form of a supersymmetric two-point function.

For the 3-point function $\Gamma_3(p,q,-p-q,\theta_1,\theta_2,\theta_3)$, a similar calculation gives
\begin{equation}
\Gamma_3=e^{\frac13 X\cdot(p\cdot X_{13}+q\cdot X_{23})}F_3(X_{13},X_{23},p,q) ,
\end{equation}
where we have defined $X\equiv
\sum_{i=1}^3 \theta_i$. The expansion of $F$ now contains eight terms:
\begin{equation}\label{eq:general_3pt}
\Gamma_3=e^{\frac13 X\cdot(p\cdot X_{13}+q\cdot X_{23})}
\left(\mc{A}-i\mc{B}_1 X_{13}^+X_{13}^- -i \mc{B}_2X_{23}^+X_{23}^-+
\left(X_{13}^+,X_{13}^-\right)\cdot \mathbf{\mc{B}}\cdot \begin{pmatrix}
X_{23}^+\\
X_{23}^-
\end{pmatrix} -
\mc{C}X_{13}^+X_{13}^-
X_{23}^+X_{23}^-\right),
\end{equation}  
with $\mathcal{B}_{\alpha\beta}$ a $2\times 2$ matrix. 

Given a 3-point function, we can isolate each of these terms. As an example, consider a free theory of $N$ fields $\Phi_a$ with $a=1,\cdots,N$. Defining $J=\bar{\Phi}^a\Phi_a$, the 3-point function $\langle J(q) J(l) J(-l-q) \rangle$ is:
\begin{equation}\label{eq:free_3pt_decomposition}
	\begin{split}
		\mc{A}&=2\frac{N}{8|q||l||q+l|}\\
		\mc{B}_1=\mc{B}_2&=0\\
		\mc{B}_{\gra\grb}&=2\left(\frac{N}{3}\frac{(l-q)_{\gra\grb}}{8|q||l||q+l|}-N\mc{I}_{\gra\grb}\right) \\
		\mc{C}&=2\left(-\frac{N}{9}\frac{(l-q)^2}{8|q||l||q+l|}-\frac{N}{8|l+q|}+\frac{2N}{3}(l-q)\cdot \mc{I}\right),
	\end{split}
\end{equation}
where we defined \cite{Aharony:2018pjn}:
\begin{equation} \mc{I}_{\gra\grb}= \int \dpi{p}\frac{p_{\gra\grb}}{p^2(p-l)^2(p+q)^2}=
\frac{\frac{l_{\alpha\beta}}{|l|}-\frac{q_{\alpha\beta}}{|q|}}{16|q+l|}
\frac{|q|+|l|-|q+l|}{|q||l|-q\cdot l}.
\end{equation}

\section{The Beta Function at $\lambda=0$ For $N_f=1$}\label{sec:Beta_function_lambda0}

As explained in Section \ref{sec:lagrangian}, at $\lambda=0$ the Lagrangian of the theory reduces to that of $\Phi^4$ theory (with $\Phi$ a superfield):
\begin{equation}\label{eq:Phi4_theory}
\mc{L}=\bar\grF^a  D^2 \grF_a + \frac{\tilde\omega}{N}(\bar\grF^a\grF_a)^2,
\end{equation}
with $a=1,\cdots,N$.
In this section we find the $\beta$ function for $\wt$ and the $\gamma$ functions for $\Phi$ and $J\equiv \bar{\Phi}^a\Phi_a$ at large $N$ and at $\lambda=0$ (to all orders in $\wt$). At leading order for small $\wt$, this was calculated in \cite{Avdeev:1992jt}, and we compare the results in this limit at the end of this section.

\subsection{Classification of Diagrams Contributing to $\Phi^4$ at large $N$}

In order to find the beta function, we must study the corrections to the $\Phi^4$ vertex in \eqref{eq:Phi4_theory}. In this section we classify the diagrams which contribute to the 4-point function at leading and subleading order in $\frac1N$. 

A general diagram for the 4-point function $\langle \bar \grF^a \grF_a \bar \grF^b \grF_b \rangle$ includes two external index lines and various internal index loops. Denote the number of vertices along the two external index lines by $v_1$ and $v_2$, and the number of internal index loops with $i$ vertices by $n_i$ ($i=2,3,...$). Since we will be using dimensional regularization, we have taken $n_1=0$ since single-vertex index loops will vanish.

In a diagram with $V$ $\Phi^4$ vertices, each involving two index lines, we have
\begin{equation}
2 V = v_1 + v_2 + 2 n_2 + 3 n_3 + 4 n_4 + \cdots,
\end{equation}
and the power of $N$ associated with this diagram is
\begin{equation}
\Delta = n_2 + n_3 + n_4 + \cdots - V.
\end{equation}
Furthermore, we always have 
\begin{equation}
v_1,v_2 \geq 1.
\end{equation}

\begin{figure}
	\centering
	\includegraphics[width=0.7\linewidth]{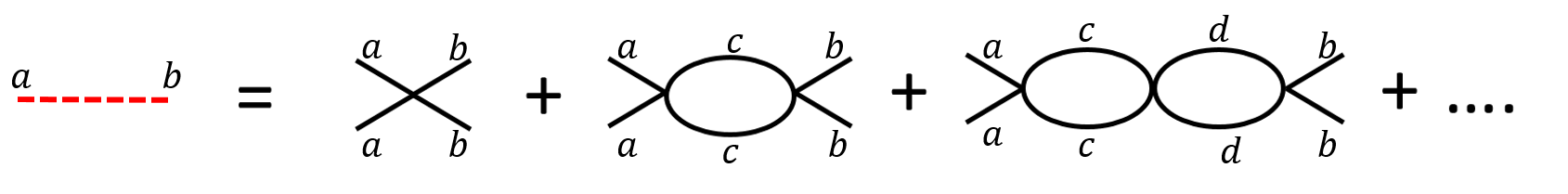}
	\caption{A chain}
	\label{fig:chain}
\end{figure}

First, note that in terms of counting powers of $N$ we can just shrink all 2-vertex index loops to zero, since removing one such loop and one vertex does not change anything (they have $\Delta=0$). Conversely, in each 4-point vertex, we can insert in the middle any number of index loops (a ``chain", which will be denoted by a red line, see Figure \ref{fig:chain}) without changing the power of $N$. So it is enough to classify diagrams with $n_2=0$, and then blow up each vertex into a chain with any number of index loops.

The leading diagram has $V=1$, $v_1=v_2=1$ with all others vanishing, and so has $\Delta=-1$.
To get other diagrams with $\Delta=-1$ we must have 
\begin{equation}
n_3+n_4+... = V-1 = (v_1+v_2)/2 + (3/2) n_3 + 2 n_4 + ... - 1,
\end{equation}
so
\begin{equation}
v_1 + v_2 + n_3 + 2 n_4 + ... - 2 = 0.
\end{equation}
The only solution is $v_1+v_2=2$ and the others vanishing, giving the chains discussed above. 

\begin{figure}
	\centering
	\begin{subfigure}{0.3\textwidth}
		\centering
		\includegraphics[width=\textwidth]{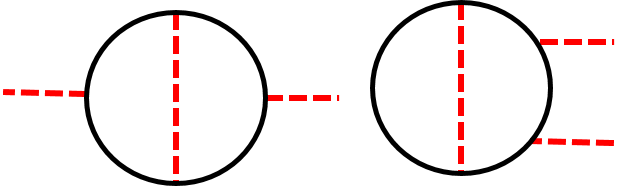}
		\caption*{Type (a)}
	\end{subfigure}
	\qquad\qquad
	\begin{subfigure}{0.23\textwidth}
		\centering
		\includegraphics[width=\textwidth]{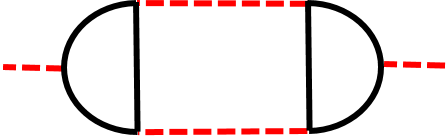}
		\caption*{Type (b)}
	\end{subfigure}
	\qquad\qquad
	\begin{subfigure}{0.17\textwidth}
		\centering
		\includegraphics[width=\textwidth]{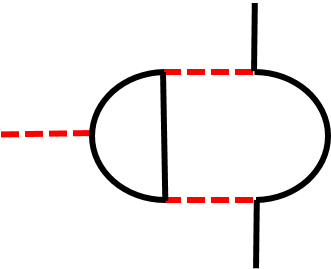}
		\caption*{Type (c)}
	\end{subfigure}
		\qquad
	\begin{subfigure}{0.13\textwidth}
		\centering
		\includegraphics[width=\textwidth]{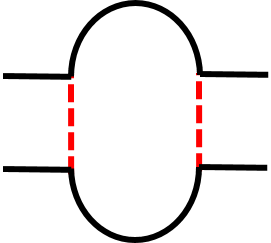}
		\caption*{Type (d)}
	\end{subfigure}
		\qquad\qquad
	\begin{subfigure}{0.32\textwidth}
		\centering
		\includegraphics[width=\textwidth]{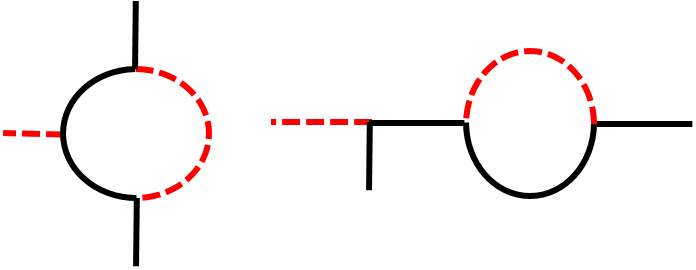}
		\caption*{Type (e)}
	\end{subfigure}
	\caption{Diagrams contributing to the beta function at order $1/N$.}
	\label{fig:diags}
\end{figure}

The next order is $\Delta=-2$. At this order we must have
\begin{equation}
v_1 + v_2 + n_3 + 2 n_4 = 4
\end{equation}
(all the higher $n_i$'s must clearly vanish). The possible solutions (up to exchanging the external index lines) are:
\begin{enumerate}[label=(\alph*)]
	\item $v_1=1, v_2=1, n_4=1$
	\item $v_1=1, v_2=1, n_3=2$
	\item $v_1=1, v_2=2, n_3=1$
	\item $v_1=2, v_2=2$
	\item $v_1=1, v_2=3$
\end{enumerate}
So all of the diagrams that contribute at subleading order in $N$ are of one of the types $(a)$-$(e)$, where we replace vertices with chains. We draw all of the possible diagrams in Figure \ref{fig:diags}.

\subsection{Calculation of Beta and Gamma Functions}\label{direct_calculation}

We start with the calculation of a chain $\Delta_{\Phi^2\Phi^2}$ (Figure \ref{fig:chain}), which is the leading order (in $1/N$) contribution to the $\Phi^4$ vertex. The result at this order can be written as an effective action term contributing to the 4-point function, of the form 
\begin{equation} \label{leading}
\frac{1}{N}(\bar\Phi^a\Phi_a)(p,\theta_1) \left(a_0+a_1\frac{D^2}{|p|}\right)(\bar\Phi^b\Phi_b)(-p,\theta_2)
\end{equation}
(here we take the product of the superfields $\bar\Phi^a$ and $\Phi_a$, which are at different points, and denote its momentum by $p$ and its fermionic superspace coordinate by $\theta_1$, and similarly for the product of $\bar\Phi^b$ and $\Phi_b$; the $D^2$ operator acts on the latter superspace coordinate, and there is no dependence on other combinations of momenta).
Note that each index loop adds a factor of $N$ while each vertex adds a factor of $\frac1N$, so that all of the diagrams in a chain come with an overall factor of $\frac1N$. The chain is given by summing the diagrams of Figure \ref{fig:chain}:
\begin{equation}\label{eq:chain}
	\Delta_{\Phi^2\Phi^2}\equiv 
		\frac{1}{N}\left(a_0+a_1\frac{D^2}{|p|}\right) =
	\frac{2\wt}{N} \sum_{n=0}^{\infty} \left(\wt\frac{D^2}{4|p|}\right)^n=
	\frac{1}{N}\frac{2\wt}{1-\wt\frac{D^2}{4|p|}}=
	\frac{1}{N}\frac{8\wt}{\wt^2+16} \frac{4|p|+\wt D^2}{|p|},
\end{equation} 
so that
\begin{equation}
a_0 = \frac{32 \wt}{\wt^2+16},\qquad\qquad a_1 = \frac{8 \wt^2}{\wt^2+16}.
\end{equation}

Next we discuss the calculation of the diagrams $(a)$-$(e)$, which are of subleading order in $1/N$. Since we are interested in the $\beta$ function, we calculate only the logarithmically diverging parts of these diagrams; these must take the same form as \eqref{leading} above, since the naive divergences must be canceled by the cutoff-dependence of the leading order terms. The explicit calculation of these terms using dimensional regularization with $d=3-\gre$ appears in Appendix \ref{app:diagrams}. The results are summarized in the following table:
\begin{center}
	\begin{tabular}{cc}
		Type & Result  \\ 
		\hline
		$(a)$	& $-\frac{a_1}{4\pi^2\gre}
		\frac{D^2}{|p|}\Delta_{\Phi^2\Phi^2}^2$ \\ 
		$(b)$	& $-\frac{1}{32\pi^2\gre}\left( a_0+a_1\frac{D^2}{|p|}
		\right)^2 \frac{D^2}{|p|}
		\Delta_{\Phi^2\Phi^2}^2$\\
		$(c)$	& $-\frac{1}{N}\frac{1}{4\pi^2\gre}\left(a_0+a_1\frac{D^2}{|p|}
		\right)^2\Delta_{\Phi^2\Phi^2}$ \\ 
		$(d)$	& $-\frac{1}{N^2}\frac{2 a_1 a_0}{\pi^2\gre}$ \\ 
		$(e)$	& $-\frac{1}{N}\frac{3a_1}{\pi^2\gre}\Delta_{\Phi^2\Phi^2}$ \\ 
	\end{tabular} 
\end{center}
Note that these are all of order $1/N^2$. 

\begin{figure}
	\centering
	\includegraphics[width=0.23\linewidth]{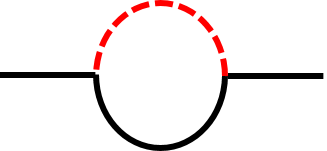}
	\caption{The leading-order contribution to the gamma function of $\Phi$}
	\label{fig:gamma}
\end{figure}

We can now find $\beta_{\wt}$ and $\gamma_{\Phi}$. $\gamma_{\Phi}$ is obtained from the diagram in Figure \ref{fig:gamma}, which gives the correction at order $1/N$ to the propagator $\langle \bar \grF^a \grF_a \rangle$.
The contribution is
\begin{equation}
\frac{a_1}{N} \int \dpi{k} \frac{D^2}{|k|(p-k)^2}=
\frac{a_1D^2}{2N\pi^2\gre}.
\end{equation}
We thus find
\begin{equation}
\gamma_{\Phi}=
\frac{1}{N}\frac{a_1}{4\pi^2}=
\frac{1}{N}\frac{2\wt^2}{\pi^2(\wt^2+16)}.
\end{equation} 

$\beta_{\wt}$ is obtained from the Callan-Symanzik equation
\begin{equation}\label{eq:CS_equation}
\left(\mu\frac{\partial}{\partial\mu}+\beta_{\wt}\frac{\partial}{\partial\wt}+4\gamma_{\Phi}\right)\mc{O}_4=0,
\end{equation}
where $\mc{O}_4$ is the connected four-point function. Using the $1/N$ expansion we can write this as
\begin{equation} \label{eq:CSexplicit}
\left(\beta_{\wt}\frac{\partial}{\partial\wt}+4\gamma_{\Phi}\right)\Delta_{\Phi^2\Phi^2}+\left((a)+(b)+(c)+(d)+(e)\right)=0,
\end{equation}
dropping the $\gre$ factor from the values of the diagrams given in the table above, where
$\beta_{\wt}$ and $\gamma_{\Phi}$ are given at leading order in $1/N$.
We find:
\begin{equation} \label{eq:betaresult}
\beta_{\wt}=-\frac{1}{N}\frac{16 \wt ^3 \left(\wt ^2-48\right)}{\pi ^2 \left(\wt ^2+16\right)^2}.
\end{equation} 
Note that equation \eqref{eq:CSexplicit} contains two terms, proportional to $1$ and to $D^2/|p|$, and both of them vanish for this value, giving a consistency check for our computation. Also note that the $\mc{N}=2$ point is at $\tilde\omega=0$, and the beta function vanishes there as expected.

\subsubsection{Additional Correlators}\label{bootstrap_beta_lambda0}

We can actually use the same results to compute also additional correlation functions, at leading and subleading order in $1/N$.
Specifically, define the operator $J(x)=\frac{1}{N}\bar\Phi^a(x) \Phi_a(x)$. There are three correlation functions we can study: the 4-point function
$\left<(\bar\Phi\Phi)(\bar\Phi\Phi)\right>$, the 3-point function $\left<J (\bar\Phi\Phi)\right>$, and the 2-point function $\left<J J\right>$,
which are all given by (some of) the same diagrams that we computed in the previous subsection, replacing if necessary two external $\Phi$'s emanating from the same vertex (which gives an external chain) by an insertion of $J$.

Start with the leading order in $1/N$. For all of these correlators we just have a generalization of the chains discussed above:
\begin{align}
\left<(\bar\Phi \Phi)(p) (\bar\Phi \Phi)(-p)\right>|_{1/N}&=\Delta_{\Phi^2\Phi^2}\\
\left<J(p) (\bar\Phi \Phi)(-p) \right>|_{1/N}&=\Delta_{J\Phi^2}=\frac{1}{2\omega}\Delta_{\Phi^2\Phi^2}\\
\left<J(p) J(-p)\right>|_{1/N}&=\Delta_{J J}=\frac{D^2}{16\omega|p|}\Delta_{\Phi^2\Phi^2}
\end{align}
Note that all terms are of order $1/N$. The first line is precisely the chain \eqref{eq:chain}, while the other two are chains with external $\Phi$'s replaced by insertions of $J$.
Each of these correlators has a term going as $1$ and another going as $\frac{D^2}{|p|}$, and naively we can use the logarithmic corrections to each of these correlators at order $1/N$ to independently compute the beta function.
However, in the 2-point correlator $\left<J(p) J(-p)\right>=b_0+b_1\frac{D^2}{|p|}$, $b_0$ is actually a contact term (with a value depending on $\omega$). At the next order in $1/N$, as in a similar discussion in \cite{Aharony:2018pjn}, a part of this can remain a contact term, and a part of this can become the 2-point function of $J$ at separate points (which becomes non-trivial in momentum space once $J$ has an anomalous dimension at order $1/N$), and a priori it is not clear which part remains a contact term and which part does not. Thus the $b_0$ part of $\left<J J\right>$ cannot be used in the following to find the $\beta,\gamma$ functions, and so we only use the $b_1$ part in the following.

Consider next the subleading order. Denote the contribution of the diagram types above (without the external leg factors which gave some powers of $\Delta_{\Phi^2 \Phi^2}$ in the table above) by $a,b,c,d,e$. We find that the logarithmic terms at order $1/N$ are given by:
\begin{align}
\left<(\bar\Phi \Phi)(p) (\bar\Phi \Phi(-p) \right>|_{1/N^2}&=(a+b)\Delta_{\Phi^2\Phi^2}^2+(c+e)\Delta_{\Phi^2\Phi^2}+d\\
\left<J(p) (\bar\Phi \Phi)(-p) \right>|_{1/N^2}&=(a+b)\Delta_{\Phi^2\Phi^2}\Delta_{J\Phi^2}+\frac12(c+e)\Delta_{J\Phi^2}\\
\left<J(p) J(-p)\right>|_{1/N^2}&=(a+b)\Delta_{J\Phi^2}^2,
\end{align}
where the first line is just a rewriting of the results of the previous subsection.

Now we can calculate the beta and gamma functions. The RG equations for the correlation functions above depend on $\beta_{\omega}$, on $\gamma_{\Phi}$ (which we already know from other considerations) and on the anomalous dimension of $J$, $\gamma_{J}=\gamma_{\Phi^2}$. We have 5 separate RG equations (the $1$ and $D^2/|p|$ terms from the first two correlators, and the $D^2/|p|$ term from the second), with two unknowns, so we have an overconstrained system, and the inputs must satisfy three constraints. Let us write
\begin{align}
a+b&=O_1+O_2\frac{D^2}{|p|}\\
c+e&=\frac1N\left(O_3+O_4\frac{D^2}{|p|}\right)\\
d&=\frac{1}{N^2}\left(O_5+O_6\frac{D^2}{|p|}\right)
\end{align}
Note that all $O_i$'s are of order 1. 
Since we have three constraints, it is enough to calculate three of the $O_i$'s to find the rest (along with the beta and gamma functions). In terms of $O_1,O_3,O_5$ we find:
\begin{align}
O_2&=\frac{O_3}{8}\\
O_4&=\frac{O_5}{4}\\
O_6&=0\\
\beta_{\wt}&=-\frac1N\frac{32 \wt  (2 \wt O_1  +O_3)+O_5 \left(\wt ^2+16\right)}{32}\\
\gamma_J&=-\frac1N\frac{64\wt O_1  +16 O_3+\wt O_5  }{32}
\end{align}
Note that both $\beta_{\wt}$ and $\gamma_J$ are of order $1/N$.
Plugging in our results from Section \ref{direct_calculation}, we find that our results for $O_2$, $O_4$ and $O_6$ are consistent with these equations, and we obtain the beta function \eqref{eq:betaresult}, and
\begin{equation}
\gamma_{J}=-\frac{16 \wt ^2 \left(\wt ^2-16\right)}{\pi ^2 \left(\wt ^2+16\right)^2}\frac{1}{N}.
\end{equation}

The discussion above explains some properties of the results from the direct calculation; for instance, it explains why $(e)$ does not have a term that is proportional to $\frac{D^2}{|p|}$. If $(e)$ had such a term, it would appear in $O_4$ with one factor of either $a_0$ or $a_1$ (since the $(e)$-type diagrams have only one internal chain). However, we see that $O_4=\frac{O_5}{4}$. $O_5$ comes from the $(d)$-type diagrams, which have two internal chains, so that it only contain terms of the form $a_0^2,a_0a_1$ and $a_1^2$. For generic $a_0$ and $a_1$ we cannot have that a term linear in $a_i$ will be equal to a term quadratic in $a_i$, and so the $\frac{D^2}{|p|}$ term in $O_4$ must vanish.

\subsubsection{Summary}\label{beta_Summary}

To summarize, at leading order in $1/N$ we have:
\begin{align}
\beta_{\wt}&=-\frac{16 \wt ^3 \left(\wt ^2-48\right)}{\pi ^2 \left(\wt ^2+16\right)^2}\frac{1}{N} \label{eq:final_beta}\\
\gamma_{J}&=-\frac{16 \wt ^2 \left(\wt ^2-16\right)}{\pi ^2 \left(\wt ^2+16\right)^2}\frac{1}{N}\label{eq:final_gamma_phi2}\\
\gamma_{\Phi}&=\frac{2\wt^2}{\pi^2(\wt^2+16)}\frac{1}{N}\label{eq:final_gamma}
\end{align}

By expanding the results above in $\tilde{\omega}$, one can compare the beta and gamma function found at leading (two-loop) order in $\tilde\omega$ in \cite{Avdeev:1992jt}. Due to the use of different conventions, we find $\gamma_{\Phi,us}=2\gamma_{\Phi,them}$,  $\gamma_{J,us}=4\gamma_{J,them}$ and $\beta_{us}=4\beta_{them}$. These factors remain consistent also in the $N_f>1$ result, see Section  \ref{sec:beta_Nf_many}.

Our results exhibit attractive RG fixed points (where the superpotential is irrelevant in the IR) at $\wt=0$ and $\wt=\infty$, and repulsive RG fixed points (where the superpotential is relevant in the IR) at $\wt = \pm \sqrt{48}$. We postpone further discussion of the beta function, and of the meaning of the fixed point at $\wt=\infty$, until Section \ref{sec:qualitative_lambda_positive}.

\section{First Steps Towards The Beta Function at $\lambda\neq 0$}\label{first_steps}

We would like to compute the beta function for $\omega$ also for non-zero values of the CS coupling. Unfortunately, we were not able to do this. For $\lambda \neq 0$, the correlation functions of $\Phi$'s are not gauge-invariant\footnote{It may be possible to extract information from such correlators by computing them in a specific gauge, but since the gauge we use breaks Lorentz-invariance, non-Lorentz-invariant counter-terms may be needed, and we do not discuss this here.}, so out of the correlators of the previous section, we can only compute $\langle J J \rangle$, which, as discussed above, gives us just a single equation for $\beta_{\omega}$ and $\gamma_J$. Moreover, we were not able to completely compute the $1/N$ correction to $\langle J J \rangle$. In this section we describe some contributions to this correction (and thus to the beta function) explicitly. These contributions require computing correlation functions of two and three $J$'s, which we compute explicitly, and which are interesting in their own right. In addition, our considerations will enable us to constrain the form of the beta function at leading order in $1/N$ for all $\lambda$, limiting the number of its zeros, as we will discuss in the next section.

The main results of this section are the two and three-point correlation functions of $J=\bar\Phi\Phi$ in the \rq t Hooft limit. Readers interested in the RG flows can skip this section and move on to Section \ref{sec:qualitative_lambda_positive}.

\subsection{General Considerations}\label{sec:beta_func_general}

In this subsection we present a general method of obtaining some constraints on the beta function for the full theory \eqref{eq:action}, systematically for all orders in $1/N$. We follow the method described in \cite{Aharony:2018pjn}. Start with the $\mc{N}=1$ Lagrangian discussed in Section \ref{sec:lagrangian}:
\begin{equation}
\mathcal{L}_{CS}+\left(\nabla^\gra \grF \right)^2+\frac{\pi\lambda\omega}{N}(\bar\Phi\Phi)^2 .
\end{equation}
We use the standard Hubbard-Stratonovich transformation to do large $N$ computations; first, we rewrite the Lagrangian using auxiliary fields $\Lambda,\Sigma$ as:
\begin{equation}
\mathcal{L}=\mathcal{L}_{CS}+\left(\nabla^\gra \grF \right)^2+\frac{\pi\lambda\omega_0}{N}(\bar\Phi\Phi)^2+\grL(\bar\Phi \Phi-N\grS)+\pi\lambda N(\omega-\omega_0)\grS^2,
\end{equation}
where we arbitrarily separated the superpotential into a term proportional to $\omega_0$ and another proportional to $(\omega-\omega_0)$. This choice is arbitrary: we can work with any convenient value of $\omega_0$, and the results cannot depend on it. Next, we integrate out the matter fields $\Phi$ and the gauge fields; in their path integral $\grL$ behaves as a source for the operator $J=\bar\Phi \Phi$ (note that from here on we use this normalization for $J$, which differs from the one of the previous section). This leaves us with an effective action for $\grL,\Sigma$:
\begin{equation}\label{eq:LargeNLagrangian}
	\mathcal{L}=-N\grL\grS+\pi \lambda N(\omega-\omega_0)\grS^2+\frac12\grL G_2\grL+\frac{1}{3!}G_3\grL^3+\frac{1}{4!}G_4\grL^4+\cdots,
\end{equation}
where the $G_n$ are the $n$-point functions of $J$ in the CS-matter theory with a superpotential coefficient $\omega_0$. Here, the notation $G_3 \grL^3$ is short-hand for three integrations over superspace of \begin{equation}G_3(x_1,\theta_1,x_2,\theta_2, x_3, \theta_3) \grL(x_1,\theta_1) \grL(x_2, \theta_2) \grL(x_3, \theta_3)\end{equation}
and similarly for the other terms.

\begin{figure}
	\centering
	\includegraphics[width=0.6\linewidth]{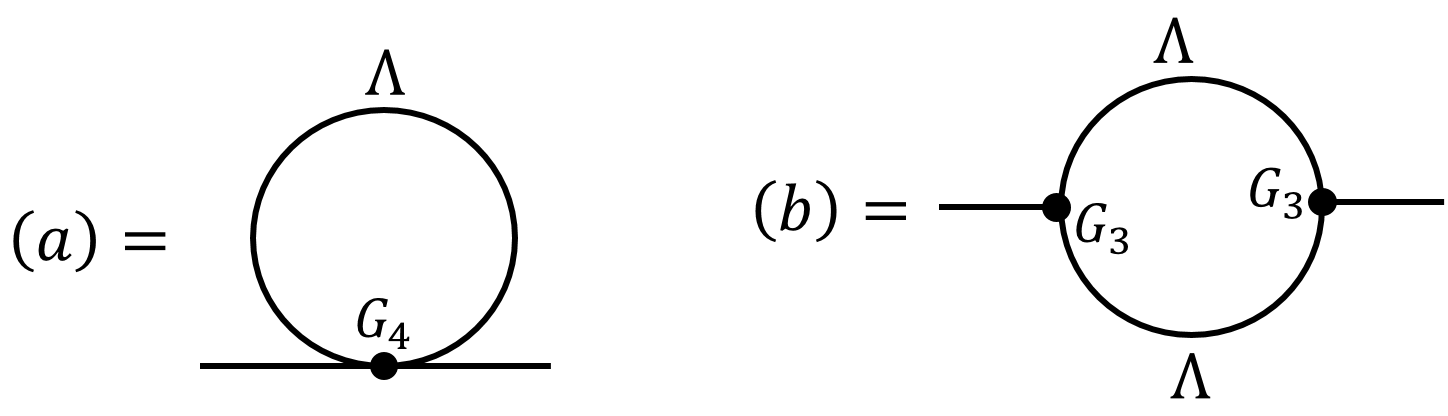}
	\caption{Diagrams contributing to the beta function of $\tilde\omega$ at order $1/N$.}
	\label{fig:a_and_b_type}
\end{figure}

Note that the $G_n$ are of order $N$ in the large $N$ limit (though they are generally corrected at higher orders in $1/N$). If we keep only the term of order $N$, then $N$ appears as a coefficient in front of the full Lagrangian \eqref{eq:LargeNLagrangian}, and it does not appear anywhere else, so the $1/N$ expansion is the same as a loop expansion with this Lagrangian (the higher order terms in $1/N$ give corrections to this). For instance, we can now find the beta function for $\omega$ by finding the 1-loop corrections to the $\Sigma^2$ term. In other words, we must study the 1-loop correction to the $\Sigma$ propagator (together with corrections at order $1/N$ to the tree-level result, from the higher order terms in $G_2$). In order to find this to order $1/N$, we need to know $G_2$, $G_3$ and $G_4$ at leading order in $1/N$ (see Figure \ref{fig:a_and_b_type}), and $G_2$ also at the first subleading order, but we do not need to know any of the higher $G_n$, so we will not write them from here on (they will be needed in order to obtain higher orders in $1/N$).
 
Let us start by repeating the calculation of the beta function at $\lambda=0$ using this formalism. We rewrite the relevant terms in the Lagrangian \eqref{eq:LargeNLagrangian} as
\begin{equation}
\mathcal{L}=\frac12\grL G_2(\tilde\omega_0)\grL+\frac{1}{3!}G_3(\tilde\omega_0)\grL^3+\frac{1}{4!}G_4(\tilde\omega_0)\grL^4-N\grL\grS+N(\tilde\omega-\tilde\omega_0)\grS^2,
\end{equation}
and choose $\tilde\omega_0=0$, so that the correlators $G_n$ are computed in the free field theory with no superpotential. A direct calculation gives $G_2(\tilde\omega_0)=\frac{N}{8}\frac{D^2}{|p|}$ (with no corrections in $1/N$). Inverting the quadratic terms in the Lagrangian, one can then find the propagators for $\Lambda,\Sigma$ (denoted by $\Delta_{\Lambda},\Delta_{\Sigma}$) and the two-point function $\langle\Lambda\Sigma\rangle$ (denoted $\Delta_{\Lambda\Sigma}$). For instance, we find
\begin{equation}
\grD_\Lambda=
\frac1N\left(\frac{32 \tilde\omega }{\tilde\omega ^2+16}+\frac{8 \tilde\omega ^2}{\tilde\omega ^2+16}\frac{D^2}{|p|}\right),
\end{equation}
which is just the chain \eqref{eq:chain}, and $\Delta_{\Sigma}$ is by construction proportional to $\langle J J \rangle$, since $\Sigma = J / N$.
We now find the beta function by finding the (diverging) quantum corrections to these propagators. The 1-loop corrections to $\Delta_\Sigma$ that we need to calculate appear in Figure \ref{fig:a_and_b_type}, where the external lines are $\Delta_{\Lambda\Sigma}$ propagators, and the internal lines are $\Lambda$ propagators (note that these are the same as the $(a)$ and $(b)$-type diagrams defined in Section  \ref{sec:Beta_function_lambda0}, which appear in Figure \ref{fig:diags}). Using these diagrams, we can get an equation for $\beta_{\wt}$ and $\gamma_{\Sigma}=\gamma_J$ from the Callan-Symanzik equation\footnote{The Callan-Symanzik equation appearing here is obtained using a $1/N$ expansion of the full Callan-Symanzik equation, see the discussion around \eqref{eq:CS_equation}.} for the two-point function $\langle \Sigma\Sigma\rangle$:
\begin{equation}
	\left(\beta_{\wt}\frac{\partial}{\partial \wt}+2\gamma_\Sigma\right)\Delta_\Sigma+\left((a)+(b)\right)\Delta_{\Lambda\Sigma}^2=0.
\end{equation}
This is manifestly the same as one of the equations we discussed in the previous section; because the term proportional to $\delta^{(2)}(\theta_1-\theta_2)$ is a contact term, it gives us a single non-trivial 
relation between $\beta_{\wt}$ and $\gamma_J$. If we denote
\begin{equation}
(a)+(b)=O_1+O_2\frac{D^2}{|p|}
\end{equation}
as above, we have
\begin{equation}\label{eq:beta_gamma_relation}
\beta_{\wt}=\frac{\gamma_J  N \left(\wt ^2+16\right)+32 O_1 \wt -4 O_2 \left(\wt ^2-16\right)}{N \wt }.
\end{equation}
Using the results of the computations of the previous section gives the following relation between $\beta_{\wt},\gamma_J$:
\begin{equation}
\beta_{\wt}=\frac{\pi ^2 \gamma_J  N \left(\wt ^2+16\right)^3+256 \wt ^2 \left(3 \wt ^2-16\right)}{\pi ^2 N \wt  \left(\wt ^2+16\right)^2},
\end{equation}
in agreement with the direct calculation from Section \ref{sec:Beta_function_lambda0}. We thus find that while this method is not enough in order to find both $\beta_{\wt}$ and $\gamma_J$, one can still obtain a relation between them; additional correlators are needed to separate the two.

In the rest of this section we perform the analysis for general $\lambda$, in the hopes of finding a similar relation. 
For any $\lambda$ we can obtain a relation between  $\beta_\omega,\gamma_J$ by calculating loop corrections to the $\Sigma^2$ term in \eqref{eq:LargeNLagrangian}. For general $\lambda$ there is nothing special about the point $\omega_0=0$ which is no longer free; in fact, it is convenient to use the value $\omega_0=1$, since then the correlators $G_n$ are computed in the theory with enhanced $\mc{N}=2$ supersymmetry. At this value we know that $J$ does not have an anomalous dimension, since it is in the same $\mc{N}=2$ multiplet as the global symmetry current, and we also know that the beta function for $\omega$ vanishes, and thus $G_2$ does not have any logarithmic terms at any order in $1/N$.
Thus, the beta function at order $1/N$ can be obtained from the same diagrams as before, appearing in Figure \ref{fig:a_and_b_type}. We find the 2 and 3-point functions, $G_2$ and $G_3$, in the next section. We then use them to calculate the $(b)$-type diagram. However, we will not be able to find the result for the 4-point function $G_4$ for general $\lambda$, and so we will not be able to calculate the $(a)$-type diagram.

\subsection{Correlation Functions of $J=\bar{\Phi}\Phi$}\label{sec:corr_funcs_of_J}

In this section we compute the two and three-point correlation functions of $J=\bar{\Phi}\Phi$ for all $\lambda$ and $\omega$, at leading order in 
$1/N$, generalizing the non-supersymmetric computations of \cite{Aharony:2012nh,GurAri:2012is}.
We start by calculating the general four-point function $\left<\bar\Phi\Phi\bar\Phi\Phi\right>$ for colinear momenta (this was done for the $\mc{N}=1$ CS-matter theories in \cite{Inbasekar:2015tsa}). This result is then used to find the vertex $\left<J\bar\Phi\Phi\right>$ for colinear momenta, which is then used to calculate $\left<JJ\right>$ and $\left<JJJ\right>$ for general momenta. We will use our results to suggest a possible generalization of the results of Maldacena-Zhiboedov \cite{Maldacena:2011jn,Maldacena:2012sf}.

In order to perform the various 3d integrals that appear in this section, we use the standard formalism \cite{Aharony:2012nh} in which we use a cutoff in the 1-2 plane and dimensional regularization in the 3 direction. In particular, we define $p_s=\sqrt{p_1^2+p_2^2}$. For other conventions see Appendix \ref{app:conventions}.

\subsubsection{Four-Point Function of $\Phi$ at Large $N$}\label{sec:four_pt}

\begin{figure}
	\centering
	\includegraphics[width=0.3\linewidth]{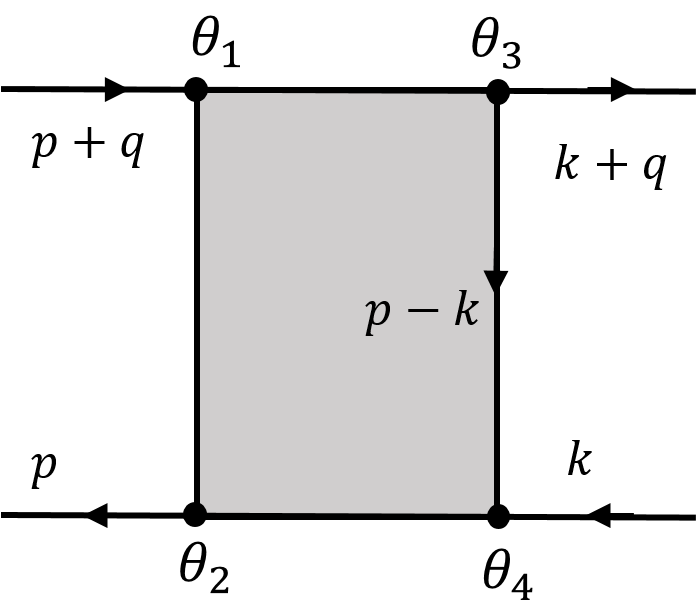}
	\caption{The leading order four-point function}
	\label{fig:4pt}
\end{figure}

The four-point function of $\Phi$ in our theories was found, for some values of the momenta, in \cite{Inbasekar:2015tsa} (in the light-cone gauge $\Gamma_-=0$; note that the result is not gauge-invariant).
At leading order in $1/N$, the correction to the four-$\Phi$ vertex in the effective action appears diagrammatically in Figure \ref{fig:4pt}.
It can be viewed as coming from a four-point vertex in the effective Lagrangian:
\begin{equation}
\mathcal{L}_4=\frac12 V(\theta_i,p,q,k)\grF_a(-p-q,\theta_1)\bar\grF^a(p,\theta_2)\bar\Phi^b(k+q,\theta_3)\Phi_b(-k,\theta_4) .
\end{equation}
The authors of \cite{Inbasekar:2015tsa} found, for $q$ in the $x_3$ direction ($q_+=q_-=0$):
\begin{equation}\label{4pt_vertex}
\begin{split}
V&=\exp\left(\frac14 X\cdot (p\cdot X_{12}+q\cdot X_{13}+k\cdot X_{43})\right) F(X_{12,}X_{13},X_{43},p,q,k),\\
F(X_{12,}X_{13},X_{43},p,q,k)&=X_{12}^+X_{43}^+\left(AX_{12}^-X_{43}^-X_{13}^+X_{13}^-+BX_{12}^-X_{43}^-+CX_{12}^-X_{13}^++DX_{13}^+X_{43}^-\right),
\end{split}
\end{equation}
with $A,B,C,D$ functions of the momenta $p,k,q$, which can be computed from a Schwinger-Dyson equation. Here, $X_{ij}=\theta_i-\theta_j$ and $X=\sum_{i=1}^4\theta_i$. 
In this work we will only need the expressions at the $\mc{N}=2$ point $\omega=1$. Defining $T(x)=e^{2i\grl\tan^{-1}(\frac{2\sqrt{x^2+m^2}}{q_3})}$,
 these are given by:
\begin{equation}
A=-\frac{2i\pi }{\grk}\frac{T(k_s)}{T(p_s)},\qquad
B=0,\qquad
C=D=\frac{2A}{(k-p)_{-}}.
\end{equation}

\subsubsection{Computation of $\left<J\bar\Phi\Phi\right>$}\label{sec:JPhiPhi}

\begin{figure}
	\centering
	\includegraphics[width=0.6\linewidth]{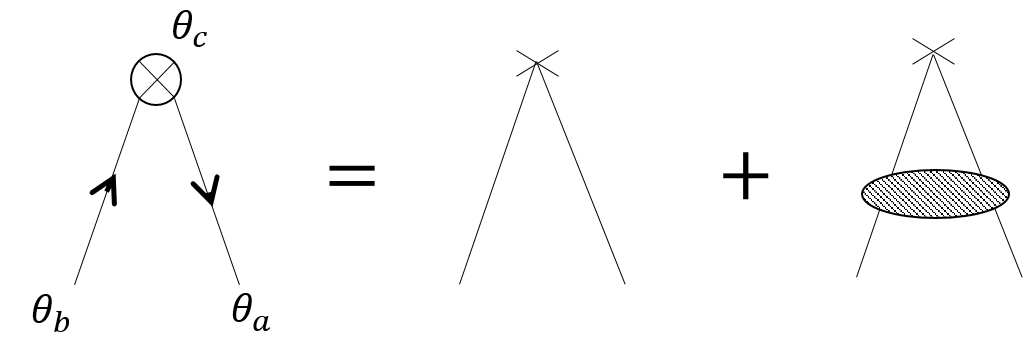}
	\caption{Diagrams contributing to $\langle J \bar\Phi\Phi\rangle$. A cross denotes a $J$ insertion, and the shaded area corresponds to the general $\Phi$ 4-point function discussed in Section \ref{sec:four_pt}.}
	\label{fig:Jphiphi_diagram}
\end{figure}

We can now calculate $\left<J\bar\Phi\Phi\right>$ in the colinear limit. Explicitly, we consider a correlation function of the form
\begin{equation}
\left<J(-q,\theta_c)\bar\Phi(p+q,\theta_b)\Phi(-p,\theta_a)\right>
\end{equation}
where we take $q_+=q_-=0$. Diagramatically, $\left<J\bar\Phi\Phi\right>$ is given by the diagrams shown in Figure \ref{fig:Jphiphi_diagram}.
The first diagram is the contribution from the free theory, while the second diagram includes all of the interaction terms.

The first diagram is simple, and it contributes
\begin{equation} \delta^2(\theta_a-\theta_c)\delta^2(\theta_b-\theta_c). 
\end{equation}
In terms of the decomposition of a general three-point function described in equation \eqref{eq:general_3pt}, this corresponds to $\mc{C}=1$ with all other coefficients vanishing.

\begin{figure}
	\centering
	\includegraphics[width=0.45\linewidth]{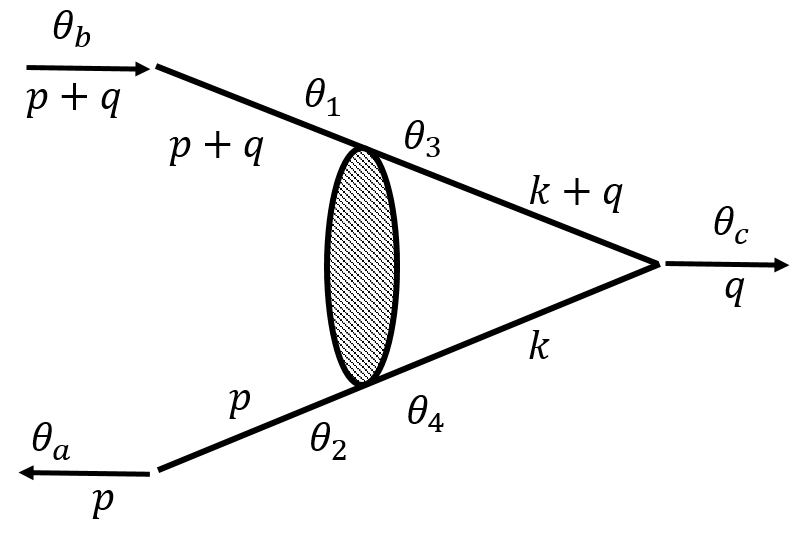}
	\caption{Interacting contribution to $\langle J \bar\Phi\Phi\rangle$}
	\label{fig:jphiphicloseup}
\end{figure}

A more detailed version of the second diagram appears in Figure \ref{fig:jphiphicloseup}.
This contributes:
\begin{equation}
\left<J\bar\Phi\Phi\right>=N\int d^2\theta_3 d^2\theta_4 \int\dpi{k}V(\theta_b,\theta_a,\theta_3,\theta_4,p,q,k)
\frac{D^2\delta(\theta_c-\theta_3)}{(k+q)^2}
\frac{D^2\delta(\theta_4-\theta_c)}{k^2}.
\end{equation}
We can now plug in the result \eqref{4pt_vertex} for the 4-point vertex $V$ in terms of the $\theta$'s, and also replace $D^2_{\theta,k}\grd^2(\theta-\theta')=e^{-\theta\cdot k\cdot \theta'}$. This allows us to do the integrals. 

Summing the two contributions, we find the general result (decomposing the 3-point function according to the general form described in equation \eqref{eq:general_3pt}):
\begin{equation}
\begin{split}
\mc{A}&=0\\
\mc{B}_1&=	\mc{B}_2=-\frac{i e^{-2 i \lambda  \tan ^{-1}\left(\frac{2 p_s}{q_3}\right)} \left(-(\omega-1) e^{i \lambda  \left(2 \tan ^{-1}\left(\frac{2 p_s}{q_3}\right)+\pi  \text{sgn}(q_3)\right)}+(\omega+3) \left(e^{2 i \lambda  \tan ^{-1}\left(\frac{2 p_s}{q_3}\right)}-e^{i \pi  \lambda  \text{sgn}(q_3)}\right)+\omega-1\right)}{ q_3 ((\omega-1) (\omega+3) \cos (\pi  \lambda  \text{sgn}(q_3))-\omega (\omega+2)-5)}\\
\mc{B}_{11}&=\frac{2 e^{i \pi  \lambda  \text{sgn}(q_3)} \left(1-e^{-2 i \lambda  \tan ^{-1}\left(\frac{2 p_s}{q_3}\right)}\right)}{p_{-} \left((\omega+3) e^{i \pi  \lambda  \text{sgn}(q_3)}-\omega+1\right)}\\
\mc{B}_{12}&=-\mc{B}_{21}=i\mc{B}_{1}\\
\mc{B}_{22}&=0\\
\mc{C}&=1+\frac{e^{-i \lambda  \left(2 \tan ^{-1}\left(\frac{2 p_s}{q_3}\right)+\pi \right)}}{6 ((\omega-1) (\omega+3) \cos (\pi  \lambda )-\omega (\omega+2)-5)}\bigg[2 \left(-1+e^{2 i \pi  \lambda }\right) \text{sgn}(q_3) \left((\omega-1) e^{2 i \lambda  \tan ^{-1}\left(\frac{2 p_s}{q_3}\right)}-\omega-5\right)+\\
&\qquad\qquad(\omega-1) (3 \omega+7) \left(-e^{2 i \lambda  \left(\tan ^{-1}\left(\frac{2 p_s}{q_3}\right)+\pi \right)}\right)-(\omega-1) (3 \omega+7) e^{2 i \lambda  \tan ^{-1}\left(\frac{2 p_s}{q_3}\right)}+\\
&\qquad\qquad 2 (\omega (3 \omega+4)+9) e^{i \lambda  \left(2 \tan ^{-1}\left(\frac{2 p_s}{q_3}\right)+\pi \right)}-4 e^{i \pi  \lambda } ((\omega+5) \cos (\pi  \lambda )-\omega+3)\bigg]
\end{split}
\end{equation}
In particular, we find a relatively simple expression at the $\mc{N}=2$ point $\omega=1$:
\begin{equation}
\begin{split}
\mc{A}&=0\\
\mc{B}_1&=	\mc{B}_2=\frac{i \left(1-e^{i \left(\pi  \lambda  \text{sgn}(q_3)-2 \lambda  \tan ^{-1}\left(\frac{2 p_s}{q_3}\right)\right)}\right)}{2 q_3}\\
\mc{B}_{11}&=-\frac{-1+e^{-2 i \lambda  \tan ^{-1}\left(\frac{2 p_s}{q_3}\right)}}{2 p_{-}}\\
\mc{B}_{12}&=-\mc{B}_{21}=i\mc{B}_{1}\\
\mc{B}_{22}&=0\\
\mc{C}&=\frac{1}{6} \left(e^{-2 i \lambda  \tan ^{-1}\left(\frac{2 p_s}{q_3}\right)}+3 e^{i \left(\pi  \lambda  \text{sgn}(q_3)-2 \lambda  \tan ^{-1}\left(\frac{2 p_s}{q_3}\right)\right)}+2\right)
\end{split}
\end{equation}
Due to the symmetry under exchanging the two $\Phi$ legs together with charge conjugation, we expect $\mc{B}_1=\mc{B}_2$ and\footnote{The precise relation is $\mc{B}_{12}(-q,p+q,-p)=-\mc{B}_{21}(-q,-p,p+q)$, but since $q_+=q_-=0$ and since $\mc{B}_{1}$ depends only on $p_s$, this reduces to the form above.} $\mc{B}_{12}=-\mc{B}_{21}$, and this is manifest above. The result has the correct limit as $\lambda\rightarrow 0$ (which is $\mc{C}=1$, with all other terms vanishing). 

\subsubsection{Computation of $\left<JJ\right>$}\label{sec:JJ}

\begin{figure}
	\centering
	\includegraphics[width=0.3\linewidth]{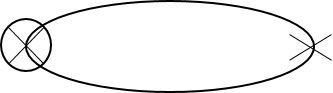}
	\caption{Diagrams contributing to $\langle JJ\rangle$}
	\label{fig:JJ_diagram}
\end{figure}

Next, we calculate the two-point function $\left<J(q)J(-q)\right>$. 
The diagram we need to calculate is shown in Figure \ref{fig:JJ_diagram}, 
which can be calculated using the results of the previous section. Note that the results of the previous section only give us this two-point function when the momentum $q$ obeys $q_-=q_+=0$, but the general result can be immediately found using Lorentz invariance. From the general form of the two-point function in equation \eqref{eq:general_2pt}, we find that the contribution from this diagram should be of the form 
\begin{equation}\label{eq:P1}
\left<J(q,\theta)J(-q,\theta')\right>=
\left(A_0(q)+A_1(q) \frac{D^2}{|q|}\right)\grd^2(\theta-\theta'),
\end{equation}
so that there are only two unknown functions we must calculate, $A_0$ and $A_1$. We find
\begin{equation}\label{eq:2pt_func_C1C2}
\begin{split}
A_0&=\frac{N  (\omega+1) \sin ^2\left(\frac{\pi  \lambda }{2}\right)}{\pi\grl  \left((\omega-1) (\omega+3) \cos (\pi  \lambda )-\omega (\omega+2)-5\right)} \\
A_1&=-\frac{N  \sin (\pi  \lambda )}{\pi\grl  \left((\omega-1) (\omega+3) \cos (\pi  \lambda )-\omega (\omega+2)-5\right)}
\end{split}
\end{equation}
We now perform some consistency checks on this result.

\begin{figure}
	\centering
	\includegraphics[width=0.8\linewidth]{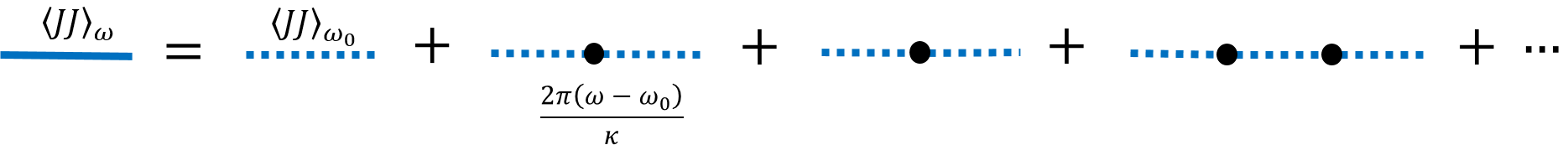}
	\caption{Diagrammatic relation between $\langle JJ\rangle_{\omega_0}$ and $\langle JJ\rangle_{\omega}$. Dashed blue lines correspond to $\langle JJ\rangle_{\omega_0}$, solid blue lines to $\langle JJ\rangle_{\omega}$, and black vertices to $\frac{\pi(\omega-\omega_0)}{\kappa}$ (along with the corresponding symmetry factor). The "chain" $\Delta^{\omega_0}_{\omega}$ can be defined as $\frac{\langle JJ\rangle_{\omega}}{\langle JJ\rangle_{\omega_0}}$.}
	\label{fig:chainsw0tow1}
\end{figure}

\begin{enumerate}
\item In the limit of $\lambda \to 0$ with fixed $\tilde\omega$ we reproduce the 2-point function of the previous section.
	\item $A_1$ is duality invariant. Note that $A_0$ is not duality invariant; however, it is a contact term which vanishes at separated points, so this just indicates that the duality transformation should be accompanied by adding an appropriate contact term (depending on $\lambda$ and $\omega$) for this 2-point function. A similar contact term appears in the 3-point function of scalar operators in the duality map between CS-fermion and CS-critical-scalar theories \cite{GurAri:2012is,Aharony:2018pjn}.
	\item We can find $\left<JJ\right>_\omega$ for general $\omega$ by starting with $\langle JJ\rangle_{\omega=\omega_0}$ for any value of $\omega_0$, and treating $\frac{\pi (\omega-\omega_0)}{\grk}(\bar{\Phi}\Phi)^2$ as a perturbation. In this case, we should have (at leading order in $1/N$)
	\begin{equation}\label{chaindef}
	\left<JJ\right>_\omega= \left<JJ\right>_{\omega_0}\Delta^{\omega_0}_{\omega}
	\end{equation}
	where the ``chain" $\Delta^{\omega_0}_{\omega}$ takes us from $\omega_0$ to $\omega$, as shown in Figure \ref{fig:chainsw0tow1}. Summing this series, one finds
	\begin{equation}\label{eq:Deltaw0w1}
	\Delta^{\omega_0}_{\omega}=  \sum_{n=0}^\infty\left(\frac{2\pi (\omega-\omega_0)}{\kappa}\left<JJ\right>_{\omega_0}\right)^n=
	\frac{1}{1-\frac{2\pi (\omega-\omega_0)}{\kappa}\left<JJ\right>_{\omega_0}}.
	\end{equation}
	A direct calculation shows that \eqref{eq:2pt_func_C1C2} indeed satisfies \eqref{chaindef} for all $\omega$ and $\omega_0$ (and for any $\lambda$).
	\item Finally, we can compare our result to the known results in CS-matter theories with only bosons \cite{Aharony:2012nh} or only fermions \cite{GurAri:2012is}. This is done by noticing that at $\omega=-1$, the term $(\phi^2)(\psi^2)$ which mixes fermions and bosons in the Lagrangian \eqref{eq:L_int} vanishes. In this case, the diagram in Figure \ref{fig:JJ_diagram} contributing to the top component of $\langle JJ\rangle$ has only fermions running in the loop, while the diagram contributing to the bottom component only has bosons running in the loop. This means that the results for the two-point functions of these components should agree with the fermion-only and boson-only calculations. We have verified that this is indeed the case.
\end{enumerate}

\subsubsection{Computation of $\left<JJJ\right>$}\label{sec:JJJ}

\begin{figure}
	\centering
	\includegraphics[width=0.35\linewidth]{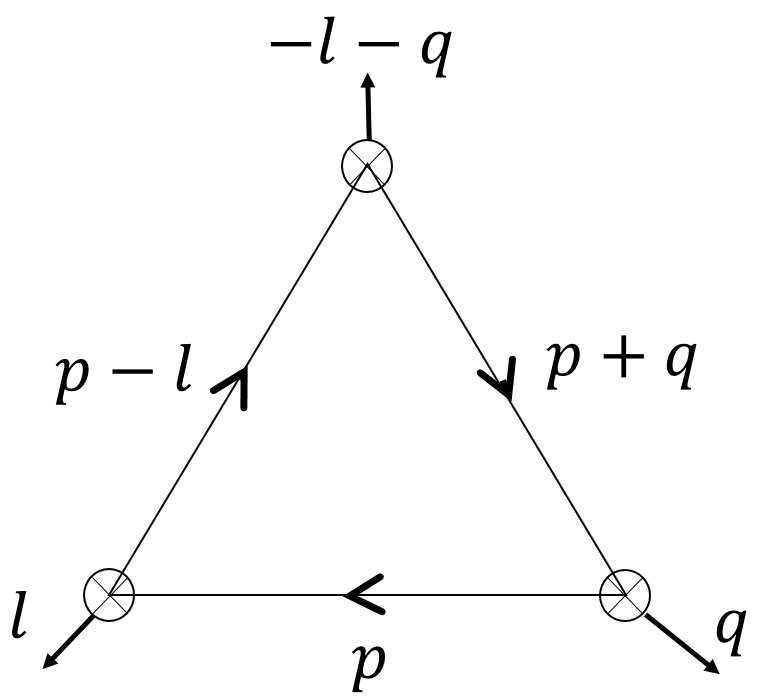}
	\caption{Diagrams contributing to $\langle JJJ\rangle$}
	\label{fig:JJJ_diagram}
\end{figure}

Next we calculate the 3-point function $\langle J(q)J(l)J(-q-l)\rangle_\omega$
, for which the relevant diagram appears in Figure \ref{fig:JJJ_diagram}. Using the $\langle J\bar\Phi\Phi\rangle$ vertex calculated in Section \ref{sec:JPhiPhi}, one can calculate $\langle JJJ\rangle_\omega$ for colinear momenta. However, it is not immediately obvious how to generalize this result to any external momenta, since we do not know the most general form of the 3-point function of superfields that is allowed by superconformal invariance. We now describe the method we used to solve this issue.

Given the three-point function $\langle JJJ\rangle_{\omega_0}$ for some $\omega_0$ (for instance, at the $\mc{N}=2$ point $\omega_0=1$), one can find the 3-point function at any other $\omega$ by multiplying each external leg by the chain \eqref{eq:Deltaw0w1}. By performing this calculation for colinear momenta, we find that there are 3 special values of $\omega$ (denoted $\omega_\lambda^{(i)}$ with $i=1,2,3$) for which $\langle JJJ\rangle_{\omega_\lambda^{(i)}}$ with colinear momenta is proportional to the free theory result (i.e. the result for $\lambda=0$). This fact should also be correct for general momenta\footnote{One might worry about the possibility of an additional structure which vanishes in the colinear limit but not in general, but this cannot be the case here. We can show this by using the effective action formalism. SUSY constrains the effective action giving the 3-point function of $J$'s, when written in superspace, to have eight possible terms that couple three $J$'s (in agreement with the decomposition in Section \ref{sec:general_3pt_func}). These terms come in two forms; terms whose bottom component has three scalars ($JJJ$, $J (D^2 J) J$, $J J (D^2 J)$ or $J(D^2 J)( D^2 J)$) and terms whose bottom component has one scalar and two fermions (e.g. $JD_\alpha J D_\beta J$). For three scalars there is only one possible structure, which we can see in the colinear limit. Similarly, for scalar-fermion-fermion diagrams there are two possible structures \cite{Iliesiu:2015akf} (an odd and an even one), and our colinear results see two structures, so neither of them can vanish there. This leaves only the possibility of contact terms, but these depend at most on a single momentum, so they cannot vanish in the colinear limit. Thus, there cannot be any additional structures that vanish in the colinear limit of the 3-point function but not in general.}, and so at these $\omega_\lambda^{(i)}$ the full result for $\langle JJJ\rangle_{\omega_\lambda^{(i)}}$ for general momenta is just the free result up to an overall factor. Then, applying the chains \eqref{eq:Deltaw0w1} once again to this result, we can find $\langle JJJ\rangle_{\omega}$ for general $\omega$ and general momenta.

Explicitly, the $\omega_\lambda^{(i)}$ for which the three-point functions are proportional to the free theory result are given by:
\begin{equation}\label{eq:omega_lambda}
\omega_\lambda^{(i)}=1-\frac{4}{2 \cos \left(\frac{\pi  (\lambda -2i+2)}{3}\right)+1}
\end{equation}
for $i=1,2,3$. Focusing on the simplest one, $\omega_\lambda^{(1)}$, we find:
\begin{equation}\label{eq:JJJomega_lambda}
\langle JJJ\rangle_{\omega_\lambda^{(1)}} =  \frac{\sin \left(\frac{\pi  \lambda }{3}\right)+\sin \left(\frac{2 \pi  \lambda }{3}\right)}{\pi  \lambda  \left(1-2 \cos \left(\frac{\pi  \lambda }{3}\right)\right)^2}
\langle JJJ\rangle_{\text{free}}
\end{equation}
Then, multiplying by the chains \eqref{eq:Deltaw0w1}, one can find $\langle JJJ\rangle_{\omega}$ for general $\omega$.
We will only explicitly write down the result at the $\mc{N}=2$ point, $\langle JJJ\rangle_{\omega=1}$. In terms of the decomposition discussed in equation \eqref{eq:general_3pt}, we find:
\begin{equation}\label{eq:3pt_func_A}
\begin{split}
\mc{A}&=\frac{ \sin (2 \pi  \lambda )}{2 \pi \grl}\mc{A}^{free}\\
\mc{B}_1&=-2\frac{N \sin ^2(\pi  \lambda)}{ \pi  \lambda|l||l+q|}\\
\mc{B}_2&=-2\frac{N \sin ^2(\pi  \lambda)}{ \pi  \lambda|q||l+q|}\\
\mc{B}_{\gra\grb}&=\frac{ \sin (2 \pi  \lambda )}{2 \pi \grl}	\mc{B}_{\gra\grb}^{free}+
2\frac{N \sin ^2(\pi  \lambda ) }{ 16\pi  \lambda}
\frac{|l|+|q|-|l+q|}{ |l||q||l+q| \left(|l||q|-l\cdot q\right)}
\left(q_{\alpha}^{\;\;\gamma} l_{\gamma\beta}+C_{\alpha\beta}|l||q|\right)\\
\mc{C}&=\frac{ \sin (2 \pi  \lambda )}{ 2\pi\grl}\mc{C}^{free}
\end{split}
\end{equation}
As a consistency check, one finds that this has the correct limit when $\grl\rightarrow 0$. Also note that symmetry under interchanging the $q,l$ legs demands that $\mc{B}_1(q,l)=\mc{B}_2(l,q)$ and $\mc{B}_{12}(q,l)=-\mc{B}_{21}(l,q)$, as is apparent in the terms above. We have also emphasized the appearance of the free theory terms $\mc{A}^{free},\mc{B}_{1}^{free},\cdots,\mc{C}^{free}$ from Section \ref{sec:general_3pt_func}.

There are two more nontrivial checks we can do. The first is to check that the result is duality invariant, which indeed it is (we have also checked this for the more general case $\omega\neq 1$, and the result remains duality invariant). Another check is to compare the top and bottom components of our result to results in theories with only fermions \cite{GurAri:2012is} or only bosons \cite{Aharony:2012nh} at $\omega=-1$, as was done for the two-point function in Section \ref{sec:JJ}. Again, we find that the results agree\footnote{The fermionic result agrees only up to an overall sign, which is due to a different convention than the one used in \cite{GurAri:2012is}.}.

As a final comment, we note that there only appear two structures in our result \eqref{eq:3pt_func_A} at $\omega=1$: a ``free" structure appearing in the terms whose coefficient is proportional to $\sin(2\pi \lambda) / \lambda$, and an ``odd" structure appearing in the terms whose coefficient is proportional to $\sin^2(\pi \lambda) / \lambda$. Surprisingly, this three-point function is composed of the same two structures (with coefficients depending on $\omega$ and $\lambda$) for all other values of $\omega$ as well. We discuss this further in the next section. 

\subsubsection{Discussion}

Above we have calculated the two and three-point correlation functions of $J=\bar{\Phi}^a\Phi_a$ for an $\mc{N}=1$ CS-matter theory at leading order in $1/N$ (these are equations \eqref{eq:2pt_func_C1C2} and \eqref{eq:3pt_func_A} respectively). There are several important properties of these correlation functions that we now discuss.

First, all of the correlation functions above (at separated points) are duality-invariant under the duality described in Section \ref{sec:lagrangian}. We consider this as further evidence for these dualities.

Second, we emphasize that the three-point function becomes proportional to the free result for some values of the couplings (denoted $\omega_{\lambda}^{(i)}$). This behavior can be traced back to the fact that the three-point function is made up of only two structures with some coefficients, and so it is reasonable that one can tune the couplings $\omega$ and $\lambda$ so that the coefficient of the ``interacting" structure vanishes, leaving only the free structure. This fact simplifies any calculation which requires the three-point function, since the three-point function at any $\omega$ is related to the three-point function at $\omega_{\lambda}^{(i)}$ through the chains \eqref{eq:Deltaw0w1}. 

Next, we comment on the calculation of the four-point function, which we need in order to obtain the full one-loop correction to the $\Sigma$ propagator as discussed above. For the Chern-Simons-fermion theory, this was computed in \cite{Turiaci:2018nua} (following \cite{Bedhotiya:2015uga})  for colinear momenta, and in principle we can perform a similar computation also in our case. However, apart from the added technical difficulties in this calculation, there is an additional difficulty in inferring the result for general momenta from the result for colinear momenta, which requires additional information. In the CS-fermion theory, \cite{Turiaci:2018nua} showed that the general result for the 4-point function is determined by the inversion formula \cite{Caron-Huot:2017vep} (see also \cite{Simmons-Duffin:2017nub}) up to a finite number of coefficients, and that the colinear limit is enough to fix these coefficients. A similar analysis may be possible also in our case, but the inversion analysis is much more complicated (the CS-fermion analysis was simplified by the fact that all 3-point functions of two $J$'s with other operators are proportional to their value in the free theory, which is not true for our theories). We leave this to future work.

Finally, our results for the 3-point functions above allow us to conjecture a possible generalization of the results of Maldacena and Zhiboedov \cite{Maldacena:2011jn,Maldacena:2012sf} to $\mc{N}=1$ supersymmetric theories. We review these results and our conjecture for the $\mc{N}=1$ generalization in Appendix \ref{app:MZ}. 
We conjecture there that any three-point function of the approximately-conserved higher spin superfields $J_s$ in these CS-matter theories is of the following form:
\begin{equation}
\langle J_{s_1}J_{s_2}J_{s_3} \rangle = \alpha_{s_1s_2s_3}\langle J_{s_1}J_{s_2}J_{s_3} \rangle_{free}+\beta_{s_1s_2s_3}\langle J_{s_1}J_{s_2}J_{s_3} \rangle_{odd}
\end{equation} 
where the first structure is the result in the free theory ($\lambda=\omega=0$) of a single matter multiplet. Some constraints on the coefficients $\alpha_{s_1s_2s_3},\beta_{s_1s_2s_3}$ are discussed in Appendix \ref{app:MZ}.

\subsection{Beta Function for $\lambda\neq 0$}

We now return to the calculation of the beta function. We use the method described in Section \ref{sec:beta_func_general}, for which we are required to find $G_2,G_3,G_4$. 
Using our result for $\langle JJ\rangle$, at leading order in $1/N$ we find
\begin{equation}
G_2(\omega_0=1)=\frac{N}{8\pi\lambda}\left(  (\cos (\pi  \lambda )-1)+\sin (\pi  \lambda )\frac{D^2}{|q|}\right) .
\end{equation}
This allows us to calculate the $\Sigma$ propagator:
\begin{equation}\label{eq:Delta_Sigma}
	\Delta_{\Sigma}=\frac{1}{\pi  \lambda  N ((\omega-1) (\omega+3) \cos (\pi  \lambda )-\omega (\omega+2)-5)}\left((\omega+1) \sin ^2\left(\frac{\pi  \lambda }{2}\right)-\sin\left(\pi  \lambda \right)\frac{D^2}{|p|}\right).
\end{equation}  
Note that this reduces to the result for the chain \eqref{eq:chain} in the limit $\lambda\to 0$ (when rewriting the result in terms of $\tilde{\omega}=\pi\omega\lambda$).

We need to calculate the two diagrams in Figure \ref{fig:a_and_b_type}. We can calculate the $(b)$-type diagram (which requires only $G_3$ that we computed above), but not the $(a)$-type diagram (which requires $G_4$, the $J$ four-point function, which we do not have).

\subsubsection{$(b)$-Type Diagram}
As discussed in Section \ref{sec:JJJ}, For any two values of the coupling $\omega_{1},\omega_2$, we can relate the corresponding three-point functions by multiplying their external legs by the chains \eqref{eq:Deltaw0w1}. Schematically, we write
\begin{equation}
\langle JJJ \rangle_{\omega_1}=\langle JJJ \rangle_{\omega_2}\left(\Delta_{\omega_1}^{\omega_2}\right)^3 .
\end{equation}
Furthermore, we found that for specific values of $\omega=\omega_{\lambda}^{(i)}$ given in \eqref{eq:omega_lambda}, the three-point function $\langle JJJ\rangle$ is proportional to the free theory result $\langle JJJ\rangle_{free}$. We thus find that \textit{any} three-point function is proportional to the free result, up to multiplication by external legs:
\begin{equation}\label{eq:prop_to_free}
\langle JJJ \rangle_{\omega}=C\left(\omega_\lambda^{(i)}\right)\langle JJJ \rangle_{free}\left(\Delta_{\omega}^{\omega_\lambda^{(i)}}\right)^3 .
\end{equation}
The exact value for one such proportionality constant $C\left(\omega_\lambda^{(i)}\right)$ is given in \eqref{eq:JJJomega_lambda}, and we will be using this value from now on.
\begin{figure}
	\centering
	\begin{subfigure}{0.3\textwidth}
		\centering
		\includegraphics[height=1.2in]{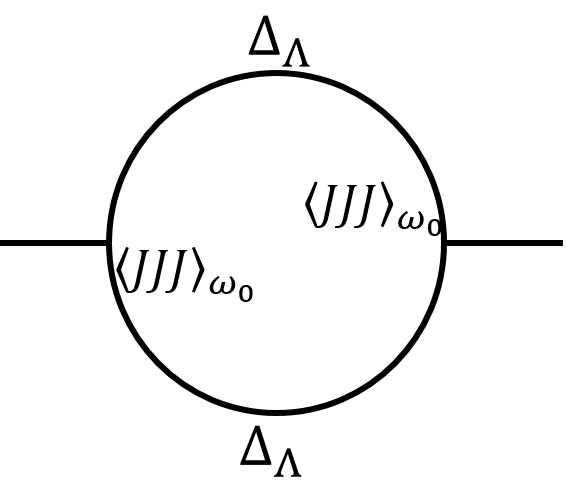}
		\caption{}
		\label{fig:btype1}
	\end{subfigure}%
	~ 
	\begin{subfigure}{0.3\textwidth}
		\centering
		\includegraphics[height=1.2in]{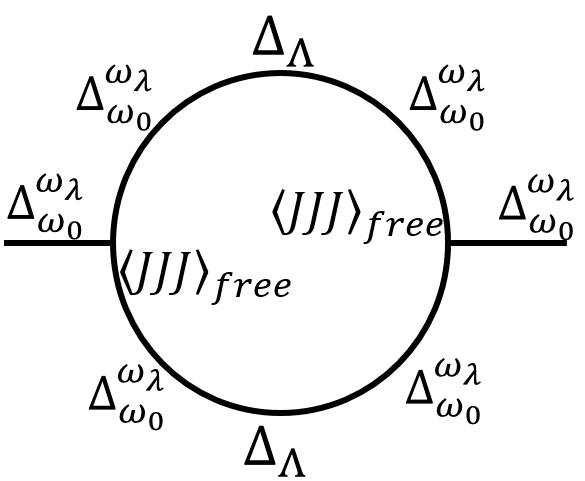}
		\caption{}
		\label{fig:btype2}
	\end{subfigure}
	~ 
	\begin{subfigure}{0.3\textwidth}
		\centering
		\includegraphics[height=1.2in]{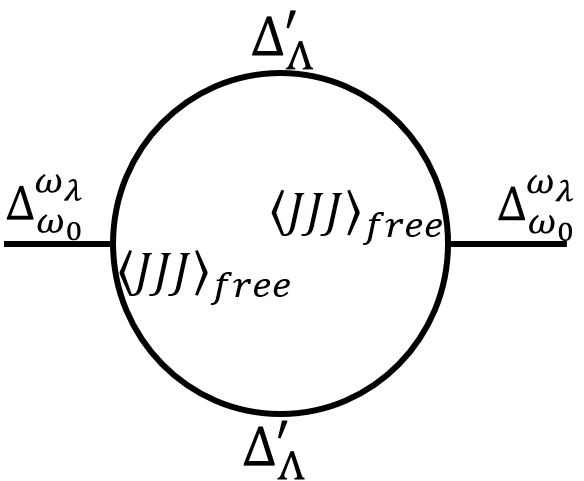}
		\caption{}
		\label{fig:btype3}
	\end{subfigure}
	\caption{Simplifying the $(b)$-type diagram}
	\label{fig:btype}
\end{figure}
We can now easily calculate the $(b)$-type diagram, which appears in Figure \ref{fig:btype1}. Using the identity \eqref{eq:prop_to_free}, we can rewrite this diagram as in Figure \ref{fig:btype2}. Finally, defining 
\begin{equation}\label{eq:delta_prime}
\Delta_{\grL}'=\left(\Delta_{\omega_0=1}^{\omega_\lambda}\right)^2\Delta_{\Lambda}
\end{equation} 
we find that this is equivalent to the diagram in Figure \ref{fig:btype3}. However, note that this final diagram is precisely the same as the $(b)$-type diagram in the $\lambda=0$ case appearing in Figure \ref{fig:diags}, apart from the fact that we have different values for $a_0,a_1$ on the internal ``chain'' (which can be read off from \eqref{eq:delta_prime}).

We thus find that this diagram is almost identical to the one we calculated for the $\lambda=0$ theory in Section \ref{direct_calculation}. In the $\lambda=0$ theory, the result was (ignoring external leg factors of $\Delta_{\Phi^2\Phi^2}$):
\begin{equation}
(b)_{\lambda=0}=-\frac{1}{32\pi^2\gre}\left( a_0+a_1\frac{D^2}{|p|}
\right)^2 \frac{D^2}{|p|} .
\end{equation}
The result for general $\lambda$ should be identical, up to the following changes:
\begin{itemize}
	\item $a_0,a_1$ of \eqref{eq:chain} must be changed to the corresponding values coming from $\Delta_{\Lambda}'$.
	\item The three-point function is not equal to the free value, but only proportional to it, with proportionality constant $\frac{\sin \left(\frac{\pi  \lambda }{3}\right)+\sin \left(\frac{2 \pi  \lambda }{3}\right)}{\pi  \lambda  \left(1-2 \cos \left(\frac{\pi  \lambda }{3}\right)\right)^2}$. We must multiply each three-point function by this proportionality constant.
	\item We must multiply the external legs by $\Delta_{\omega_0=1}^{\omega_\lambda}$.
\end{itemize}
In summary, we find for the logarithmically diverging parts:
\begin{equation}\label{eq:b_type}
	(b)=\left(\frac{\sin \left(\frac{\pi  \lambda }{3}\right)+\sin \left(\frac{2 \pi  \lambda }{3}\right)}{\pi  \lambda  \left(1-2 \cos \left(\frac{\pi  \lambda }{3}\right)\right)^2}\right)^2
	\left(-\frac{1}{32\pi^2\gre}\left(\Delta_\Lambda'
	\right)^2 \frac{D^2}{|p|}\right) \left(\Delta_{\omega_0=1}^{\omega_\lambda}\right)^2 .
\end{equation} 
As a consistency check, this has the correct $\lambda\to 0$ limit and is duality-invariant.

\subsection{Summary}

Putting together the $(a)$ and $(b)$-type diagrams, we can solve the Callan-Symanzik equation:
\begin{equation}\label{eq:full_CS}
\left(\beta_\omega \frac{\partial}{\partial \omega}+2\gamma_J\right) \Delta_\Sigma+((a)+(b)) \Delta_{\Lambda\Sigma}^2=0
\end{equation}
The $(b)$ contribution appears in equation \eqref{eq:b_type}, $\Delta_\Sigma$ appears in equation \eqref{eq:Delta_Sigma}, and we are missing the $(a)$ contribution. As discussed above, only the $\frac{D^2}{|p|}$ term in the equation above is physical, meaning that even if we compute this contribution, we also need $\gamma_J$ at order $1/N$ in order to find $\beta_{\omega}$ at this order.

\section{Qualitative Behavior of $N_f=1$ Fixed Points for $\lambda\neq 0$}\label{sec:qualitative_lambda_positive}

In the previous section we did not manage to find the explicit solution for the beta function for all $\lambda$. However, in this section we will show that our results above (with some additional arguments) are enough in order to conjecture the qualitative behavior of the fixed points for all $\lambda$. We start by showing that the beta function must have at most $6$ roots for all values of $\lambda$. We then discuss all of the exact results one can obtain by combining the results of Section \ref{sec:Beta_function_lambda0}, the duality \eqref{eq:duality} and general considerations from the symmetries of the theories. We then conjecture the qualitative form of the fixed points for general $\lambda$. Finally, we discuss some interesting consequences of these results.

\subsection{Upper Bound on Number of Fixed Points}

We now use the calculation outlined in Section \ref{sec:beta_func_general} to show that there are at most $6$ fixed points for all values of $\lambda$. This is done by showing that the beta function must be of the form 
\begin{equation}\label{eq:general_beta}
\beta_\omega=\frac{P_6(\omega,\lambda)}{N Q(\omega,\lambda)^2},
\end{equation} 
where 
\begin{equation} 
Q(\omega,\lambda)=(\omega-1) (\omega+3) \cos (\pi  \lambda )-\omega (\omega+2)-5.
\end{equation}
Note that $Q(\omega,\lambda) \leq 0$, and it vanishes only for $\lambda=1$ and $\omega=-1$; our discussion below is relevant away from this value.
Here and in the rest of this section, $P_n(\omega,\lambda)$ stands for an arbitrary $n$-th order polynomial in $\omega$, and not any specific polynomial.

First, we show that some combination of $\beta_{\omega}$ and $\gamma_J$ is of the form \eqref{eq:general_beta}. In equation \eqref{eq:full_CS} we have found a Callan-Symanzik equation which relates $\beta_{\omega},\gamma_J$:
\begin{equation}\label{eq:CS_again}
 \left(\beta_\omega \frac{\partial}{\partial \omega}+2\gamma_J\right) \Delta_\Sigma+((a)+(b)) \Delta_{\Lambda\Sigma}^2=0,
\end{equation}
where $(a)$,$(b)$ appear in Figure \ref{fig:a_and_b_type}. We can obtain most of the $\omega$ dependence of this expression from the discussion above. We start by writing the propagators explicitly:
\begin{align}
\Delta_{\Lambda\Sigma}&=\frac{2 (\omega+3-(\omega-1) \cos (\pi  \lambda ))-2 (\omega-1) \sin (\pi  \lambda )\frac{D^2}{|p|}}{NQ(\omega,\lambda)}\\
\Delta_{\Sigma}&=\frac{(\omega+1) \sin ^2\left(\frac{\pi  \lambda }{2}\right)+\sin (\pi  \lambda )\frac{D^2}{|p|}}{NQ(\omega,\lambda)}\\
\Delta_{\Lambda}&=-\frac{4 \pi  \lambda  (\omega-1) ((\omega-1) \cos (\pi  \lambda )-\omega-3)+4 \pi  \lambda  (\omega-1)^2 \sin (\pi  \lambda )\frac{D^2}{|p|}}{NQ(\omega,\lambda)}
\end{align}
Now, note that the $(a)$-type diagram has a single $\Delta_{\Lambda}$ propagator \eqref{eq:Delta_Sigma}, which gives its full $\omega$-dependence, while the $(b)$-type diagram has two such propagators. In total, we find that the $\frac{D^2}{|p|}$ component of the Callan-Symanzik equation \eqref{eq:CS_again} reduces to
\begin{equation} \left(\beta_\omega \frac{\partial}{\partial \omega}+2\gamma_J\right) \frac{\sin (\pi  \lambda )}{Q(\omega,\lambda)}+\frac{P_6(\omega,\lambda)}{NQ(\omega,\lambda)^4}=0,
\end{equation}
which can be simplified to
\begin{equation}\label{eq:CS_constraint_1}
\beta_\omega (\omega+1) f(\lambda)+2 Q(\omega,\lambda) \gamma_J+\frac{P_6(\omega,\lambda)}{NQ(\omega,\lambda)^2}=0,
\end{equation}
and so we have found one combination of $\beta_{\omega}$ and $\gamma_J$ which is of the form \eqref{eq:general_beta}.

\begin{figure}
	\centering
	\begin{subfigure}{0.23\linewidth}
		\centering
		\includegraphics[width=\textwidth]{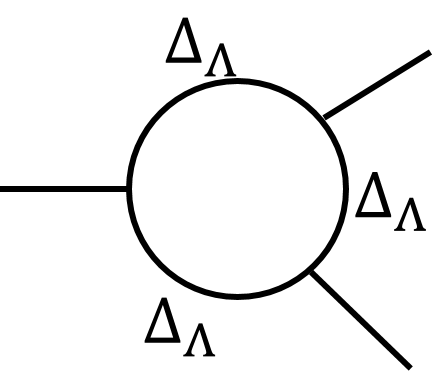}
		\caption{$\mc{O}_3$}
		\label{fig:O3O4O5a}
	\end{subfigure}
	\qquad
	\begin{subfigure}{0.25\linewidth}
		\includegraphics[width=\textwidth]{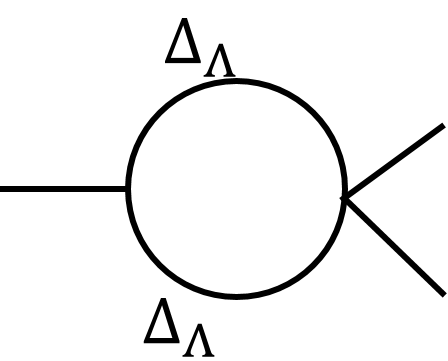}
		\caption{$\mc{O}_4$}
	\end{subfigure}
	\qquad	
	\begin{subfigure}{0.20\linewidth}
		\includegraphics[width=\textwidth]{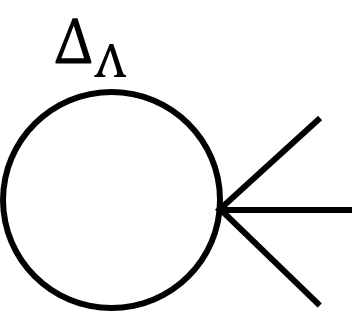}
		\caption{$\mc{O}_5$}
	\end{subfigure}
	\caption{Diagrammatic expressions for $\mc{O}_3$,$\mc{O}_4$,$\mc{O}_5$. Each external leg can be either a $J$ or a $J_1$, with a $\Delta_{\Lambda \Sigma}$ connected to the $J$ leg.}
	\label{fig:O3O4O5}
\end{figure}

In order to constrain $\beta_\omega$ on its own, we need one more such combination. In analogy to the non-SUSY version of these CS-matter theories, the $\mc{N}=1$ CS-matter theories contain an infinite tower of approximately-conserved higher-spin superfields $J_s$ (by which we mean that their twists differ from 1 by terms of order $\frac1N$). Let us consider the Callan-Symanzik equation for the three-point function $\langle J J_1J_1\rangle$, where $J_1$ is the approximately-conserved superfield whose lowest component has spin one (see \cite{Nizami:2013tpa} for a precise definition). An argument similar to the one made around equation \eqref{eq:Deltaw0w1} shows that at leading order in $1/N$ we can calculate $\langle JJ_1J_1\rangle_\omega$ by calculating $\langle JJ_1J_1\rangle_{\omega=1}$ and then multiplying by a single chain. Schematically we write
\begin{equation} 
\langle JJ_1J_1\rangle_\omega=\langle JJ_1J_1\rangle_{\omega=1}\Delta^{\omega=1}_\omega .
\end{equation}
This chain is just $\Delta^{\omega=1}_\omega=\frac{\Delta_{\Sigma}(\omega)}{\Delta_{\Sigma}(\omega=1)}$, which means that we have the entire $\omega$-dependence of this three-point function:
\begin{equation} \label{zerooneone}
\langle JJ_1J_1\rangle_\omega=g(\lambda,\theta_i)\Delta_{\Sigma}(\omega,\lambda) ,
\end{equation}
where we have emphasized the fact that the function $g$ depends on the superspace coordinates. We can now write a Callan-Symanzik equation for this three-point function at order\footnote{Note that $J_1$ is conserved  and so its anomalous dimension vanishes.} $1/N$:
\begin{equation} 
\left(\beta_\omega \frac{\partial}{\partial \omega}+\gamma_J\right) \langle JJ_1J_1\rangle+(\mc{O}_3+\mc{O}_4+\mc{O}_5) \Delta_{\Lambda\Sigma}=0,
\end{equation}
where the 1-loop contributions to $\mc{O}_3,\mc{O}_4,\mc{O}_5$ appear in Figure \ref{fig:O3O4O5}. These diagrams are written as we would compute them using an effective action similar to the one of Section \ref{sec:beta_func_general}, with sources also for $J_1$ and not just for $\Sigma \propto J$.
Again, using the $\omega$-dependence of the various propagators (and focusing on any contribution to the 3-point function which at leading order in $1/N$ depends only on the $\frac{D^2}{|p|}$ component of $\Delta_{\Sigma}$ in \eqref{zerooneone}), this Callan-Symanzik equation reduces to 
\begin{equation}\label{eq:CS_constraint_2} 
\beta_\omega (\omega+1) f(\lambda)+\gamma_J Q(\omega,\lambda)+\frac{P_7(\omega,\lambda)}{NQ(\omega,\lambda)^2}=0,
\end{equation}
and so we have another independent combination of $\beta_{\omega}$ and $\gamma_J$ which is similar to the form \eqref{eq:general_beta}.

Comparing the $\omega$-dependence of equations \eqref{eq:CS_constraint_1},\eqref{eq:CS_constraint_2} we find that we must have
\begin{equation}
\beta_\omega (\omega+1)=\frac{P_7(\omega,\lambda)}{NQ(\omega,\lambda)^2}, 
\end{equation}
or
\begin{equation}\label{eq:almost_final_beta}
\beta_\omega =\frac{P_7(\omega,\lambda)}{N(\omega+1)Q(\omega,\lambda)^2} .
\end{equation}
As a final step, we will show that the factor of $(\omega+1)$ in the denominator must cancel with the numerator. Note that if $P_7$ does not have a root at $\omega=-1$ for generic $\lambda$, then for generic $\lambda$'s our expression \eqref{eq:almost_final_beta} has a pole at $\omega=-1$. However, the duality transformation \eqref{eq:duality} relates $\omega=-1$ and $\omega=\infty$, and the beta function \eqref{eq:almost_final_beta} for the dual coupling has no pole at $\omega=\infty$; since $Q$ is quadratic in $\omega$, the right-hand side of \eqref{eq:almost_final_beta} grows at most as $\omega^2$ for large $\omega$, so the beta function for $1/\omega$ grows at most as a constant. We thus find that in order for the result \eqref{eq:almost_final_beta} to be duality invariant, the numerator must always have a root at $\omega=-1$. We can then write
\begin{equation}
\beta_\omega =\frac{P_6(\omega,\lambda)}{NQ(\omega,\lambda)^2} 
\end{equation}
which agrees with the desired result \eqref{eq:general_beta}. In particular, we learn that the beta function has at most six roots for every $\lambda$ (for large enough $N$, when we can ignore higher contributions in $1/N$).

\subsection{Exact Results and Conjecture}\label{sec:beta_conjecture}

We now outline our conjecture for the qualitative behavior of the fixed points. In Section \ref{sec:Beta_function_lambda0}, we found six fixed points for $\lambda=0$, at
\begin{equation}\label{eq:wt_fp}
\tilde\omega_c=0,0,0,\sqrt{48},-\sqrt{48},\infty.
\end{equation}
We would now like to expand this result as much as possible. We start by finding the fixed points in terms of the coupling $\omega$ in the $\mc{N}=1$ Lagrangian \eqref{eq:matter_Lagrangian}. Perturbation theory and parity considerations will then give us the behavior of the fixed points at leading order in $\lambda$. Then, using the duality, we manage to find the fixed points at strong coupling, for $1-\lambda\ll 1$. We know that the $\mc{N}=2$ point $\omega=1$ is always a fixed point. Finally, we find that the point $(\lambda=\frac12,\omega=-3)$ has an emergent time-reversal symmetry which forces it to be a fixed point. These facts allow us to give a conjecture for the behavior of the fixed points for general $\lambda$.   

We start by finding the fixed points in terms of $\omega$ instead of $\tilde\omega$.
Using the definition $\tilde\omega=\pi\lambda\omega$ and our results for the fixed points $\tilde{\omega}_c$ \eqref{eq:wt_fp}, we immediately find three fixed points at $\omega_c=\frac{\sqrt{48}}{\pi\lambda},-\frac{\sqrt{48}}{\pi\lambda},\infty$ for small $\lambda$. The fixed points at $\tilde \omega=0$ require a little more work, and cannot be found just from our results here; but they can be found from a 2-loop computation at small $\tilde\omega$ and $\lambda$, that was performed in \cite{Avdeev:1992jt}. They found three fixed points at small $\lambda$ and $\tilde\omega$, which in our normalizations are at $\omega_c=-1,-\frac13,1$. We can identify these with the three fixed points at $\tilde{\omega}=0$ that we saw for $\lambda=0$. To summarize, we find six fixed points for small $\lambda$, at
\begin{equation} \label{smalllfp}
\omega_c=-\frac{\sqrt{48}}{\pi\lambda},\;-1,\;-\frac{1}{3},\;1,\;\frac{\sqrt{48}}{\pi\lambda},\;\infty
\end{equation}
We are assuming $\lambda>0$ without loss of generality, and in general the fixed points depend only on $|\lambda|$ by parity. The fixed points at $\omega_c=1$, $\omega_c=-1$ and $\omega_c=\infty$ are attractive in the IR, and the other three are repulsive.

Next, let us discuss the meaning of the fixed point at $\omega=\infty$. We can rewrite the superpotential of our CS-matter theory $W=\frac{\pi\lambda\omega}{N}|\Phi|^4$ by introducing an auxiliary superfield $H$ and taking
\begin{equation}\label{eq:singlet_W}
W=H|\Phi|^2-\frac{N}{4\pi\lambda\omega}H^2,
\end{equation}
since integrating out $H$ leads back to the original theory. This suggests that also if we couple our CS-matter theory with $\omega=0$ to an extra dynamical superfield $H$ with the superpotential \eqref{eq:singlet_W}, then at low energies it would flow to our CS-matter theory with the parameter $\omega$, perhaps with small corrections coming from the dynamics of $H$ (which are suppressed by $1/N$); $H$ is generically very massive and can be integrated out. So we can describe our theories either in the original language, or in this new language, and in the new language the natural coupling is $1/\omega$ rather than $\omega$. This suggests that the theories with $\omega \gg 0$ and $\omega \ll 0$ are in fact similar, so that the space of couplings $\omega$ is actually topologically a circle rather than a line; for large $|\omega|$ we should use the alternative parameterization \eqref{eq:singlet_W} of the space of theories in terms of $1/\omega$. The large $\omega$ behavior of the beta function is consistent with having at most six fixed points on this circle for large $N$, where one of these may be at infinity (as we found for $\lambda=0$).

For $\lambda=0$ we can write the new parameter as $1/{\tilde \omega}$, and in the language of \eqref{eq:singlet_W} we see that if we assign odd parity to $H$ then the coupling $1/\tilde\omega$ breaks parity, so its beta function has to vanish at $\tilde\omega=\infty$, consistent with what we found above. For $\lambda \neq 0$ we have no parity symmetry, so it seems that there is nothing special about the point $\omega=\infty$, and the fixed point there can move to a finite value of $1/\omega$.

Next, let's discuss how our results \eqref{smalllfp} would change when we slightly increase $\lambda$. Note that since $\beta_\omega$ is invariant under parity, the points $-1,-1/3,1$ can only be corrected at order $O(\lambda^2)$. Similarly, the points $\pm\frac{\sqrt{48}}{\pi\lambda}$ can only be corrected at order $O(\lambda)$. Also note that the fixed point at $\omega=1$ has $\mc{N}=2$ SUSY, and so it is exact to all orders in $\lambda$ since the coupling is not renormalized. Perfoming a similar analysis also near $\omega=\infty$, we can write the full set of fixed points for small $\lambda$ as:
\begin{equation} \omega_c=-\frac{\sqrt{48}}{\pi\lambda}+O(\lambda),\;
-1+O(\lambda^2),\;
-\frac13+O(\lambda^2),\;
1,\;
\frac{\sqrt{48}}{\lambda}+O(\lambda),\;O(1/\lambda^2).
\end{equation}

We can now use the duality \eqref{eq:duality} to learn about the behavior near $\lambda=1$. The duality maps\footnote{More precisely, starting from $\lambda>0$ we use a combination of a duality transformation and a parity transformation to go to $\lambda \to 1$.} the six fixed points at $\lambda=0$ to six fixed points at $\lambda=1$:
\begin{equation} 
\omega_c=-1,\;-1,\;-1,\;1,\;5,\;\infty.
\end{equation}
For $\lambda$ close to $1$, their leading order behavior can be read off from the behavior of the corresponding point for $\lambda$ close to zero. We do not have a direct argument explaining why one of the fixed points for $\lambda=1$ is at $\omega=\infty$ (in the language of the previous section, the existence of a fixed point at infinity depends on whether the $\omega^6$ term in $P_6(\omega,\lambda)$, which can only come from the diagram of Figure \ref{fig:O3O4O5a}, is present or not).

Next, let us show that there must be a fixed point at $(\omega,\lambda)=(-3,1/2)$. First, note that at $\lambda=1/2$, which maps into itself by a duality transformation followed by a parity (time-reversal)  transformation, the duality must map the set of zeros of the beta function into itself. Since there is a zero at $\omega=1$, which is a fixed point of the duality, we learn that there must be at least one more root that sits at a fixed point of the duality. Thus, at $\lambda=1/2$ there must be at least another root either at $\omega=1$ or $\omega=-3$. Let us show that at least one additional root appears at $\omega=-3$ by studying the behavior of the point at $(\omega,\lambda)=(-3,1/2)$ under time reversal. In addition to the usual definition of the time reversal transformation $T$, we can define another time reversal transformation $T'$, which is defined as
\begin{equation}\label{eq:Tprime}
	T'=T\circ D
\end{equation}
where $T$ is the standard time reversal transformation, and $D$ is the action of the duality \eqref{eq:duality}. It is easy to see that at the point $(\omega,\lambda)=(-3,1/2)$, $T'$ is an emergent symmetry (this is also true at the self-dual $\mc{N}=2$ point $(\omega,\lambda)=(1,1/2)$). If we deform the theory at $(\omega,\lambda)=(-3,1/2)$ by a small deformation $\delta\omega$, we find $T':\delta\omega\to-\delta\omega$, so that it breaks the $T'$ symmetry. We conclude that we cannot generate a $|\Phi|^4$ term in the superpotential along the RG flow that starts at this point, since this would break the symmetry $T'$. This is thus a fixed point of the RG flow.

\begin{figure}
	\centering
	\includegraphics[width=0.7\textwidth]{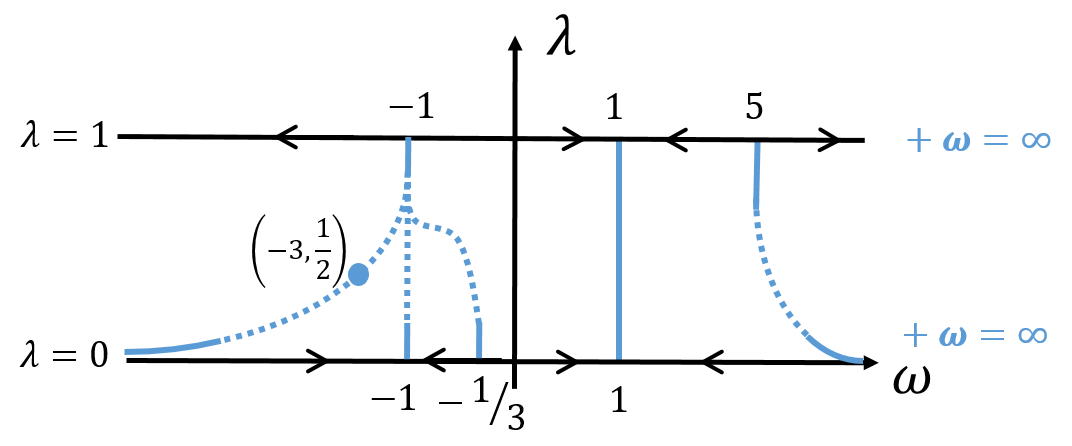}
	\caption{Qualitative RG flows of the $\mc{N}=1$ $N_f=1$ CS-matter theories at large $N$. The solid lines (and the point $(-3,\frac12)$) represent exact results that follow from a direct analysis of the $\beta$ function for small $\lambda$. The dashed lines are the conjectured behavior for all $\lambda$. There is an additional fixed point at $\omega=\infty$ for $\lambda$ close to zero and close to one, whose behavior for general $\lambda$ cannot be found using our analysis. The results for negative $\lambda$ can be obtained by using a parity transformation $\lambda\to -\lambda$, under which the set of roots of the beta function is invariant. Note that we draw $\lambda$ as a continuous variable, even though for finite $N$ it is discrete.}
	\label{fig:RootPlot}
\end{figure}

We present our exact results for the RG flows discussed above using solid lines in Figure \ref{fig:RootPlot}.
Using these results, we conjecture a qualitative picture of the RG flows points for general $\lambda$ using dashed lines in Figure \ref{fig:RootPlot}, for large values of $N$. While this conjecture is the simplest way in which we can connect the RG flows at small $\lambda$ with those at $\lambda$ close to one, it can fail in a number of ways. However, since we have shown that for large $N$ there are at most six roots for the beta function, there cannot be additional pairs of roots which appear for $0<\lambda<1$ without some other pair annihilating. In the simplest conjecture there are three stable fixed points for all values of $\lambda$. One of these is the $\mc{N}=2$ point which maps to itself under the duality, and the two other stable fixed points are exchanged by the duality. Similarly there are three unstable fixed points (which have two $\mc{N}=1$-preserving relevant operators) for all $\lambda$, one which maps to itself under the duality (and which includes the point $(\omega,\lambda)=(-3,1/2)$) and two which are exchanged by it.

Finally, we can discuss deformations of these fixed points which don't preserve $\mc{N}=1$ SUSY. For $\lambda$ close to zero, these deformations were studied in \cite{Avdeev:1992jt} for the three fixed points close to the origin ($\omega=-1,-\frac13,1$). We will follow the notation of this paper. There are four classically marginal operators:
\begin{align}
\mc O_{\alpha}& = (\bar{\psi}\psi)(\bar{\phi}\phi),\nonumber  \\
\mc O_{\beta}& = (\bar{\psi}\phi)(\bar{\phi}\psi),\nonumber  \\
\mc O_{\gamma}& = \frac14\left((\bar{\psi}\phi)(\psi^*\phi)+(\bar{\psi^*}\bar\phi)(\psi\bar{\phi})\right),\nonumber  \\
\mc O_{h}& = -\frac{2\pi}{\kappa}(\bar{\phi}\phi)^3.
\end{align}
In particular, the combination
\begin{equation}
\mc{O}_\omega=2\mc{O}_\alpha+2\mc{O}_\beta+4\mc{O}_\gamma+\omega\mc{O}_h
\end{equation}
preserves $\mc{N}=1$ SUSY (at large $N$). The RG flow eigenvectors and their anomalous dimensions at the three fixed points appear in the following table:
\begin{center}
	\begin{tabular}{ |c|c||c|c||c|c| }
		\hline
		\multicolumn{2}{|c||}{$\omega=1$} & \multicolumn{2}{|c||}{$\omega=-\frac13$}& \multicolumn{2}{|c|}{$\omega=-1$} \\
		\hline
		$\mc{O}_\omega$  & $2$  &  $\mc{O}_\omega$  & $-\frac23$ & $\mc{O}_\omega$ & $1$ \\
		$\mc{O}_\gamma$   &  $2$  & $a_1\mc{O}_\alpha+a_2\mc{O}_\beta+a_3\mc{O}_\gamma+\mc{O}_h$ & $\frac29(\sqrt3+1)$  & $\mc{O}_h$ & $3$ \\
		$\mc{O}_h$ &   $6$  &  $b_1\mc{O}_\alpha+b_2\mc{O}_\beta+b_3\mc{O}_\gamma+\mc{O}_h$   & $-\frac29(\sqrt3-1)$ & $ -\frac12\mc{O}_\beta+2\mc{O}_\gamma$ & $\frac16$ \\
		$-\mc{O}_\alpha+\mc{O}_\beta$ & $\frac23$ & $\mc{O}_h$ & $-\frac43$  &   $-14\mc{O}_\alpha+7\mc{O}_\beta+14\mc{O}_\beta+\mc{O}_h$ & $-\frac12$ \\
		\hline
	\end{tabular}
\end{center}
All anomalous dimensions come with an additional factor of $\frac{(2\pi)^2\lambda^2}{N}$, and
\begin{align}
a_1& = -\frac{2 \left(29 \sqrt{3}-27\right)}{7 \sqrt{3}+15} & a_2 &=-\frac{4 \left(\sqrt{3}+30\right)}{7 \sqrt{3}+15 } & a_3 &=\frac{4 \left(31 \sqrt{3}+33\right)}{7 \sqrt{3}+15} \\
b_1& =-\frac{2 \left(29 \sqrt{3}+27\right)}{7 \sqrt{3}-15}  & b_2 &=-\frac{4 \left(\sqrt{3}-30\right)}{7 \sqrt{3}-15} & b_3 & =\frac{4 \left(31 \sqrt{3}-33\right)}{7 \sqrt{3}-15}
\end{align}
The first row for each fixed point is the direction in which $\mc{N}=1$ SUSY is preserved, and which appears in Figure \ref{fig:RootPlot}. As emphasized in \cite{Avdeev:1992jt}, the $\mc{N}=2$ fixed point is stable also to all non-SUSY deformations, but the other ones are not.

\subsection{Phase Diagram}

The theories we discuss all have at least one $\mc{N}=1$-preserving relevant deformation -- a mass term $W = \mu |\Phi|^2$ -- and for the three stable fixed points described above, we conjecture that this is the only $\mc{N}=1$-preserving relevant deformation. Turning it on, with either sign for $\mu$, the theory develops a mass gap and flows to a pure CS theory in the IR; the point $\mu=0$ is generally a point of a second order phase transition between the $\mu<0$ and $\mu>0$ phases (we study theories which do not break supersymmetry, so the vacuum energy is always zero and all phase transitions are second order). Starting from an $SU(N)_{\kappa}$ CS-matter theory, there are four phases that can naturally appear. In our conventions, integrating out the massive matter field leads to an $SU(N)_{\kappa}$ or $SU(N)_{\kappa-1}$ pure CS theory at low energies, depending on the sign of $\mu$. Classically these are the only options for $\mu \omega > 0$, but for $\mu \omega < 0$ there are also classical Higgsed vacua where $\Phi$ obtains an expectation value, and the low-energy CS theory is $SU(N-1)_{\kappa}$ or $SU(N-1)_{\kappa-1}$. So classically there is one type of SUSY vacuum for $\mu \omega > 0$, and two types for $\mu \omega < 0$. Quantum mechanically this picture is modified as $\lambda$ is turned on, with boundaries between phases at values of $\omega$ that depend on $\lambda$. The precise phase structure at infinite $N$ (for a more general theory with arbitrary scalar and fermion couplings, not necessarily supersymmetric) was found in \cite{Dey:2019ihe}, by analyzing the effective potential; we use their notation for the phases, described in Table \ref{table:Phases}. As in the classical analysis, there is always one sign of $\mu$ leading to a single SUSY vacuum, and another leading to vacua in two different phases, but the identity of the phases changes as $\omega$ and $\lambda$ are varied.
\begin{table}[!h]
	\centering
	\begin{tabular}{ c c c c }
		Phase & Fermion & Boson & Low Energy Theory\\
		\hline 
		(+,+) & +ve mass & UnHiggsed & $SU(N)_\kappa$\\  
		(-,+) & -ve & UnHiggsed & $SU(N)_{\kappa-1}$ \\ 
		(+,-) & +ve & Higgsed & $SU(N-1)_{\kappa}$\\
		(-,-) & -ve & Higgsed & $SU(N-1)_{\kappa-1}$\\  
	\end{tabular}
	\caption{Low-energy Phases of $SU(N)_k$ with one fundamental matter field}
	\label{table:Phases}
\end{table}

\begin{figure}
	\centering
	\includegraphics[width=0.7\textwidth]{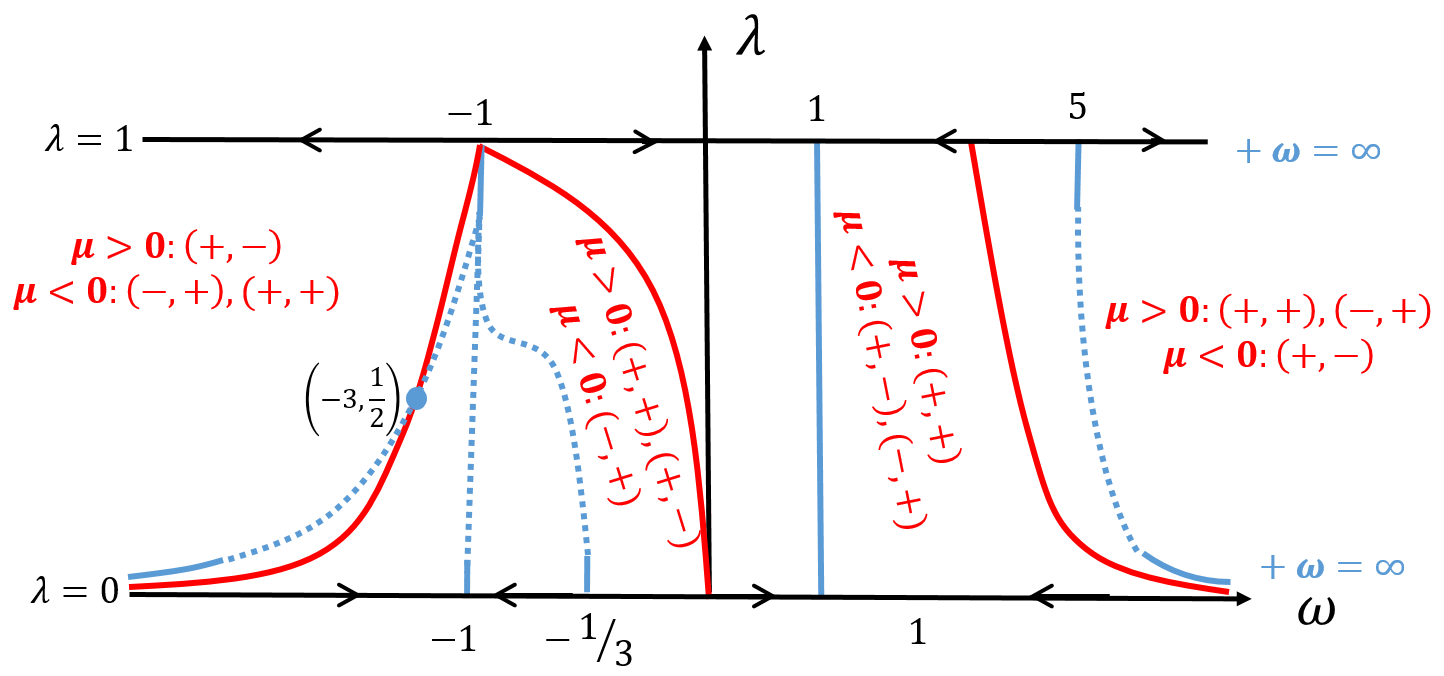}
	\caption{Superimposing the phase diagram on the RG flow diagram.}
	\label{fig:PhasesPlot}
\end{figure}

The result for the phase diagram, for $0 \leq \lambda \leq 1$, appears superimposed on the result for the roots of the beta function in Figure \ref{fig:PhasesPlot}. The red lines represent lines (found in \cite{Dey:2019ihe}) across which the phase structure jumps. In between any two red lines, we write down the phases when $\mu$ is positive and negative using the notation in Table \ref{table:Phases}. For example, for $\omega$ close to 1, we find that for positive mass the only vacuum is $SU(N)_{\kappa}$, while for negative mass there are two vacua: $SU(N-1)_{\kappa}$ and $SU(N)_{\kappa-1}$.

We see that in the simplest assumption for the evolution of the fixed points, each of the three stable fixed points -- the one at $\omega=1$, the one near $\omega=-1$, and the one near $\omega=\infty$ -- has a different phase structure close to it.  Note that in Figure \ref{fig:PhasesPlot} it seems that the phase structure jumps across $\omega=\infty$, but in fact, as discussed above, near this point it is better to think of the theory by adding another singlet superfield and using the superpotential \eqref{eq:singlet_W}. In this language, near $\omega=\infty$ the operator $|\Phi|^2$ is set to zero by the equation of motion, so it is better to describe the mass deformation by adding 
\begin{equation}
W = {\hat \mu} H \equiv \frac{N \mu}{2\pi \lambda \omega} H, 
\end{equation}
which gives the same action upon integrating out $H$. In terms of this more appropriate parameterization by ${\hat \mu}$, the phase structure is continuous as we cross $\omega=\infty$, with $(+,+)$ and $(-,+)$ phases for ${\hat \mu} > 0$, and a $(+,-)$ phase for ${\hat \mu} < 0$, consistent with our picture of the space of couplings as a circle. For non-zero values of $\lambda$, this circle is divided into three regions with different phases, and we find one large $N$ stable fixed point in each of these phases.

\subsection{Discussion of Dualities and Exact Moduli Spaces}\label{sec:beta_discussion}

Let us summarize our results. For large but finite $N$, we find six fixed points for small and large $\lambda$. We find that for any $\lambda$ there are at most six fixed points, and we conjecture that there are exactly six fixed points for all $\lambda$ (see Figure \ref{fig:RootPlot}). There is a duality between the six fixed points with $\lambda$ and those with $\lambda'=\lambda-{\rm sign}(\lambda)$, which at leading order in $1/N$ relates them according to equation \eqref{eq:duality}. From each of the three unstable fixed points we can flow to the two adjacent stable fixed points, so the duality between the unstable fixed points implies the duality between the stable ones. In addition, we can describe each theory both in the original CS-matter language, and in the language of adding an extra singlet superfield as in \eqref{eq:singlet_W}. For large $N$ the two descriptions clearly flow to the same fixed points, and we conjecture that also for finite $N$ there is a duality between them, namely the CS-matter theories with some range of values of $\omega$ in the UV flow to the same fixed points as the $H |\Phi|^2$ theories with some range of values of the $H^2$ superpotential in the UV (and with the same gauge group and level). All in all, each fixed point thus has four dual descriptions (with $\lambda$ or with $\lambda'$, and with or without the singlet $H$).

At finite $N$ the precise values of $\omega_c$ will be corrected, but we conjecture that, at least for large enough $N$, the number of fixed points and the dualities between them persist. At finite $N$ we have several different dualities, whose precise form may be found by demanding level-rank duality of the low-energy theories resulting after mass deformations: a fixed point of the $SU(N)_{k+\frac{N-1}{2}}$ theory maps to one of the $U(k)_{-N-\frac{k-1}{2},-N+\frac{1}{2}}$ theory, a fixed point of the $U(N)_{k+\frac{N-1}{2},k-\frac{1}{2}}$ theory maps to one of the $SU(k)_{-N-\frac{k-1}{2}}$ theory, and a fixed point of the $U(N)_{k+\frac{N-1}{2},k-\frac{1}{2}\pm N}$ theory maps to one of the $U(k)_{-N-\frac{k-1}{2},-N+\frac{1}{2}\mp k}$ theory. As $N$ decreases, some of the fixed points may disappear or others may appear, and the various dualities may or may not persist for small $N$; it would be interesting to study down to which value of $N$ the various fixed points and the dualities between them survive.

The dualities of the stable fixed points have already appeared in the literature before, but here we clarify which UV theories flow to the corresponding fixed points, and add the duality between the unstable fixed points as well.

The duality of the $\mc{N}=2$ point at $\omega=1$ is a special case of $U(N)$ and $SU(N)$ $\mc{N}=2$ dualities discussed in \cite{Benini:2011mf,Aharony:2013dha,Aharony:2014uya}, and is believed to hold for all values of $N$ and $k$. The $\mc{N}=2$ fixed point $\omega=1$ is attractive for $\lambda$ close to $0$ or $1$, and it  was shown to be attractive for finite $N$ and large $k$ as well in \cite{Avdeev:1992jt}. It is thus reasonable to assume that this point is attractive for all $N$ and $k$. This is in agreement with comments appearing in \cite{Choi:2018ohn,Bashmakov:2018ghn}.

It is natural to assume that also the large $N$ phase structure that we discussed near this point persists for finite values of $N$ and $k$. Note that, at least for large $N$, the phase structure that appears at the $\mc{N}=2$ point is unique to this fixed point -- all other fixed points have a different phase diagram. This phase structure was shown to appear for the theories studied in \cite{Choi:2018ohn,Bashmakov:2018ghn}, and in particular for domain walls of four dimensional $\mc{N}=1$ $N_f=1$ SQCD. This is evidence for the fact that the theories studied in these papers do indeed flow to the $\mc{N}=2$ fixed point, and so it is evidence for the various patterns of SUSY enhancement discussed in these papers.

The authors of \cite{Benini:2018umh} proposed the following duality (and related $SU-U$ and $U-U$ dualities) for the case $N_f=1$:
\begin{equation}\label{eq:BeniniBenvenutiDuality}
\begin{split}
U(k)_{N+\frac{k}{2}-\frac12,N-\frac12}+\Phi\\W=-\frac{1}{4}(\bar \Phi \Phi)^2
\end{split}
\quad\longleftrightarrow\quad
\begin{split}
SU(N)_{-k-\frac{N}{2}+\frac12}+\tilde \Phi
\\W= H(\bar{\tilde{\Phi}}\tilde \Phi)-\frac{1}{3}H^3
\end{split}
\end{equation} 
The operator $H^3$ was added on the right-hand side in order to make the classical phase structure match on the two sides of the duality as one deforms by a mass $\mu$; this was also the reason for the choice of sign of the superpotential on the left-hand side (its precise value plays no role).
We recognize that the superpotential on the right-hand side, without the $H^3$ term (which is irrelevant in the IR anyway) is the same as the one we used in \eqref{eq:singlet_W} to obtain a different description of our CS-matter theories, which is more convenient near $\omega=\infty$.
Thus, we interpret this duality as a finite $N$ version of the duality we found connecting the IR-stable fixed point that we found near $\omega=-1$, with the fixed point close to $\omega=\infty$,
\begin{equation}
\begin{split}
U(k)_{N+\frac{k}{2}-\frac12,N-\frac12}+\Phi\\ \omega\sim-1\quad\quad
\end{split}
\quad\longleftrightarrow\quad
\begin{split}
SU(N)_{-k-\frac{N}{2}+\frac12}+\tilde \Phi
\\ \omega\sim\infty\qquad
\end{split}
\end{equation}
There is a large range of values of $\omega$ which flows to each of these fixed points, either in its original description or in the one with the extra singlet field $H$ (where it is natural to flow to the large $\omega$ fixed point if one starts with no $H^2$ term in the superpotential at high energies). All theories in this range are conjectured to be IR-dual.
Note that the phases appearing for these two fixed points in Figure \ref{fig:PhasesPlot} exactly match the phases required in the proposed duality \eqref{eq:BeniniBenvenutiDuality}, without the need to consider the extra $H^3$ term in $W$ (which can be added in the UV, but does not affect the IR).

Finally, we discuss exact moduli spaces. We argued above that the theories with $(\omega,\lambda)=(-3,1/2)$ are self-dual at large $N$, and that this leads to an emergent time-reversal symmetry, which is a combination of time-reversal with a duality transformation. Under this emergent time-reversal symmetry the operator $|\Phi|^2$ maps to itself, but the superpotential has to be odd, so this prevents any superpotential from being generated, so that there must be an exact moduli space. The corresponding theories for finite values of $N$ and $k$ were already discussed in \cite{Choi:2018ohn}; in particular they mentioned the emergent time-reversal symmetry of the theories
\begin{equation}\label{eq:emergent_T_dualities}
\begin{split}
U(N)_{\frac{3}{2}N-\frac12 N_f,2N-\frac12 N_f}+N_f\;\Phi\\
U(N)_{\frac{3}{2}N-\frac12 N_f,-\frac12 N_f}+N_f\;\Phi
\end{split}
\end{equation}
for all values of $N,N_f$; for $N_f=1$ these are special cases of the dualities we wrote above, with $k=N$. One can ask whether they all have an exact moduli space. Indeed, the simplest examples of the theories in \eqref{eq:emergent_T_dualities} (given by $N=N_f=1$) have been found to have moduli spaces\footnote{The theory $U(1)_{1/2}+\Phi$ is dual to a free matter multiplet and so has a moduli space \cite{Bashmakov:2018wts}. The theory $U(1)_{3/2}+\Phi$ is dual to $U(1)_{0}+\Phi_2$ where $\Phi_2$ has charge 2 under the gauge symmetry, and this was also shown to have a moduli space \cite{Gaiotto:2018yjh}.}. For all values of $N$ it is natural to assume that the theories \eqref{eq:emergent_T_dualities} discussed in \cite{Choi:2018ohn} are the finite $N$ versions of the large $N$ $(\omega,\lambda)=(-3,1/2)$ fixed points, with the same action of the emergent time-reversal symmetry on $|\Phi|^2$. This implies that all these theories, at the corresponding fixed point, have an exact moduli space. Since our arguments depend on the emergent time-reversal symmetry, we do not expect an exact moduli space in the corresponding Yang-Mills-Chern-Simons theories, which flow to these fixed points; it is just a property of the fixed point conformal field theories. Note that according to our discussion we expect these fixed points to be unstable, namely to have two relevant operators, a $|\Phi|^2$ and a $(|\Phi|^2)^2$ superpotential (at least for large enough $N$). We will perform a similar analysis in the next section for the case $N_f>1$, and we will find a similar result, where there exists a fixed point with an exact moduli space in the large-$N$ limit.

\section{Theories with $N_f>1$}\label{sec:Nf_many}

In this section we consider the general theory of section \ref{sec:many_matter_fields} with $N_f>1$. We start by generalizing the duality transformation \eqref{eq:duality}. We then calculate the $\beta$ and $\gamma$ functions at $\lambda=0$, generalizing our $N_f=1$ computation. We manage to map only some of the fixed points to their value at strong coupling. We discuss various generalizations of some phenomena that appeared for $N_f=1$, like exact moduli spaces and the behavior at infinite $\omega_i$.

\subsection{Duality}\label{sec:Nf_many_duality}

We start by generalizing the large $N$ duality transformation \eqref{eq:duality} to the case $N_f>1$; the action of the duality on $\lambda$ is the same for all $N_f \ll N$, but we need to understand how the duality acts on the superpotential couplings. Consider the theory given in \eqref{eq:matter_lagrangian_Nf}, with an additional mass term which is not necessarily $SU(N_f)$-invariant, namely
\begin{equation}\label{eq:W_Phi}
W_\Phi=m_{0,ij}(\bar{\Phi}_i\Phi_j) +\frac{\pi\omega_0}{\kappa}\left(\bar{\Phi}^i\Phi_i\right)^2+\frac{\pi\omega_1}{\kappa}\left(\bar{\Phi}^i\Phi_j\right) \left(\bar{\Phi}^j\Phi_i\right),
\end{equation}
with $i,j=1,\cdots,N_f$. Let us rewrite this in a simpler form. Define $J_{ij}\equiv \bar\Phi_i\Phi_j$, and decompose $J_{ij}$ into its trace and adjoint parts:
\begin{equation}
\mc{J} = J_{kk},\quad \mc{J}_{ij}=J_{ij}-\frac{\delta_{ij}}{N_f}\mc{J},
\end{equation}
with $\tr(\mc{J}_{ij})=0$. Similarly, define
\begin{equation}
M_0=\tr (m_{0,ij}),\quad M_{0,ij}=m_{0,ij}-\frac{\delta_{ij}}{N_f}M_0 ,
\end{equation}
so that $\tr (M_{0,ij})=0$. It is easy to show that the Lagrangian \eqref{eq:W_Phi} can be rewritten as
\begin{equation}\label{eq:new_couplings}
W_\Phi=M_{0}^{ij}\mc{J}_{ij}+\frac{M_0}{N_f}\mc{J}+\frac{1}{N_f}\frac{\pi\lambda}{N}\omega_2 \mc{J}^2+\frac{\pi\lambda}{N}\omega_1 \mc{J}^2_{ij}
\end{equation}
where we have defined $\omega_2\equiv N_f \omega_0+\omega_1$.

We propose that the generalization of the duality \eqref{eq:duality} to $N_f>1$ is given by
\begin{equation}\label{eq:dual_Nf}
M_{0,ij}'=-\frac{2M_{0,ij}}{1+\omega_1},\quad
M_0'=-\frac{2M_0}{1+\omega_2},\quad
\omega_1'=\frac{3-\omega_1}{1+\omega_1},\quad  
\omega_2'=\frac{3-\omega_2}{1+\omega_2}.
\end{equation}
This can be thought of as two copies of the $N_f=1$ transformation \eqref{eq:duality}, one for the flavor-singlet and one for the flavor-adjoint sector, where $\omega_1,\omega_2$ transform independently. 
Equation \eqref{eq:dual_Nf} implies that under the duality $\mc{J}'_{ij} = -\frac{1+\omega_1}{2} \mc{J}_{ij}$, $\mc{J}' = -\frac{1+\omega_2}{2} \mc{J}$.
For completeness we write the transformation of $\omega_0$ as well:
\begin{equation} \omega_0'=-\frac{4\omega_0}{(1+\omega_1)(1+\omega_2)}  .
\end{equation}
As an initial consistency check, this transformation is consistent with the fact that $\omega_1=1,\omega_0=0$ has $\mc{N}=2$ SUSY and so is a self-dual point\footnote{The full set of self-dual points are
\begin{equation} (\omega_1,\omega_2)=(1,1),\;(1,-3),\;(-3,1),\;(-3,-3) ,
\end{equation}
or equivalently
\begin{equation} (\omega_1,\omega_0)=(1,0),\;\left(1,-\frac{4}{N_f}\right),\;\left(-3,\frac{4}{N_f}\right),\;(-3,0) .\end{equation}}.

We now perform two independent calculations which test this proposal: the zero-temperature pole masses in the unHiggsed phase (obtained from the gap equations) and the two and three-point correlation functions of $\mc{J}_{ij}$ and $ \mc{J}$.

\subsubsection{Gap Equations in unHiggsed Phase}
 
Using an immediate generalization of the mass gap calculation performed in \cite{Inbasekar:2015tsa}, we can find the physical masses of the mass-deformed theory for the case $N_f>1$. As a reminder, in \cite{Inbasekar:2015tsa} the mass gap equations were found by solving the Schwinger-Dyson equation for the propagator of $\Phi$ assuming no Higgsing, with the result
\begin{equation}\label{eq:mass_gap_Nf1}
 m= \frac{ 2m_0}{\lambda \text{sign}(m) (\omega_1-1)+2}.
\end{equation}
Under the duality the physical mass transforms as
\begin{equation} m'=-m \end{equation}
as expected.

We repeat the calculation for the case $N_f>1$ in Appendix \ref{app:gap_eqs}. The result, when the mass matrix $M_{0,ij}$ is diagonal, is
\begin{equation}\label{eq:Nf_mass}
m_{ij}= \frac{ 2M_{0,ij}}{\lambda \text{sign}(m_{ij}) (\omega_1-1)+2}+
	\frac{\frac{2}{N_f}M_0\delta_{ij}}{ \lambda \text{sign}(m_{ij}) (\omega_2-1)+2}.
\end{equation}
As an immediate generalization of the $N_f=1$ result \eqref{eq:mass_gap_Nf1}, we find that under the proposed duality transformation \eqref{eq:dual_Nf} we have $m'_{ij}=-m_{ij}$ as expected.

\subsubsection{Correlation Functions of $\mc{J},\mc{J}_{ij}$}

The two and three-point correlation functions of $J$ for the case $N_f=1$ appear in Section \ref{sec:corr_funcs_of_J}. These were found by using the general four-point function of $\Phi$'s, see \cite{Inbasekar:2015tsa}. Thus, in order to find the result for $\mc{J}$ and $\mc{J}_{ij}$, we must generalize the calculation of the four-point function.

Luckily, if the interactions are written as in \eqref{eq:new_couplings}, this is relatively simple. We find that the Schwinger-Dyson equation for the four point function $(\bar{\Phi}^i\Phi_j) (\bar{\Phi}^j\Phi_i)$ with $i\neq j$ is exactly the same as the equation for $N_f=1$, with $\omega$ replaced by $\omega_1$. Similarly, the equation for the four point function $(\bar{\Phi}^i\Phi_i) (\bar{\Phi}^j\Phi_j)$ is the same as the equation for $N_f=1$, with $\omega$ replaced by $\omega_2$ (up to some overall factors of $N_f$). 
	
As a result, the two and three-point functions are an immediate generalization of the results of Section \ref{sec:corr_funcs_of_J} for the two and three-point functions of $J$ when $N_f=1$. We find
\begin{equation}
\begin{split}
\langle \mc{J}_{ij}^{2}\rangle=\langle J^2\rangle|_{\omega\to\omega_1},\\
\langle \mc{J}_{ij}^{3}\rangle=\langle J^3\rangle|_{\omega\to\omega_1},
\end{split}
\qquad
\begin{split}
\langle \mc{J}^2\rangle=N_f\langle J^2\rangle|_{\omega\to\omega_2}\\\langle \mc{J}^3\rangle=N_f\langle J^3\rangle|_{\omega\to\omega_2}
\end{split}
\end{equation}
where the two-point function $\langle J^2\rangle$ is given in equations \eqref{eq:P1},\eqref{eq:2pt_func_C1C2}, and the three-point function $\langle J^3\rangle$ is given in equation \eqref{eq:3pt_func_A} for $\omega=1$ (but as discussed above this equation, the result for general $\omega$ can be obtained immediately from this result by multiplying it by chains).

We can now check our proposal for the duality transformation. This check is immediate, since $\langle J^2\rangle,\langle J^3\rangle$ are themselves duality-covariant according to the original $N_f=1$ duality, and each sector transforms independently.

\subsection{Beta and Gamma Functions for $\lambda=0$}\label{sec:beta_Nf_many}

As explained in Section \ref{sec:lagrangian}, for $N_f > 1$ the Lagrangian has two classically marginal operators:
\begin{equation} \frac{\tilde\omega_0}{N}(\bar\Phi^a_i\Phi^a_i)(\bar\Phi^b_j\Phi^b_j),\qquad
\frac{\tilde\omega_1}{N}(\bar\Phi^a_j\Phi^a_i)(\bar\Phi^b_i\Phi^b_j) .
\end{equation}
From now on we use brackets to describe a trace over gauge indices, so we can write these as
\begin{equation} \frac{\tilde\omega_0}{N}(\bar\Phi_i\Phi_i)(\bar\Phi_j\Phi_j),\qquad
\frac{\tilde\omega_1}{N}(\bar\Phi_j\Phi_i)(\bar\Phi_i\Phi_j) .
\end{equation}
We can now calculate the leading order beta functions at $1/N$. Since the gauge indices are the same, we have the same diagrams as for the $N_f=1$ case, but we must follow the flavor indices more carefully now.

We have two beta functions to calculate, for $\tilde\omega_0$ and $\tilde\omega_1$, and three different types of correlators:
\begin{itemize}
	\item $
	\mc{O}_r=\langle(\bar\Phi_i\Phi_i)(\bar\Phi_j\Phi_j)\rangle$ with $i\neq j$
	\item $
	\mc{O}_g=\langle(\bar\Phi_j\Phi_i)(\bar\Phi_i\Phi_j)\rangle
	$ with $i\neq j$
	\item $	\mc{O}_b=\langle(\bar\Phi_i\Phi_i)(\bar\Phi_i\Phi_i)\rangle
	$
\end{itemize}
$r,g,b$ stand for red, green, blue in our figures.

We start by calculating the leading order in $1/N$ contributions to these correlators, which are just generalizations of the chains \eqref{eq:chain}. Next we calculate the contributions at next order in $1/N$, which are generalizations of the $(a)$-$(e)$ diagrams. This allows us to find the beta and gamma functions.

\subsubsection{Leading Order: Chains}

\begin{figure}
	\centering
	\includegraphics[width=0.5\linewidth]{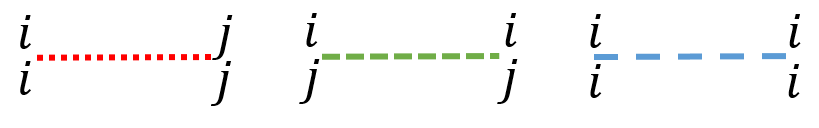}
	\caption{The chains for $N_f>1$}
	\label{fig:chains}
\end{figure}

We now have three types of chains, shown in Figure \ref{fig:chains}.
In the red and green chains we take $i\neq j$. In the limit where $N_f=1$ and either $\tilde\omega_0=0$ or $\tilde\omega_1=0$, the blue chain should correspond to the chain found in the $N_f=1$ case \eqref{eq:chain}.

The calculation of the various chains is similar to the $N_f=1$ case in Section \ref{direct_calculation}. We find:
\begin{align}
\Delta_r&=2\sum_{l=1,k=0}\left(\begin{array}{c}
l+k\\
l
\end{array}\right)\tilde\omega_{0}^{l}N_{f}^{l-1}\tilde\omega_{1}^{k}\left(\frac{D^2}{4|p|}\right)^{l+k-1}=
32 \tilde\omega_0\frac{ 16-\tilde\omega_2 \tilde\omega_1+4(\tilde\omega_2+\tilde\omega_1)\frac{D^2}{|p|}}{\left(\tilde\omega_1^2+16\right) \left(\tilde\omega_2^2+16\right)}=
A_0+A_1 \frac{D^2}{|p|}\\
\Delta_{g}&=2\tilde\omega_1 \sum_{n=0} \left(\tilde\omega_1\frac{D^2}{4|p|}\right)^n=
\frac{8\tilde\omega_1}{\tilde\omega_1^2+16} \frac{4|p|+\tilde\omega_1 D^2}{|p|}=B_0+B_1 \frac{D^2}{|p|}\\
\Delta_b&=\Delta_r+\Delta_g
\end{align}
Where we have defined $\tilde\omega_2=N_f\tilde\omega_0+\tilde\omega_1$. So at leading order in $1/N$ we find
\begin{equation} 
\mc{O}_r=\Delta_r, \qquad \mc{O}_g=\Delta_g, \qquad\mc{O}_b=\Delta_b=\Delta_r+\Delta_g 
\end{equation}

\subsubsection{Subleading Order: $(a)$-$(e)$ Diagrams}
We have three types of correlation functions to calculate, corresponding to $\mc{O}_r,\mc{O}_g,\mc{O}_b$. The diagrams contributing to these are of the same form as the $(a)-(e)$ diagrams in Figure \ref{fig:diags}, to which we must add the correct flavor index structure. Thus, each of the $(a)-(e)$ diagrams contribute to $\mathcal{O}_r$, $\mathcal{O}_g$ and $\mathcal{O}_b$ according to the structure of the flavor indices on the external legs. An explicit calculation for the logarithmically diverging terms in dimensional regularization gives:
\begin{center}
	\begin{tabular}{ccc}
		Type & Contribution to $\mc{O}_r$ & Contribution to $\mc{O}_g$  \\ 
		\hline
		$(a)$	& $ -\frac{4 \Delta_g \Delta_r (A_1+B_1 N_f)+2 \Delta_r^2 N_f (A_1+B_1 N_f)+B_1 \Delta_g^2}{8 \pi ^2}\frac{D^2}{|p|} $ &
		$ -\frac{ \Delta_g^2 (2 A_1+B_1 N_f)}{8 \pi ^2}\frac{D^2}{|p|} $ \\ 
		$(b)$	& $ -\frac{\Delta_g^4+2 \Delta_g^2 \Delta_r^2 \left(N_f^2+5\right)+2 \Delta_r^4 N_f^2+4 \Delta_g^3 \Delta_r N_f+8 \Delta_g \Delta_r^3 N_f}{64 \pi ^2}\frac{D^2}{|p|} $
		& $ -\frac{ \Delta_g^3 (4 \Delta_r+\Delta_g N_f)}{64 \pi ^2}\frac{D^2}{|p|} $\\ 
		$(c)$	& $ -\frac{\Delta_g^3+6 \Delta_g \Delta_r^2+2 \Delta_g^2 \Delta_r N_f+2 \Delta_r^3 N_f}{8 \pi ^2} $ 	& $ -\frac{\Delta_g^2 (4 \Delta_r+\Delta_g N_f)}{8 \pi ^2} $\\ 
		$(d)$	& $ -\frac{2 A_0 A_1+B_0 B_1}{\pi ^2} $	& $ -\frac{2 A_0 B_1+2 A_1 B_0+B_0 B_1 N_f}{\pi ^2} $ \\ 
		$(e)$	& $ -\frac{2 \Delta_r (A_1+B_1 N_f)+B_1 \Delta_g}{\pi ^2} $& $ -\frac{\Delta_g (2 A_1+B_1 N_f)}{\pi ^2} $ \\ 
	\end{tabular} 
\end{center}
For $\mc{O}_b$, a direct calculation shows that the logarithmically diverging terms at order $1/N$ obey the relation $\mc{O}_b=\mc{O}_r+\mc{O}_g$, as they must for the consistency of the Callan-Symanzik equation at this order.

\subsubsection{Gamma Function}

\begin{figure}
	\centering
	\includegraphics[width=0.4\linewidth]{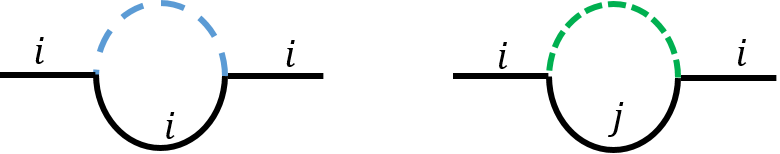}
	\caption{Diagrams contributing to the gamma function}
	\label{fig:gammanf}
\end{figure}

The diagrams contributing to $\gamma_{\Phi}$ appear in Figure \ref{fig:gammanf}.
Using our result for $N_f=1$ and just adding the flavor index sums, we find the following results for the two diagrams:
\begin{equation}
\frac{A_1+B_1}{2\pi^2\gre}D^2,\quad (N_f-1)\frac{B_1}{2\pi^2\gre}D^2 
\end{equation}
So the total contribution is
\begin{equation} 
\frac{A_1+N_f B_1}{2\pi^2\gre}D^2 
\end{equation}
The gamma function is then
\begin{equation}\label{eq:gamma_Nf}
\gamma_\Phi=
\frac12 \frac{A_1+N_f B_1}{2\pi^2}=
\frac{2}{\pi ^2 N \left(\tilde\omega_1^2+16\right)}
 \left(\frac{16 \tilde\omega_0 (N_f \tilde\omega_0+2 \tilde\omega_1)}{(N_f \tilde\omega_0+\tilde\omega_1)^2+16}+N_f \tilde\omega_1^2\right)
\end{equation} 
One can expand $\gamma_\Phi$ in small $\tilde\omega_0,\tilde\omega_{1}$, and the result agrees with that of \cite{Avdeev:1992jt} (up to the same overall factor of 2 discussed in Section \ref{beta_Summary}).

\subsubsection{Beta Functions}

In order to find the beta functions $\beta_0,\beta_1$, it is enough to consider the Callan-Symanzik equation for $\mc{O}_r$ (the Callan-Symanzik equations for $\mc{O}_g,\mc{O}_b$ can be used as consistency checks). This equation is
\begin{align}
\left(\beta_1 \frac{\partial}{\partial \tilde\omega_1}+
\beta_2 \frac{\partial}{\partial \tilde\omega_2}+4\gamma_{\Phi}\right) \Delta_r+\left((a)_r+(b)_r+(c)_r+(d)_r+(e)_r\right)=&0,
\end{align}
where we use $\tilde\omega_2=N_f\tilde\omega_0+\tilde\omega_1$ as in \eqref{eq:new_couplings}. Since each of the terms in the equation once again has two terms (one proportional to $1$ and one proportional to $\frac{D^2}{|p|}$), this Callan-Symanzik equation gives us two independent equations for $\beta_1$ and for $\beta_2$.
Solving these equations we find
\begin{align}
\begin{split}
\beta_1=&\frac{8 \tilde\omega_1}{\pi ^2 N_f \left(\tilde\omega_1^2+16\right)^2 \left(\tilde\omega_2^2+16\right)}\bigg[ \tilde\omega_1^4 \left(2 \left(\tilde\omega_2^2+48\right)-N_f^2 \left(\tilde\omega_2^2+16\right)\right)+16 \left(3 N_f^2-10\right) \tilde\omega_1^2 \left(\tilde\omega_2^2+16\right)-\\
&-2 \tilde\omega_1^5 \tilde\omega_2+64 \tilde\omega_1^3 \tilde\omega_2+1536 \tilde\omega_1 \tilde\omega_2+1024 \tilde\omega_2^2\bigg],
\end{split}\label{eq:beta1_final}\\
\begin{split}
\beta_2=&-\frac{16}{\pi ^2 N_f \left(\tilde\omega_1^2+16\right)^2 \left(\tilde\omega_2^2+16\right)^2}\bigg[ \left(N_f^2-1\right) \tilde\omega_1^3 \left(\tilde\omega_2^2-16\right) \left(\tilde\omega_2^2+16\right)^2-\\
&-32 \tilde\omega_1^2 \tilde\omega_2 \left(N_f^2 \left(\tilde\omega_2^2+16\right)^2-2 \left(\tilde\omega_2^4-8 \tilde\omega_2^2+128\right)\right)+\tilde\omega_1^4 \tilde\omega_2^3 \left(\tilde\omega_2^2-48\right)+256 \tilde\omega_2^3 \left(\tilde\omega_2^2-48\right)\bigg].
\end{split}\label{eq:beta2_final}
\end{align}
As a consistency check, we find that the solutions to the Callan-Symanzik equations for $\mc{O}_g,\mc{O}_b$ agree with the results above. Also, in the limit $N_f=1$ with either $\tilde\omega_0\to 0$ or $\tilde\omega_1\to 0$, the results agree with our beta function at $N_f=1$ \eqref{eq:final_beta}. Next, note that $\beta_1$ in \eqref{eq:beta1_final} is proportional to $\tilde\omega_1$, and so it vanishes if $\tilde\omega_1=0$. Indeed, at $\tilde\omega_1=0$ the symmetry of the Lagrangian \eqref{eq:matter_lagrangian_Nf_lambda0} is enhanced from $U(N_f)\times U(N_c)$ to $U(N_f\times N_c)$, and so we expect the value $\tilde\omega_1=0$ to be preserved under the RG flow. As a final consistency check, we note that expanding the results for the beta functions \eqref{eq:beta1_final},\eqref{eq:beta2_final} for small $\tilde\omega_0,\tilde\omega_1$ reproduces the result from \cite{Avdeev:1992jt} (again, up to the overall factor).

\subsubsection{Generalization to Finite $\lambda$ and Discussion}

\begin{figure}
	\centering
	\begin{subfigure}{0.45\linewidth}
		\centering
		\includegraphics[width=\textwidth]{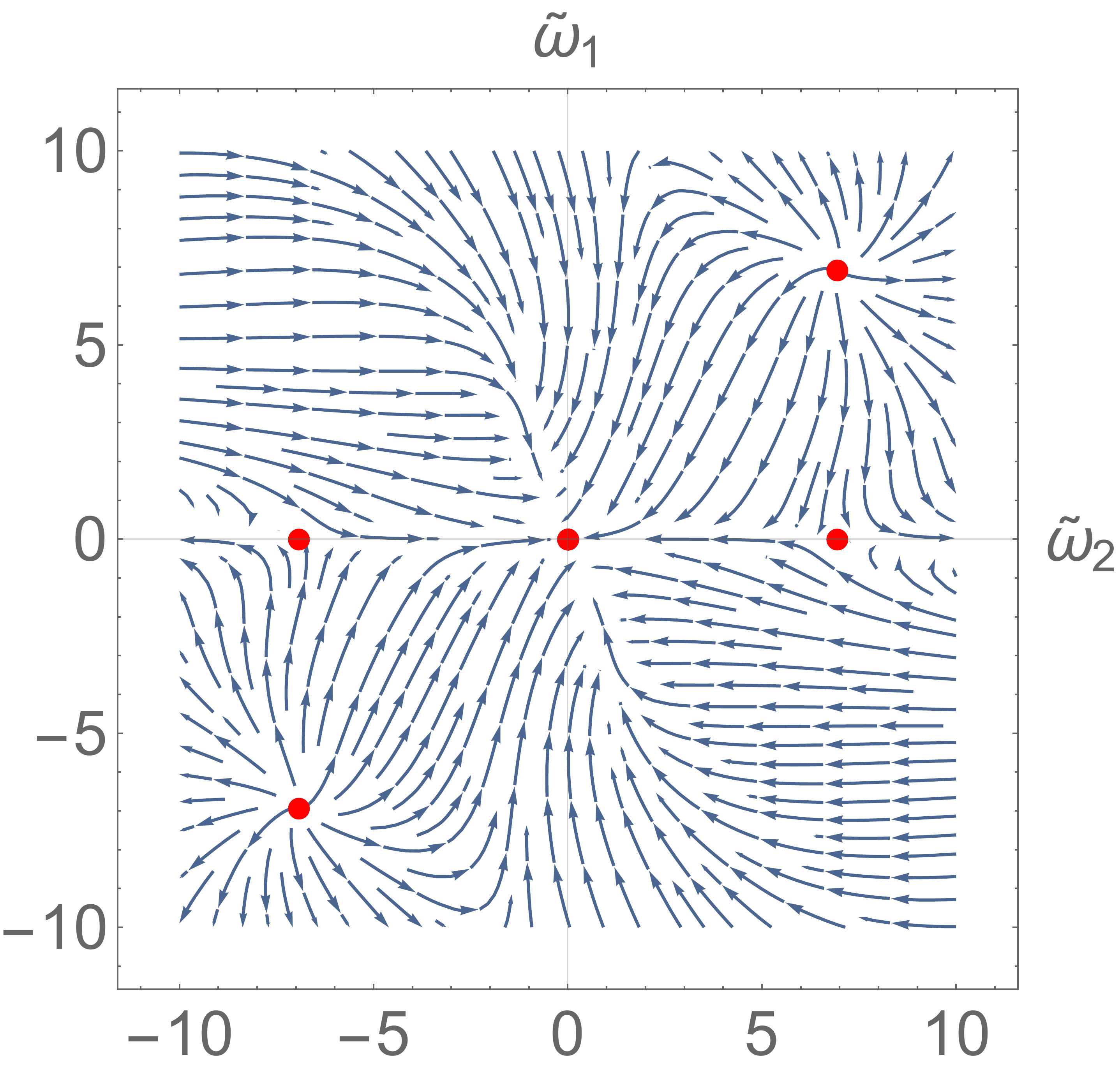}
		\caption{$N_f=2$}
	\end{subfigure}
	\qquad
	\begin{subfigure}{0.45\textwidth}
		\centering
		\includegraphics[width=\textwidth]{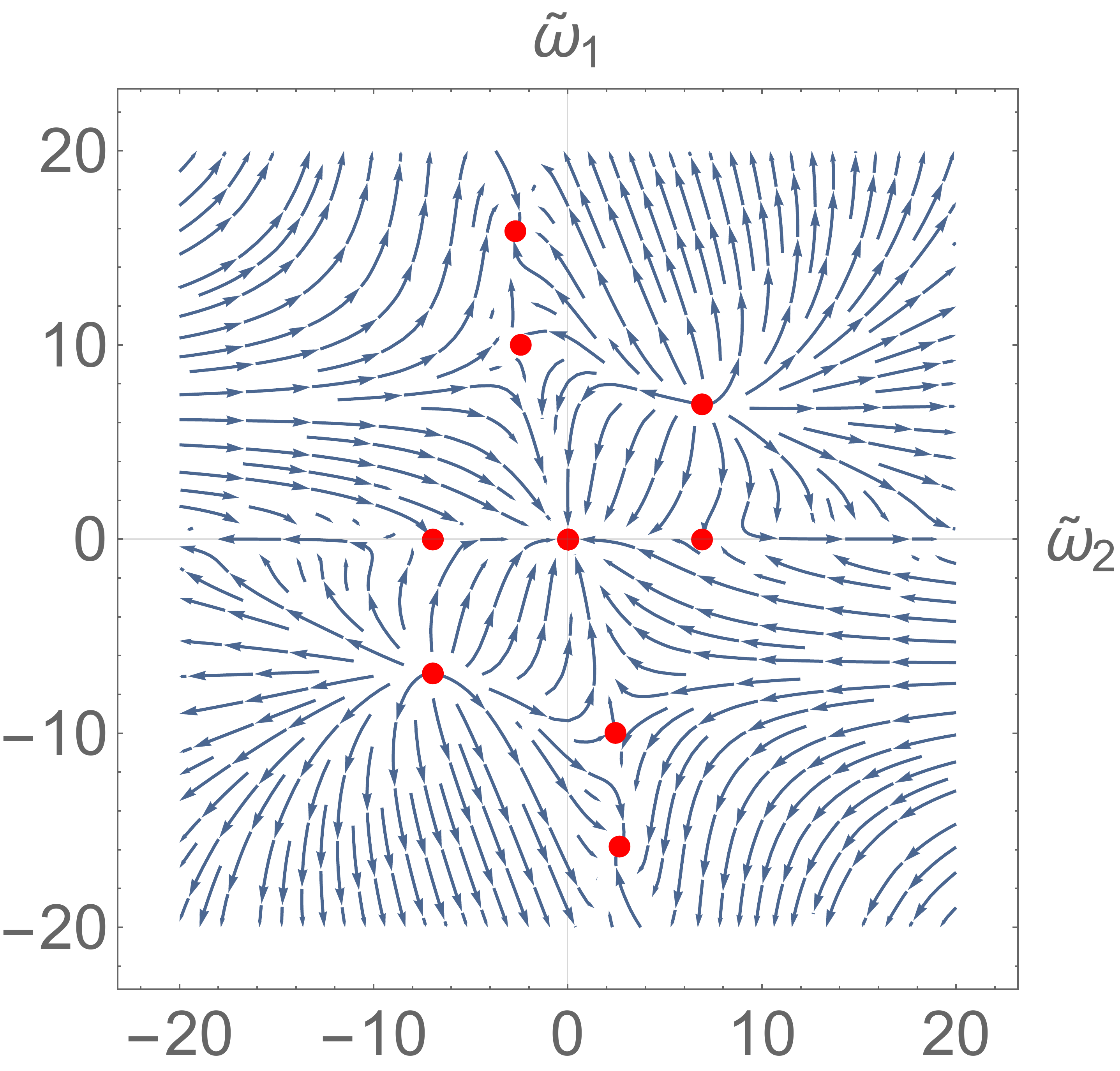}
		\caption{$N_f=3$ (general structure is generic for $N_f>2$)}
	\end{subfigure}
	\caption{Fixed points in terms of $\wt_1,\wt_2$ at $\lambda=0$.}
	\label{fig:beta_Nf}
\end{figure}

We have found $\gamma_{\Phi}$ \eqref{eq:gamma_Nf} and the two beta functions \eqref{eq:beta1_final},\eqref{eq:beta2_final}. We plot the results in Figure \ref{fig:beta_Nf}. Explicitly,
for $N_f=2$ the fixed points are:
\begin{equation}\label{eq:fixed_points_Nf2_omega2}
(\tilde{\omega}_2,\tilde{\omega}_1)=(0,0),\;\left(\pm\sqrt{48},0\right),\;\left(\pm\sqrt{48},\pm\sqrt{48}\right),\;(\infty,0),\;(0,\infty),\;(\infty,\infty).
\end{equation} 
The only attractive fixed points are $(0,0)$ and $(\infty,0)$.
For $N_f>2$ we have the same fixed points \eqref{eq:fixed_points_Nf2_omega2}, and in addition there are four extra fixed points for finite values of $\tilde\omega_i$, which are all repulsive.

As for $N_f=1$, we were not able to compute the beta functions for general values of $\lambda$, but we know some things about them.
We first attempt to generalize the results to small $\lambda$, that is, we attempt to find the fixed points in terms of the $\omega_i$. We focus on the fixed points as a function of $(\omega_2,\omega_1)$ since these are the useful values for the duality. For the fixed point at $(\tilde{\omega}_2,\tilde{\omega}_1)=(0,0)$ we can use the results of \cite{Avdeev:1992jt} in order to find the fixed points in terms of $\omega_i$ at small $\lambda$, while for the other fixed points we can use the relation $\tilde{\omega}_i=\pi\lambda\omega_i$. This is enough to determine the $O(\frac{1}{\lambda})$ contributions to the fixed points, but when these vanish we cannot determine the $O(1)$ values.

\begin{figure}
	\centering
	\begin{subfigure}{0.45\linewidth}
		\centering
		\includegraphics[width=\textwidth]{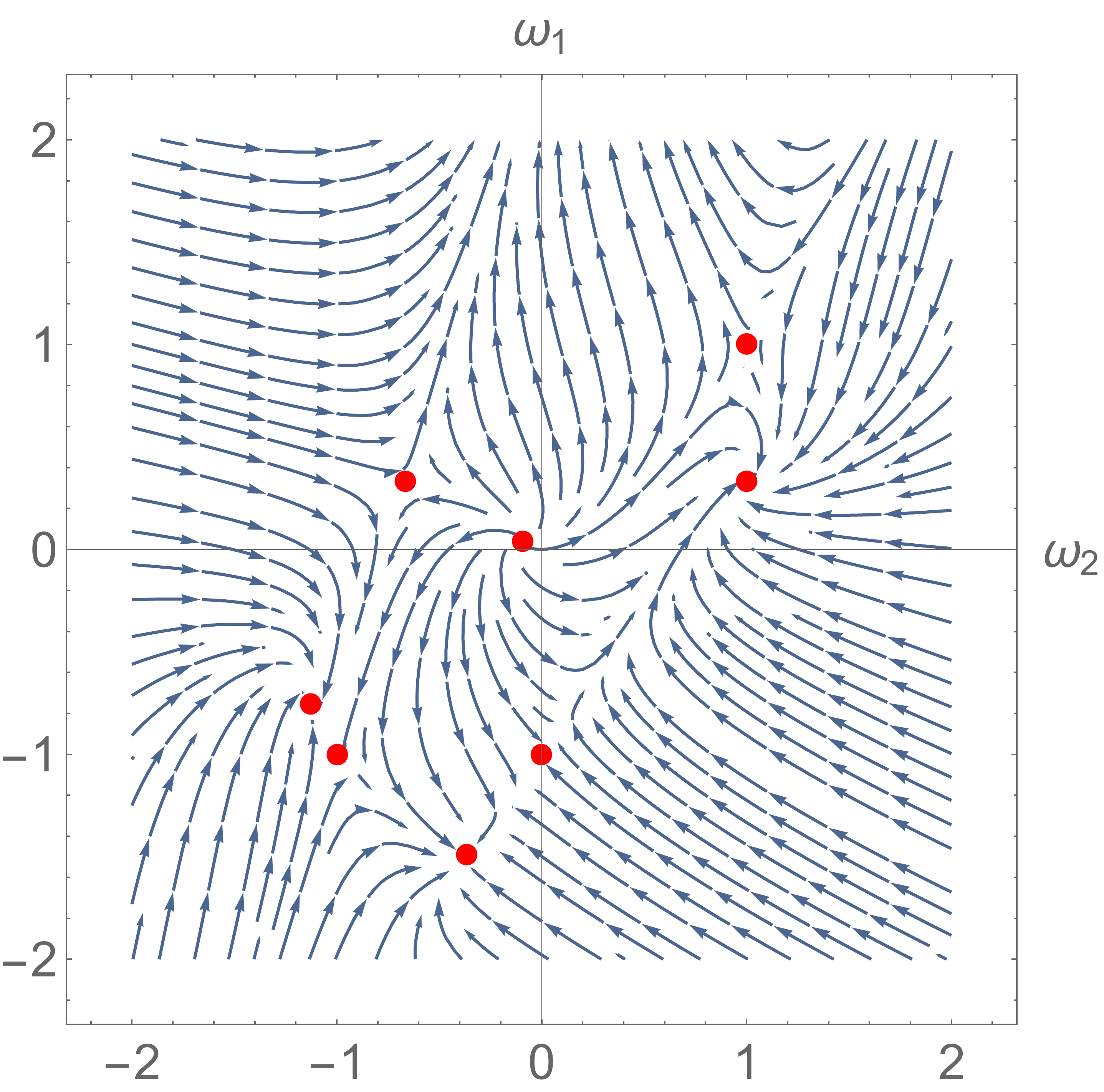}
		\caption{$N_f=2$}
		\label{fig:origin_1}
	\end{subfigure}
	\qquad
	\begin{subfigure}{0.45\textwidth}
		\centering
		\includegraphics[width=\textwidth]{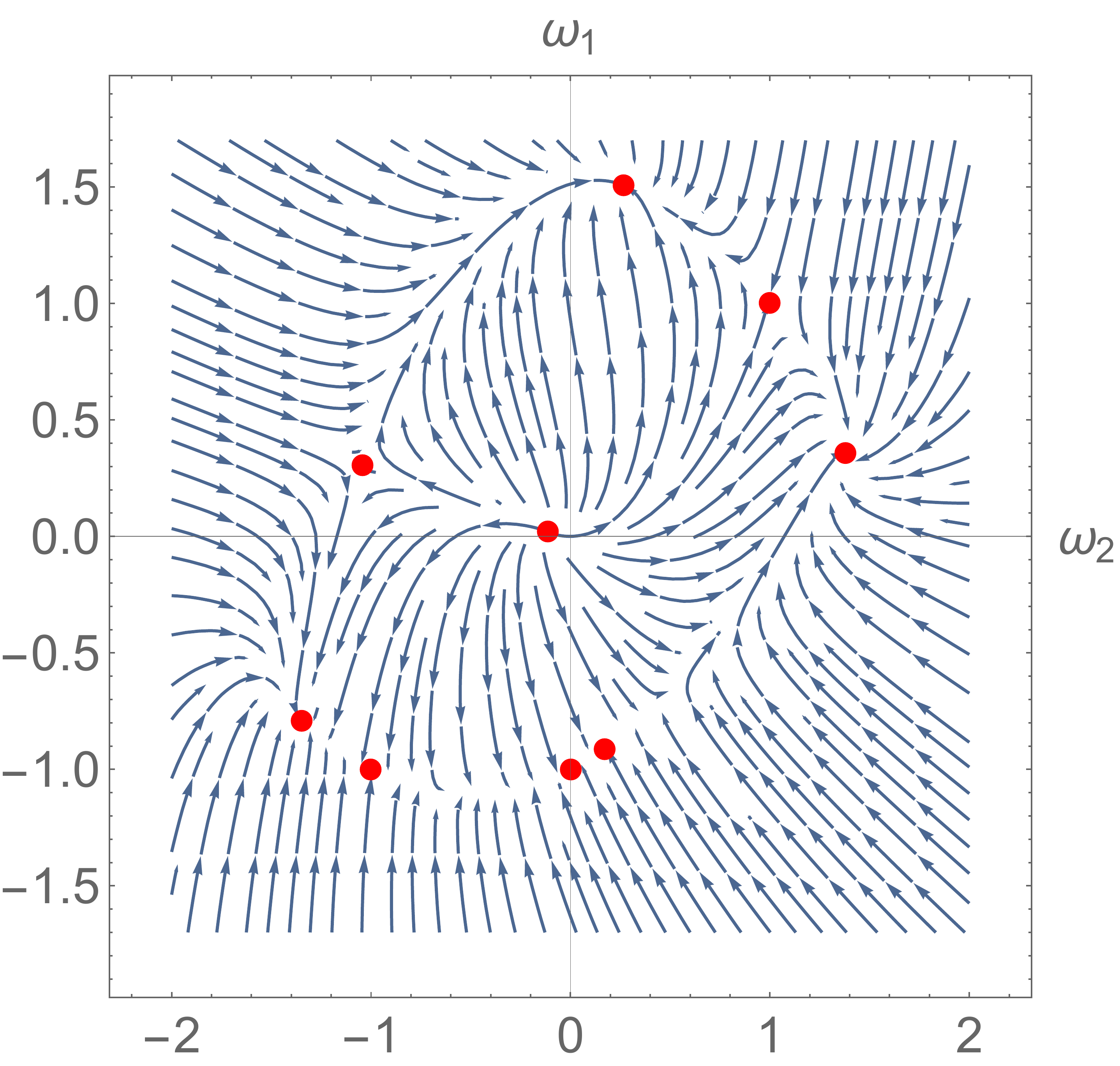}
		\caption{$N_f=3$ (picture is generic for $N_f>2$)}
		\label{fig:origin_2}
	\end{subfigure}
	\caption{Fixed points in terms of $\omega_1,\omega_2$ for small $\lambda$ near to origin.}
	\label{fig:beta_Nf_origin}
\end{figure}

For $N_f=2$, the results of \cite{Avdeev:1992jt} give eight fixed points close to the origin (shown in Figure \ref{fig:origin_1}). In addition, there are two fixed points at $(\omega_2,\omega_1)=\left(\pm\frac{\sqrt{48}}{\pi\grl},\pm\frac{\sqrt{48}}{\pi\grl}\right)$ and one at $(\omega_2,\omega_1)=(\infty,\infty)$. There are four other fixed points for which the finite value of one of the two parameters $\omega_2$,$\omega_1$ cannot be determined, which correspond to $(\tilde\omega_2,\tilde\omega_1)=(\pm\sqrt{48},0)$, $(\tilde\omega_2,\tilde\omega_1)=(\infty,0)$ and $(\tilde\omega_2,\tilde\omega_1)=(0,\infty)$.

For $N_f>2$ there are \cite{Avdeev:1992jt} nine fixed points near the origin (see Figure \ref{fig:origin_2}), and again there are two fixed points at $(\omega_2,\omega_1)=\left(\pm\frac{\sqrt{48}}{\pi\grl},\pm\frac{\sqrt{48}}{\pi\grl}\right)$, and one at $(\omega_2,\omega_1)=(\infty,\infty)$, and the same four other fixed points for which only one of the two parameters $\omega_2$,$\omega_1$ can be determined. In addition, there are four extra fixed points which are at infinity for $\lambda=0$, and for small $\lambda$ behave as $c/\lambda$ for some constant $c$.

For all $N_f > 1$ the $\mc{N}=2$ fixed point is no longer stable, but there are other stable fixed points with $\omega_i$ of order one. One of these is expected to appear on the domain walls of $4d$ SQCD with $N_f$ flavors \cite{Bashmakov:2018ghn}. The only other stable fixed point at large finite $N$ is the one that starts at $\lambda=0$ from $(\tilde{\omega}_2,\tilde{\omega}_1)=(\infty,0)$.

We can use parity arguments to constrain the leading-order behavior in $\lambda$ of the roots, just as we did in the $N_f=1$ case in Section \ref{sec:qualitative_lambda_positive}.
Furthermore, using the duality \eqref{eq:dual_Nf} one can map these fixed points into fixed points near $\lambda=1$. One can then try to connect the fixed points at $\lambda=0$ with those at $\lambda=1$ in order to try and conjecture the behavior of the fixed points for general $\lambda$. However, since we cannot find the $\omega_1$ values of all of the fixed points, we cannot find all of the fixed points at $\lambda=1$, and so it is difficult to connect the fixed points at weak and strong coupling.

As in the $N_f=1$ case, for infinite $N$ we can rewrite the theories using variables that are more appropriate for large $\omega$ by adding singlet superfields, and for finite $N$ we conjecture that the theories with the extra superfields still flow to the same fixed points, at least for large enough $N$. The form of the superpotential \eqref{eq:new_couplings} suggests that we can add an $SU(N_f)$-singlet superfield $H$ coupled to $\mc{J}$, an $SU(N_f)$-adjoint superfield $H_{ij}$ ($\tr(H_{ij})=0$) coupled to $\mc{J}_{ij}$, or both, giving three extra dual descriptions for each of our CS-matter theories. For example, the theory with both types of superfields added, which gives a smooth description of the region near the fixed point at $(\omega_2,\omega_1)=(\infty,\infty)$, is
\begin{equation}
\label{eq:singlets_W}
W=H \mc{J} -\frac{N_f N}{4\pi\lambda\omega_2}H^2 + H_{ij} \mc{J}_{ij}  -\frac{N}{4\pi\lambda\omega_1}H_{ij}^2.
\end{equation}
This makes it natural to view the space of couplings $\omega_1$, $\omega_2$ as a compact space with the topology of a product of two circles, and with some finite number of fixed points on this space. The relevant deformations in the description \eqref{eq:singlets_W} are terms in the superpotential linear in $H$ and in $H_{ij}$. These alternative descriptions make it clear that for $\lambda=0$, the fixed points at $({\tilde \omega}_2,{\tilde \omega}_1)=(0,\infty),(\infty,0),(\infty,\infty)$ have time-reversal symmetry, explaining why the beta functions vanish there.

As a final comment, we note that using time-reversal symmetry, we can once again find some of the fixed points at the self-dual value $\lambda=\frac12$, in analogy with the $N_f=1$ case in Section \ref{sec:beta_conjecture}. A similar argument to the one there shows that the points 
\begin{equation}\label{eq:emergent_T_Nflarge}
(\omega_1,\omega_2)=(-3,-3),(1,-3),(-3,1),(1,1)
\end{equation}
all have emergent time-reversal symmetry $T'$ at $\lambda=\frac12$ (note that the last point is the $\mc{N}=2$ point). Under this symmetry, a small deformation $(\delta \omega_1,\delta \omega_2)$ around any one of the four points \eqref{eq:emergent_T_Nflarge} transforms as
\begin{equation}
T'(\delta \omega_1,\delta \omega_2)=(-\delta \omega_1,-\delta \omega_2).
\end{equation}
which means that this deformation breaks the symmetry, and so it cannot be generated along the RG flow emanating from this point. In other words, the points \eqref{eq:emergent_T_Nflarge} are all fixed points of the RG flow at $\lambda=\frac12$.

We can also discuss the existence of moduli spaces of the theories \eqref{eq:emergent_T_dualities} at large $N$, as a generalization of the $N_f=1$ result discussed in Section \ref{sec:beta_discussion}. A similar argument shows that of the four fixed points \eqref{eq:emergent_T_Nflarge} with emergent time-reversal symmetry, only the point $(\omega_1,\omega_2)=(-3,-3)$ is expected to have a moduli space at large $N$, because at this point all the operators that can appear in the superpotential are even under the emergent time-reversal symmetry. Once again, we expect this moduli space to appear for finite $N$ in the corresponding fixed points of the theories \eqref{eq:emergent_T_dualities} as well.

\section*{Acknowledgements}

We would like to thank  F.~Benini, K.~Inbasekar, S.~Jain, Z.~Komargodski, S.~Minwalla, and T. Sharma for many useful discussions and for comments on a draft of this paper. This work was supported in part  by an Israel Science Foundation center for excellence grant (grant number 1989/14) and by the Minerva foundation with funding from the Federal German Ministry for Education and Research. OA is the Samuel Sebba Professorial Chair of Pure and Applied Physics.


\pagebreak

\begin{appendices}
	
	\section{Conventions}\label{app:conventions}
	We mostly follow \cite{Gates:1983nr}, and some additional conventions from \cite{Inbasekar:2015tsa}.
	
	\subsection{Lorentzian}
	The metric signature is $ \eta_{\mu\nu}=\{-,+,+\} $. Spinors are $\psi^\alpha$, and 
	spinor indices are raised and lowered with 
\begin{equation}
C_\ab=-C^\ab=  (\sigma_2)_\ab  \then C_\ab C^{\ga\de}=\de_\al{}^\de\de_\be{}^\ga-\de_\be{}^\de\de_\al{}^\ga.
\end{equation}
Explicitly, for a spinor $\psi$ we have
\begin{equation}\label{psic}
\psi^\al=C^\ab\psi_\be~~,~~\psi_\be =\psi^\al C_\ab~~,~~\psi^2\equiv \frac12\psi^\al\psi_\al=i\psi^+\psi^-\ .
\end{equation}
	Vectors are real and symmetric $V_{\alpha\beta}$ with $V_{\alpha\beta}=V_\mu \gamma^\mu_{\alpha\beta}$, with $\gamma^\mu$ the $\gamma$ matrices.
	We will be using light-cone coordinates, where
	\begin{equation} k_\pm=\frac{\pm k_0+k_1}{\sqrt 2},\qquad k_s^2=2k_+k_-=k_1^2-k_0^2,\qquad k^2=k_s^2+k_z^2.
	\end{equation}
	We choose a basis for the gamma matrices such that a spinor matrix $\mathbf{p}_{\alpha\beta}$ is related to a vector $p_\mu$ by:
	\begin{equation} \mathbf{p}_{11}=p_0+p_1=\sqrt{2}p_+,\qquad \mathbf{p}_{22}=p_0-p_1=-\sqrt{2}p_-,\qquad \mathbf{p}_{12}=p_z ,
	\end{equation}
	so that
	\begin{equation} p^2=p_z^2+2p_+p_-=\mathbf{p}_{12}^2-\mathbf{p}_{11}\mathbf{p}_{22}=-\det (\mathbf{p}), 
	\end{equation}
	\begin{equation}
	p\cdot k= p_z k_z+(p_+k_-+p_-k_+)=\mathbf{p}_{12}\mathbf{k}_{12}-\frac12(\mathbf{p}_{11}\mathbf{k}_{22}+\mathbf{p}_{22}\mathbf{k}_{11}) .
	\end{equation}
	
	\subsection{Euclidean}
	The metric signature is $ \eta_{\mu\nu}=\{+,+,+\} $. This means that we rotate $k_0\rightarrow ik_0$.
	Correspondingly, we have
	\begin{equation} k_\pm=\frac{\pm i k_0+k_1}{\sqrt 2},\qquad k_s^2=2k_+k_-=k_1^2+k_0^2,\qquad k^2=k_s^2+k_z^2.
	\end{equation}
	As before, we have:
	\begin{equation} \mathbf{p}_{11}=p_0+p_1=\sqrt{2}p_+,\qquad \mathbf{p}_{22}=\sqrt{2}p_-,\qquad \mathbf{p}_{12}=p_z. \end{equation}
	
	\subsection{Superspace Conventions}

The spinor derivatives $D_\alpha$ in superspace obey the usual algebra:
\begin{equation}\label{Dalg}
\{D_\al,D_\be\}=2i\pa_\ab,
\end{equation}
and again the convention is $D^2=\frac12D^\al D_\al$. The following identities are very useful:
\begin{equation}
\begin{gathered} \label{didents}
D_\al D_\be =i\pa_\ab -C_\ab D^2~~,~~ D^\al D_\be D_\al =0~~,~~ D^2 D_\al =- D_\al D^2 = i \pa_\ab D^\be~~,\\
\pa^\ab\pa_{\ga\be}=\de_\ga{}^\al \square ~~,~~(D^2)^2=\square~~,~~ \square \equiv \frac12 \pa^\ab\pa_\ab.
\end{gathered}
\end{equation}

The free scalar superfield action is
\begin{equation}\label{SX}
S=\frac12 \int d^2\ta ~(\Phi D^2 \Phi +m \Phi^2)=\frac12 \int d^2\ta \left(-\frac12D^\al \Phi D_\al \Phi +m\Phi^2\right)~~.
\end{equation}
This superfield can be expanded in components as 
\begin{equation} 
\Phi=\phi+\theta\psi-\theta^2 F,
\end{equation}
with $\phi$ a real boson, $\psi$ a real fermion and $F$ an auxiliary field.

The gauge multiplet is described by covariantizing the spinor and vector derivatives:
\begin{equation}\label{YM}
\{\na_\al,\na_\be\}=2i\na_\ab~~,~~ \na_\al\equiv D_\al -i \G_\al~~,
\end{equation}
where the generators are hermitian (which is why there is an $i$ in the definition of the gauge covariant derivative $\na_\al$). The gauge multiplet $\Gamma^\alpha$ can also be expanded in components:
\begin{equation} \Gamma^\alpha=\chi^\al-\theta^\al B+i\theta^\be A_\be^\al -\theta^2(2\lambda^\al-i\pa^{\al\be}\chi_\be) 
\end{equation}
with $A$ the gauge field, $\lambda^\al$ the gaugino, $B$ an auxiliary scalar and $\chi^\al$ an auxiliary fermion. 

\subsection{Chern-Simons Levels}

We now describe our regularization procedure for Chern-Simons levels. We mostly follow the notation appearing in the literature (for a detailed review see \cite{Aharony:2018pjn}), and consider only $SU(N)$ gauge theories for simplicity.

We start with the non-supersymmetric Chern-Simons theories. There are two common regularizations appearing in the literature. One is Yang-Mills regularization, by adding a Yang-Mills term for the gauge fields, making them free at high energies, for which the UV level is denoted by $k$. The other is dimensional regularization, for which the level is denoted by $\kappa$. The former is often used when discussing small $k,N$ dualities, while the latter is common when discussing the large-$N$ \rq t Hooft limit. The relation between these two is the following: a Chern-Simons theory at level $\kappa$ in dimensional regularization is the same as an identical theory with Yang-Mills regularization at level $k$, with 
\begin{equation} \kappa = \text{sign}(k)(|k|+N).
\end{equation}
 
Next, we describe the case for $\mc{N}=1$ theories. There is an additional subtlety here due to the gaugino, which must be integrated out, and the CS level shifts accordingly. If we start with a pure $\mc{N}=1$ $SU(N)_{k^{\mc{N}=1}}$ CS theory, it is equivalent to a non-supersymmetric $SU(N)$ CS theory with level\footnote{We are assuming here that $k^{\mc{N}=1}$ is large enough so that SUSY is not broken \cite{Witten:1999ds}.}
\begin{equation}
 k = \text{sign}(k^{\mc{N}=1})\left(|k^{\mc{N}=1}|-\frac{N}{2}\right),\qquad \kappa=k^{\mc{N}=1}+\text{sign}(k^{\mc{N}=1})\frac{N}{2}.
 \end{equation}
In particular, the \rq t Hooft coupling $\lambda$ for an $\mc{N}=1$ theory $SU(N)_{k^{\mc{N}=1}}$ is $\lambda=\frac{N}{\kappa}=\frac{N}{k^{\mc{N}=1}+\frac{N}{2}}$.

Our theories sometimes have enhanced $\mc{N}=2$ SUSY, and so we view them as $\mc{N}=2$ theories. In this description, there is an extra gaugino that must be integrated out, so that a pure $\mc{N}=2$ $SU(N)_{k^{\mc{N}=2}}$ CS theory is equivalent to a non-supersymmetric $SU(N)$ CS theory with level
\begin{equation} 
k = \text{sign}(k^{\mc{N}=2})\left(|k^{\mc{N}=2}|-N\right),\qquad \kappa=k^{\mc{N}=2}.
\end{equation}

Finally, we discuss the result of integrating out massive fundamental matter fields. Integrating out a massive boson does not shift the level, while each fundamental fermion of mass $m$ that is integrated out leads to a shift by $\frac{\text{sign}(m)}{2}$ in the level $\kappa$ of the low-energy CS theory.

In order to comply with the notation in the literature, in the present paper we use $\kappa$ for dualities in the large-$N$ \rq t Hooft limit, and $k^{\mc{N}=1}$ for finite-$N$ dualities (dropping the $\mc{N}=1$ superscript when the context is clear).

\section{Calculating the $(a)-(e)$ Diagrams}\label{app:diagrams}

\subsection{Quick Introduction to $3d$ $\mc{N}=1$ Supergraphs}

We study the superspace Lagrangian
\begin{equation}
{\bar \grF}_a D^2 \grF_a + \frac{\tilde\omega}{N}(|\grF_a|^2)^2,
\end{equation}
with $a=1,\cdots,N$. The $\grF$ propagator is \begin{equation}\label{eq:propagator}
\langle {\bar\grF}_a(p,\theta) \grF_b(-p,\theta') \rangle= \frac{\grd_{ab}D^2}{p^2}\grd(\theta-\theta').
\end{equation}
Interaction vertices are taken care of in a similar fashion to the non-SUSY case.

One main part of calculating supergraphs is the D-algebra. Each diagram comes with some $D$'s acting on some propagators, but only diagrams with specific configurations of $D$'s are non-vanishing. A non-vanishing diagram must have exactly one factor of $D^2$ acting inside each of its loops. This is due to the following identities:
\begin{align}
\delta(\theta-\theta')\delta(\theta'-\theta)&=0,\\
\delta(\theta-\theta')D_\alpha \delta(\theta'-\theta)&=0,\\
\delta(\theta-\theta')D^2 \delta(\theta'-\theta)&=\delta(\theta-\theta').
\end{align}
In other words, having no factors of $D$ in a loop or exactly one factor of $D$ in a loop will make it vanish (note that using the superspace identities \eqref{didents}, 3 factors of $D$ or higher can always be reduced to either 0,1 or 2 factors of $D$).

\begin{figure}[t]
	\centering
	\includegraphics[width=0.55\linewidth]{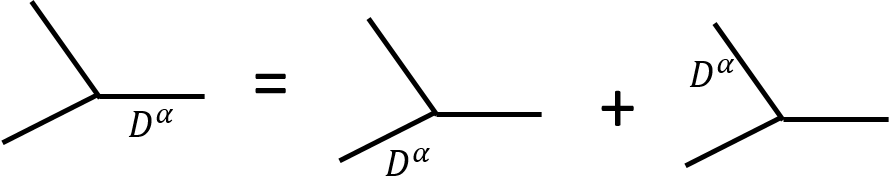}
	\caption{Integration by parts in superspace}
	\label{fig:ibp}
\end{figure}

The game is thus to put all of the factors of $D$ acting inside a certain loop on the same propagator. The diagrams that do not vanish then must have exactly one factor of $D^2$ inside each loop. To get all of the factors of $D$ to act on one propagator, we use integration by parts. This means that at a specific vertex, we can transfer a factor of $D$ that acts on one propagator to the other propagators attached to that same vertex. A diagrammatic version of this operation (for a $\grF^3$ interaction) is shown in Figure \ref{fig:ibp} (up to signs).
Using integration by parts we can get to a point where each loop has all of its $D$ factors acting on a a single propagator (or on external legs), and then we can throw out this diagram unless each loop has exactly one $D^2$.

\begin{figure}[t]
	\centering
	\includegraphics[width=0.7\linewidth]{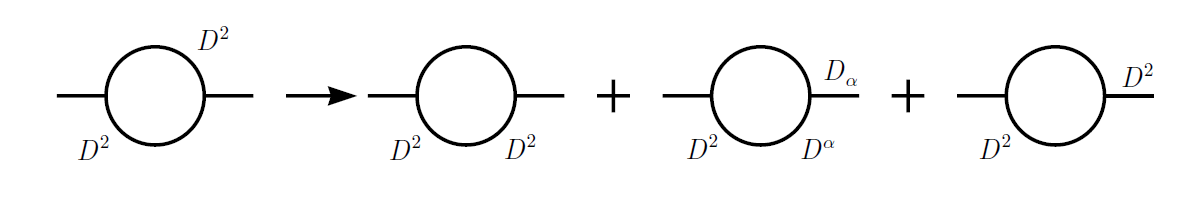}
	\caption{A one-loop diagram in $\Phi^3$ theory}
	\label{fig:example1}
\end{figure}

An example of calculating a specific diagram (again using $\Phi^3$ interactions) is shown in Figure \ref{fig:example1}.
Each (internal) propagator starts with a $D^2$ due to \eqref{eq:propagator}, and we integrate by parts the top propagator such that all $D$'s act on exactly one propagator inside the loop. We find that only the rightmost diagram contributes, since it is the only one with the correct configuration of $D$'s inside the loop.

\subsection{Comments on the Calculations}

\begin{itemize}
	\item For our calculation of log-divergent terms in $\langle \bar\grF \grF \bar\grF \grF \rangle$, we only need to calculate diagrams that have no $D$'s acting on external $\grF$ legs (since these would be corrections to the Kahler potential and not the superpotential). Note that this does not prohibit $D$'s from acting on external chains, since the chains include factors of $D$'s themselves.
	\item We keep all factors of $N$ implicit.
	\item Some integrals that appear multiple times in the calculation are:
	\begin{equation}
	\int\frac{d^{3}k}{(2\pi)^3}\frac{1}{k^2\left(k+p\right)^{2}}	=\frac{1}{8\left|p\right|},
	\end{equation}
	\begin{equation}
	\int\frac{d^3 k}{(2\pi)^3} \frac{1}{k^2\left(k+p_{1}\right)^{2}\left(k-p_{3}\right)^{2}}=\frac{1}{8\left|p_{1}\right|\left|p_{1}+p_3\right|\left|p_{3}\right|}.	\end{equation}
	We encounter only one diverging integral, of the form $\int\frac{d^3k}{(2\pi)^3}\frac{1}{|k|(p+k)^2}$. Using dimensional regularization, this is
	\begin{equation}
	\int\dpi{k}\frac{1}{|k|(p+k)^2}=\frac{1}{2\pi^2}\frac{1}{p^{\gre}\gre}(1+O(\gre)).
	\end{equation}
	\item The numerical factors we find in the following are for diagrams which contribute to $\langle |\Phi_a|^2|\Phi_b|^2 \rangle$ with $a\neq b$. In our normalization for $\tilde\omega$, the tree level result for this is $2{\tilde \omega}$.
\end{itemize}

\subsection{Calculation of the Diagrams}
\subsubsection{Type (a)}
There are two types of type $(a)$ diagrams. The D-algebra for the first type appears in Figure \ref{fig:typeanew} (here and in the following, we keep only diagrams which may diverge). Ignoring the external legs, the diverging terms in these diagrams are:
\begin{figure}[t]
	\centering
	\begin{subfigure}{0.8\linewidth}
		\centering
		\includegraphics[width=\textwidth]{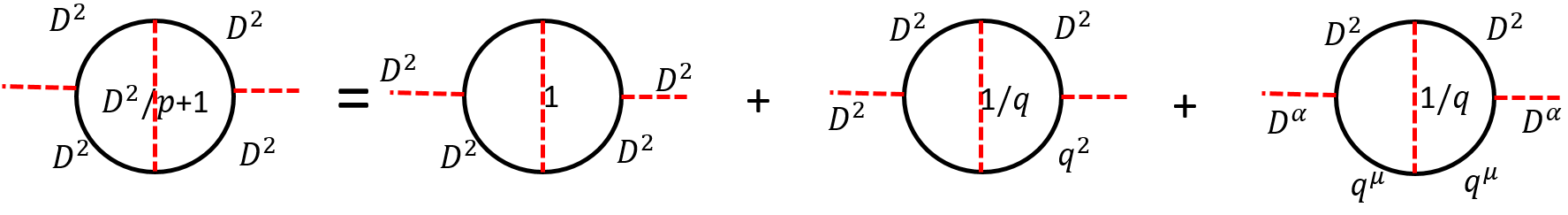}
		\caption{}
		\label{fig:typeanew}
	\end{subfigure}
	\qquad
	\begin{subfigure}{0.7\textwidth}
		\centering
		\includegraphics[width=\textwidth]{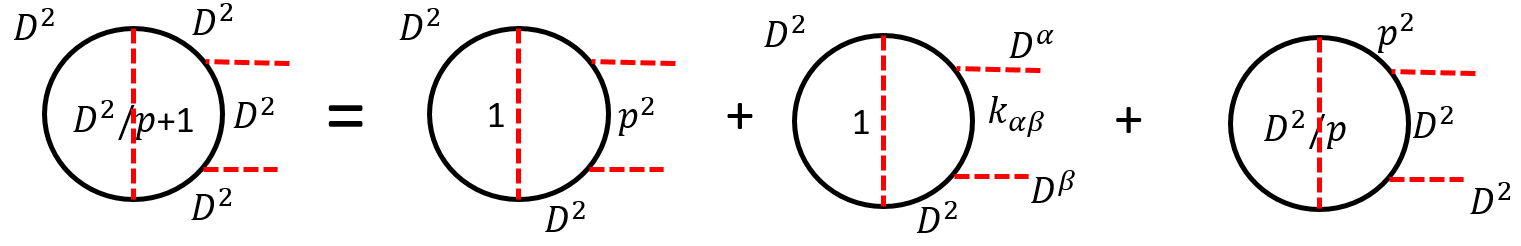}
		\caption{}
		\label{fig:typea1new}
	\end{subfigure}
	\label{fig:typea}
	\caption{Type $(a)$ diagrams.}
\end{figure}
\begin{equation}
0,\qquad -2a_1\frac{1}{8}\frac{1}{2\pi^2\gre}\frac{D^2}{|p|}, \qquad 
0.
\end{equation}
The second type appears in Figure \ref{fig:typea1new}. These contribute
\begin{equation} 0,\qquad 0,\qquad -2a_1 \frac{1}{8}\frac{1}{2\pi^2\gre}\frac{D^2}{|p|}. \end{equation}
So the full $(a)$-type contribution (including the external leg factors) is
\begin{equation} -\frac{a_1}{4\pi^2\gre}
	\frac{D^2}{|p|}\left(a_0+a_1\frac{D^2}{|p|}\right)^2. \end{equation}

\subsubsection{Type (b)}
\begin{figure}[t]
	\centering
	\begin{subfigure}{0.23\linewidth}
		\centering
		\includegraphics[width=\textwidth]{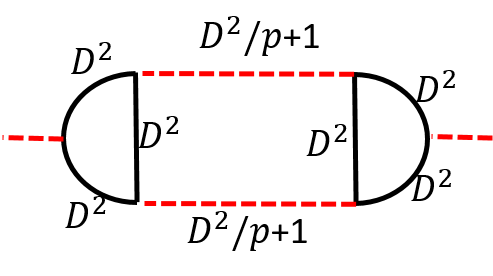}
		\caption{}
		\label{fig:typeb1}
	\end{subfigure}
	\qquad
	\begin{subfigure}{0.55\textwidth}
		\centering
		\includegraphics[width=\textwidth]{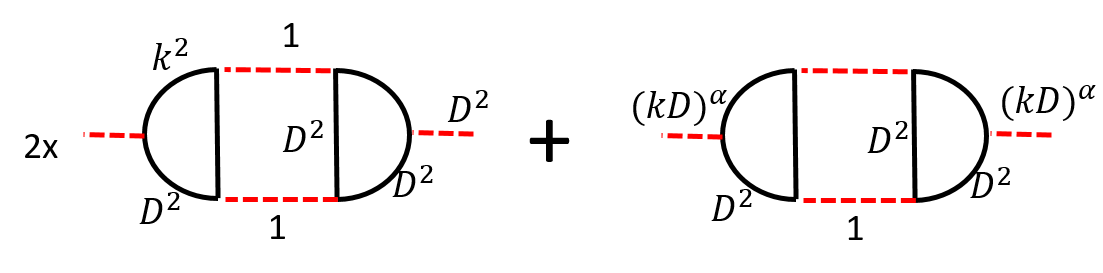}
		\caption{}
		\label{fig:typeb11}
	\end{subfigure}
	\qquad
	\begin{subfigure}{0.67\textwidth}
	\centering
	\includegraphics[width=\textwidth]{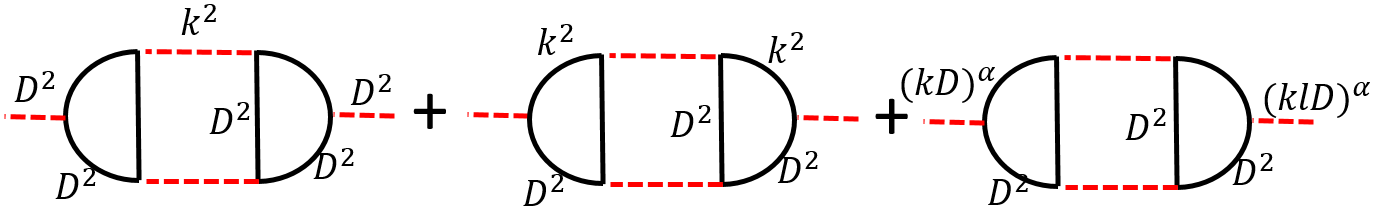}
	\caption{}
	\label{fig:typeb21}
	\end{subfigure}
	\qquad
	\begin{subfigure}{1\textwidth}
	\centering
	\includegraphics[width=\textwidth]{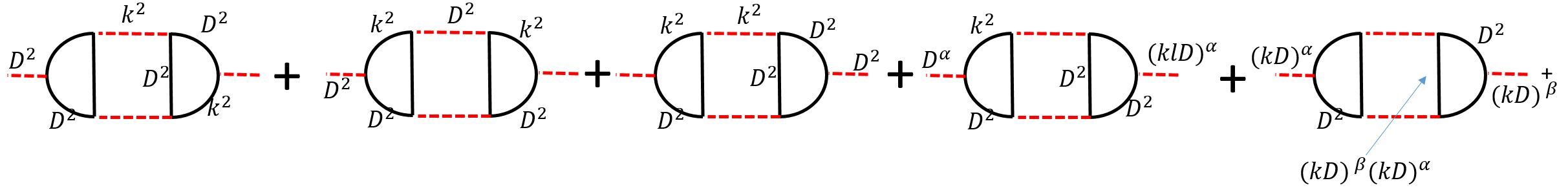}
	\caption{}
	\label{fig:typeb31}
	\end{subfigure}
	\label{fig:typeb}
	\caption{Type $(b)$ diagrams.}
\end{figure}

These appear in Figure \ref{fig:typeb1}.
There are 4 types of diagrams, where in each type we choose either the 1 or the $D^2$ from the top and bottom chains. We do each type separately. 

The $(b)_{(1,1)}$ appear in Figure \ref{fig:typeb11} and contribute:
\begin{equation}
-2\cdot 2a_0^2\frac18\frac18\frac{D^2}{2\pi^2\gre|p|},\qquad 0.
\end{equation}
The $(b)_{(1,D^2)}$
diagrams appear in Figure \ref{fig:typeb21} and contribute:
\begin{equation}
2a_0a_1\frac18\frac18\frac{1}{2\pi^2\gre},\qquad 2a_0a_1\frac18\frac18\frac{1}{2\pi^2\gre}, \qquad 0.
\end{equation}
The $(b)_{(D^2,1)}$ Diagrams are the same as the $(b)_{(1,D^2)}$ diagrams.
Finally, the $(b)_{(D^2,D^2)}$ diagrams appear in Figure \ref{fig:typeb31}. These contribute
\begin{equation}
2a_1^2\frac{D^2}{128\pi^2\gre|p|}, \qquad 
0, \qquad
0, \qquad
2\frac{a_1^2}{128\pi^2\gre}\frac{D^2}{|p|} , \qquad
0.
\end{equation}

So the full $(b)$-type contribution is (including the two external chains)
\begin{equation} 
-\frac{1}{32\pi^2\gre}\left( a_0+a_1\frac{D^2}{|p|}
\right)^2 \frac{D^2}{|p|}
\left( a_0+a_1\frac{D^2}{|p|}\right)^2 .
\end{equation}

\subsubsection{Type (c)}
\begin{figure}[t]
	\centering
	\includegraphics[width=0.85\linewidth]{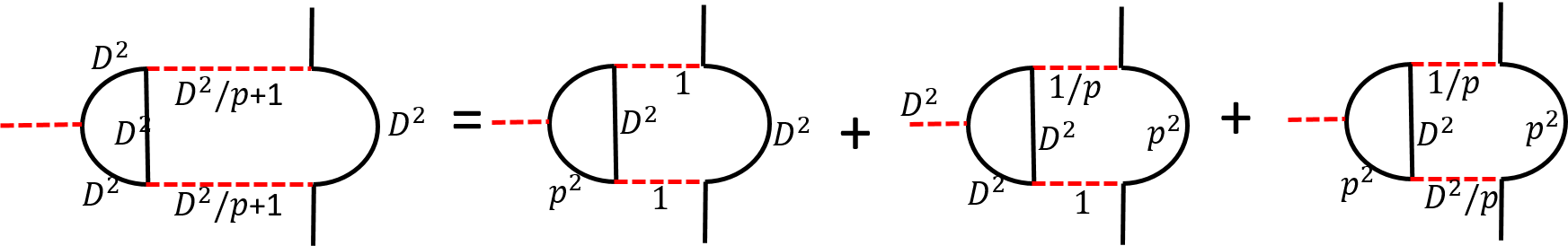}
	\caption{Type $(c)$ diagrams.}
	\label{fig:typecnew}
\end{figure}

The diagrams appear in Figure \ref{fig:typecnew}.
The diagrams contribute:
\begin{equation}
-4a_0^2\frac18\frac{1}{2\pi^2\gre},\qquad
-4\cdot 2a_0a_1\frac18\frac{1}{2\pi^2\gre}\frac{D^2}{|p|}, \qquad
4a_1^2\frac18\frac{1}{2\pi^2\gre},
\end{equation}
so the total contribution (including the external chain) is
\begin{equation}
-\frac{1}{4\pi^2\gre}\left(a_0+a_1\frac{D^2}{|p|}
	\right)^3.
\end{equation}

\subsubsection{Type (d)}
\begin{figure}
	\centering
	\includegraphics[width=0.15\linewidth]{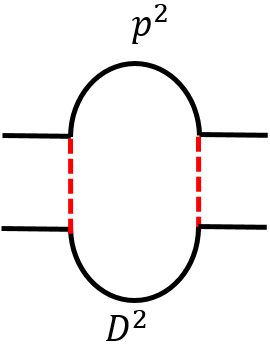}
	\caption{Type $(d)$ diagrams.}
	\label{fig:typednew}
\end{figure}

After doing the D-algebra, the only relevant diagram appears in Figure \ref{fig:typednew}. Its contribution is
\begin{equation}
 -\frac{2a_0a_1}{\pi^2\gre}.
\end{equation}

\subsubsection{Type (e)}

\begin{figure}[t]
	\centering
	\begin{subfigure}{0.15\linewidth}
		\centering
		\includegraphics[width=\textwidth]{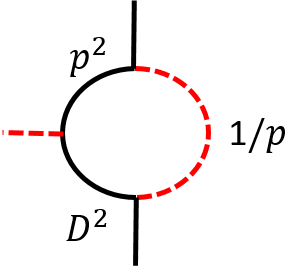}
		\caption{}
		\label{fig:typeenew_1}
	\end{subfigure}
	\qquad
	\begin{subfigure}{0.73\textwidth}
		\centering
		\includegraphics[width=\textwidth]{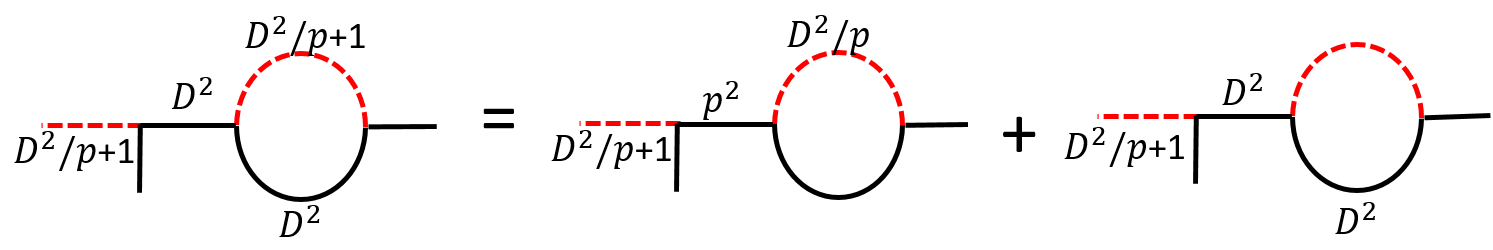}
		\caption{}
		\label{fig:typeenew_2}
	\end{subfigure}
	\label{fig:typeenew}
	\caption{Type $(e)$ diagrams.}
\end{figure}

There are two $(e)$-type diagrams. After doing the D-algebra, the diagram contributing to the logarithmic terms from the first type appears in Figure \ref{fig:typeenew_1}.
This diagram contributes
\begin{equation} 
-\frac{a_1}{\pi^2\gre}.
\end{equation}
The second type appears in Figure \ref{fig:typeenew_2}. This is a correction to the external $\Phi$ propagator, but we include it here since we consider the Callan-Symanzik equation for the full connected 4-point function. These contribute
\begin{equation} 
 -\frac{2a_1}{\pi^2\gre},\qquad
0.
\end{equation}
So the total contribution (including the external chain) is
\begin{equation}
-\frac{3a_1}{\pi^2\gre}\left(a_0+a_1\frac{D^2}{|p|}\right) .
\end{equation}

\section{Conjecture for the $\mc{N}=1$ Version of Maldacena-Zhiboedov}\label{app:MZ}

Our results for the 3-point functions in Section \ref{sec:JJJ} suggest a possible generalization of the results of Maldacena and Zhiboedov \cite{Maldacena:2011jn,Maldacena:2012sf} to $\mc{N}=1$ supersymmetric theories (such a generalization may be studied directly by their methods, but this was not yet done as far as we know). The main result of \cite{Maldacena:2012sf} can be stated as follows. Consider a $2+1d$ CFT which has a large $N$ expansion, and whose large $N$ spectrum of single-trace operators includes:
\begin{itemize}
	\item a single spin-two conserved current.
	\item a series of higher spin currents\footnote{We use lowercase $j$ to denote non-SUSY currents and uppercase $J$ to denote SUSY current multiplets.} $j_s$ with even spins $s\in 2\mb{Z}$ which are approximately conserved (by which we mean that their twists differ from $1$ by corrections of order $1/N$).
	\item a single scalar operator $j_0$. Theories in which this operator has dimension $2$ at leading order in $1/N$ are called quasi-fermion, while theories in which it has dimension 1 are called quasi-boson.
\end{itemize}
Then, at leading order in $1/N$, the three-point functions of these operators are constrained to have a specific form. In particular, these three-point functions are made up of three structures:
\begin{equation}\label{eq:MZ}
\langle j_{s_1}j_{s_2}j_{s_3} \rangle = c_1\langle j_{s_1}j_{s_2}j_{s_3} \rangle_{free,bos}+c_2\langle j_{s_1}j_{s_2}j_{s_3} \rangle_{free,fer}+c_3\langle j_{s_1}j_{s_2}j_{s_3} \rangle_{odd}
\end{equation} 
where the first two structures are the result in the free theory of a single boson and a single Majorana fermion respectively, and all the coefficients may be determined up to a single constant (for quasi-fermion theories) or up to two constants (for quasi-boson theories). When all $s_i > 0$, the coefficients are independent of the spin.

Our $\mc{N}=1$ theories with $N_f=1$ obey only the second assumption (explicitly, we have a single higher-spin approximately-conserved supercurrent multiplet $J_s$ for $s=0,\frac12,1,\cdots$ \cite{Nizami:2013tpa}). In particular, our theories include two scalar single-trace operators $\bar\phi^a\phi_a$,$\bar\psi^a\psi_a$, and two spin-2 operators which are conserved at large $N$. We thus might not expect the three-point functions to have a simple form. However, the results in the present paper suggest a generalization of the theorem of \cite{Maldacena:2012sf} to theories with $\mc{N}=1$ SUSY. In particular, if we think of $J=\bar{\Phi}^a \Phi_a$ as a supersymmetric generalization of the scalar single-trace operator $j_0$, then we have found that its large $N$ three-point functions for a wide range of theories (with $N_f=1$ and  with different values of $\omega$) are made up of only two structures:
\begin{equation} \label{eq:MZ_N0}
\langle JJJ \rangle = \alpha\langle JJJ \rangle_{free}+\beta\langle JJJ \rangle_{odd} ,
\end{equation}
where the first structure is the result in the free theory ($\lambda=\omega=0$) of a single matter multiplet. We conjecture a similar form for three-point functions of higher-spin currents as well:
\begin{equation}\label{eq:MZ_N1}
\langle J_{s_1}J_{s_2}J_{s_3} \rangle = \alpha_{s_1s_2s_3}\langle J_{s_1}J_{s_2}J_{s_3} \rangle_{free}+\beta_{s_1s_2s_3}\langle J_{s_1}J_{s_2}J_{s_3} \rangle_{odd} .
\end{equation} 
In our theories, the coefficients $\alpha_{s_1s_2s_3},\beta_{s_1s_2s_3}$ are functions of a single parameter $\lambda$ when all $s_i>0$, and may also depend on $\omega$ when at least one $s_i$ vanishes. We conjecture that general arguments of approximately-conserved higher-spin symmetry imply such a dependence on one or two parameters, that would appear in the non-conservation equations of the high-spin currents as in \cite{Maldacena:2012sf}. We will present evidence that when all spins $s_i$ are even integers, the coefficients $\alpha_{s_1s_2s_3},\beta_{s_1s_2s_3}$ are independent of the $s_i$. We conjecture that more generally, the coefficients are independent of the $s_i$ also when all $s_i> \frac{1}{2}$, but we do not yet have evidence for this more general claim.

Let us use the non-SUSY results to show that the bottom component of $\langle J_{s_1}J_{s_2}J_{s_3} \rangle$ obeys this conjecture when all spins $s_i$ are positive even integers (we use $A|$ to denote the bottom component of $A$). We can decompose $J_s|$ in terms of the corresponding currents in the bosons-only and fermions-only theories, $j_s^{b}$ and $j_s^{f}$, by $J_s|=j_s^{f}+j_s^{b}$. Setting $\omega=-1$, a discussion similar to the one in Section \ref{sec:JJ} shows that the contributions from the bosons-only and fermions-only theories detach in planar diagrams\footnote{This is clear at $\omega=-1$, but one can show that if all $s_i>0$ then these correlators do not depend on $\omega$, and so the contributions detach for all $\omega$.}, and we find 
\begin{equation}
\langle J_{s_1}J_{s_2}J_{s_3} \rangle|=
\langle j_{s_1}^{b}j_{j_2}^{b}j_{j_3}^{b} \rangle+\langle j_{s_1}^{f}j_{s_2}^{f}j_{s_3}^{f} \rangle .
\end{equation}
Taking the bottom component of our conjecture \eqref{eq:MZ_N1} and plugging in this equation, we find
\begin{equation}
\langle j_{s_1}^{b}j_{s_2}^{b}j_{s_3}^{b} \rangle+
\langle j_{s_1}^{f}j_{s_2}^{f}j_{s_3}^{f} \rangle =  \alpha_{s_1s_2s_3}\left(\langle j_{s_1}^{b}j_{s_2}^{b}j_{s_3}^{b} \rangle_{free}+
\langle j_{s_1}^{f}j_{s_2}^{f}j_{s_3}^{f} \rangle_{free}\right)+ \beta_{s_1s_2s_3}\left(\langle j_{s_1}^{b}j_{s_2}^{b}j_{s_3}^{b} \rangle_{odd}+
\langle j_{s_1}^{f}j_{s_2}^{f}j_{s_3}^{f} \rangle_{odd}\right)  
\end{equation}
In order for our conjecture to be correct, this equation for the non-SUSY three-point function must be obeyed. Using the general form \eqref{eq:MZ} and plugging in the correct coefficients $c_i$ for the bosons-only and fermions-only theories with the same value of $\lambda$ (see \cite{Maldacena:2012sf,Aharony:2012nh}), we find that this equation is indeed obeyed. Apart from giving us a nice consistency check, this calculation also allows us to find the coefficients $\alpha_{s_1s_2s_3},\beta_{s_1s_2s_3}$ for even spins $s_i$ in terms of the non-SUSY coefficients $c_i$ in \eqref{eq:MZ}. In particular, this proves that the coefficients $\alpha_{s_1s_2s_3},\beta_{s_1s_2s_3}$ are independent of the spins $s_i$ when the spins are all even integers.

Another consistency check can be performed using correlators of the form $\langle J_0 J_0 J_s\rangle$. In \cite{Nizami:2013tpa}
it was shown that superconformal invariance allows just one possible structure in this correlator for $s$ an even integer even, which is just the free structure. For $s$ an odd integer or half-integer, this correlator vanishes. This is in agreement with the conjecture above. Similarly, it was shown in \cite{Nizami:2013tpa} that correlators of the form $\langle J_1J_1J_1\rangle$ and $\langle J_{1/2}J_{1/2}J_{1/2}\rangle$ vanish, which trivially agrees with the conjecture.

It would be nice to confirm that a general analysis along the lines of  \cite{Maldacena:2012sf} indeed leads to  \eqref{eq:MZ_N0} and \eqref{eq:MZ_N1}, with the coefficients fixed in terms of one undetermined constant (two when one of the spins $s_i$ vanishes); our results express these coefficients in terms of $\lambda$ and $\omega$.
We note that for exactly conserved higher-spin currents, it was conjectured already in \cite{Nizami:2013tpa} that only two structures can appear. Our results seem to indicate that this is true also for approximately-conserved higher spin currents.

\section{The Gap Equations for $N_f>1$}\label{app:gap_eqs}

We solve the gap equations for $N_f>1$ with arbitrary masses. This analysis is a generalization of the gap equation solved in \cite{Inbasekar:2015tsa}, and we follow that calculation closely. Our superpotential is:
\begin{equation}
W_\Phi=m_{0,ij}(\bar{\Phi}_i\Phi_j) +\frac{\pi\omega_0}{\kappa}\left(\bar{\Phi}^i\Phi_i\right)^2+\frac{\pi\omega_1}{\kappa}\left(\bar{\Phi}^i\Phi_j\right) \left(\bar{\Phi}^j\Phi_i\right),
\end{equation}
and we use the $U(N_f)$ symmetry to diagonalize the mass matrix, so that the bare masses are 
\begin{equation}
m_{0,ij}=\text{diag}(m_{0,1},m_{0,2},...,m_{0,N_f}).
\end{equation}
It is useful to distinguish the symmetric-traceless part of this matrix and the trace part. Define
\begin{equation}
M_0=\tr (m_{0,ij}),\quad M_{0,ij}=m_{0,ij}-\frac{\delta_{ij}}{N_f}M_0 ,
\end{equation}
so that $\tr (M_{0,ij})=0$.

\begin{figure}[t]
	\centering
	\includegraphics[width=0.75\linewidth]{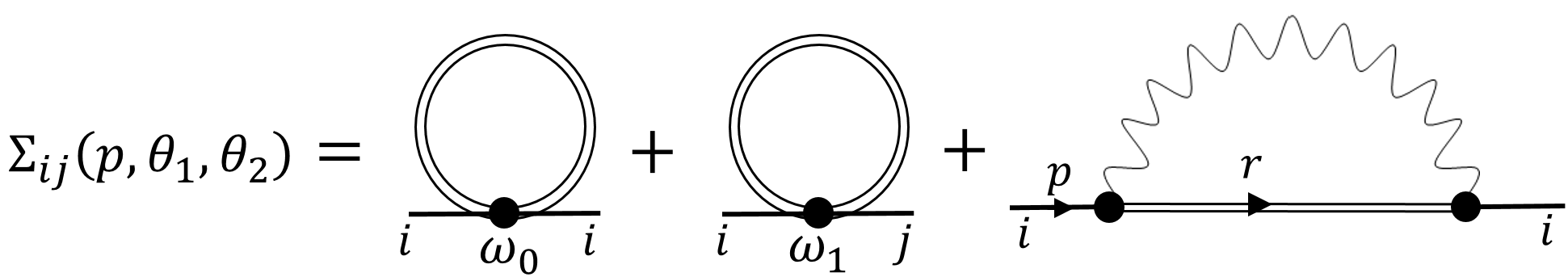}
	\caption{Diagrammatic version of the gap equation. A double line denotes the full matter propagator $ P(r,\theta_1,\theta_2)=-\frac{1}{D^2+m_0+\Sigma} $, and a wiggly line denotes the full gauge field propagator obtained in \cite{Inbasekar:2015tsa} (both propagators are at leading order in $1/N$).}
	\label{fig:GapEqNflargenew}
\end{figure}

Now we solve the gap equation for the self-energy $\Sigma_{ij}$. The diagrammatic version of the gap equation appears in Figure \ref{fig:GapEqNflargenew}.
Explicitly, following the derivation in \cite{Inbasekar:2015tsa}, this gives:
\begin{align}\nonumber
\Sigma_{ij}\left(p,\theta_{1},\theta_{2}\right)&=
2\pi\lambda\omega_0\delta_{ij}\int\frac{d^3r}{(2\pi)^3}\delta^2\left(\theta_1-\theta_2\right)\mbox{Tr}\left[P_{kl}\left(r,\theta_1,\theta_2\right)\right]+2\pi\lambda\omega_1\int\frac{d^3r}{(2\pi)^3}\delta^2\left(\theta_1-\theta_2\right)P_{ij}\left(r,\theta_1,\theta_2\right)-\\\nonumber
&\quad -2\pi\lambda\delta_{ij}\int\frac{d^3r}{(2\pi)^{3}}\delta^{2}\left(\theta_1-\theta_{2}\right)P_{ii}\left(r,\theta_1,\theta_2\right)\\
&=
2\pi\lambda\left[\omega_0\delta_{ij}\int\frac{d^{3}r}{\left(2\pi\right)^3}\delta^2\left(\theta_1-\theta_2\right)\mbox{Tr}\left[P_{kl}\left(r,\theta_1,\theta_2\right)\right]+\left(\omega_1-\delta_{ij}\right)\int\frac{d^3 r}{(2\pi)^3}\delta^2\left(\theta_1-\theta_{2}\right)P_{ij}\left(r,\theta_1,\theta_2\right)\right]
\label{gap_eqs}
\end{align}
where $P_{ij}$ is the full matter propagator. Since the final line does not depend on the momentum $p$, we guess a solution where $\Sigma_{ij}$ is constant. Furthermore, note that if we guess a diagonal solution for $\Sigma_{ij}$, then $P_{ij}$ becomes diagonal as well. We thus guess a solution of the form $\Sigma_{ij}=\text{diag}(m_1-m_{0,1},m_2-m_{0,2},...)$, which leads to $P_{ij}=\text{diag}(P_1,...,P_{N_f})$ with $P_i=\frac{D^2-m_i}{p^{2}+m_i}\delta^2\left(\theta_1-\theta_2\right)$.
The gap equations \eqref{gap_eqs} then reduce to
\begin{equation} 
m_i-m_{0,i}=2\pi\lambda\left(\omega_0\sum_j\int\frac{d^{3}r}{\left(2\pi\right)^{3}}\frac{1}{r^{2}+m_j}+(\omega_1-1)\int\frac{d^{3}r}{\left(2\pi\right)^{3}}\frac{1}{r^{2}+m_i}\right).
\end{equation}
Calculating the integrals using dimensional regularization, we find
\begin{equation}
m_i-m_{0,i}=
\left(1-\omega_1\right)\frac{\lambda\left|m_i\right|}{2}-\omega_0\frac{\lambda\sum_j\left|m_j\right|}{2}.
\end{equation}
The solution is:
\begin{equation}
m_i= \frac{ 2M_{0,i}}{\lambda \text{sign}(m_i) (\omega_1-1)+2}+
\frac{\frac{2}{N_f}M_0}{ \lambda \text{sign}(m_i) (\omega_2-1)+2},
\end{equation}
where $\omega_2=N_f\omega_0+\omega_1$. 
Note that the result has split into two separate contributions from the symmetric-traceless part and the trace part, which are each (almost) identical to the $N_f=1$ result \eqref{eq:mass_gap_Nf1}. It is thus immediate to show that this result is invariant under the proposed duality transformation \eqref{eq:dual_Nf}.

\end{appendices}

\printbibliography

\end{document}